\documentclass[12pt]{article}
\usepackage[latin1]{inputenc}
\usepackage{cite}
\usepackage{amsmath}
\usepackage{amsfonts}
\usepackage{amssymb}
\usepackage{graphicx}
\usepackage{geometry}
\usepackage{amssymb,epsfig}
\usepackage{hyperref}

\makeatletter
\renewcommand\section{\@startsection {section}{1}{\z@}%
                                 {-3.5ex \@plus -1ex \@minus -.2ex}%nn
                                   {2.3ex \@plus.2ex}%
                                   {\normalfont\large\bfseries}}
\renewcommand\subsection{\@startsection{subsection}{2}{\z@}%
                                   {-3.25ex\@plus -1ex \@minus -.2ex}%
                                     {1.5ex \@plus .2ex}%
                                     {\normalfont\bfseries}}
\renewcommand\subsubsection{\@startsection{subsubsection}{3}{\z@}%
                                   {-3.25ex\@plus -1ex \@minus -.2ex}%
                                     {1.5ex \@plus .2ex}%
                                     {\normalfont\itshape}}
\makeatother

\def\pplogo{\vbox{\kern-\headheight\kern -29pt
\halign{##&##\hfil\cr&{\ppnumber}\cr\rule{0pt}{2.5ex}&\ppdate\cr}}}
\makeatletter
\def\ps@firstpage{\ps@empty \def\@oddhead{\hss\pplogo}%
  \let\@evenhead\@oddhead % in case an article starts on a left-hand page
}%      The only change in \maketitle is \thispagestyle{firstpage} instead of \thispagestyle{plain}
\def\maketitle{\par
 \begingroup
 \def\thefootnote{\fnsymbol{footnote}}
 \def\@makefnmark{\hbox{$^{\@thefnmark}$\hss}}
 \if@twocolumn
 \twocolumn[\@maketitle]
 \else \newpage
 \global\@topnum\z@ \@maketitle \fi\thispagestyle{firstpage}\@thanks
 \endgroup
 \setcounter{footnote}{0}
 \let\maketitle\relax
 \let\@maketitle\relax
 \gdef\@thanks{}\gdef\@author{}\gdef\@title{}\let\thanks\relax}
\makeatother

\numberwithin{equation}{section}

\newcommand{\be}{\begin{equation}}
\newcommand{\bea}{\begin{eqnarray}}
\newcommand{\ee}{\end{equation}}
\newcommand{\eea}{\end{eqnarray}}

\newcommand{\Tr}{{\rm Tr}}

\newcommand{\yu}{y_u}
\newcommand{\yd}{y_d}
\newcommand{\yl}{y_l}
\newcommand{\yud}{y_u^\dagger}
\newcommand{\ydd}{y_d^\dagger}
\newcommand{\yld}{y_l^\dagger}
\newcommand{\gHd}{g_{H_2}}
\newcommand{\gHcd}{g_{H^*_2}}
\newcommand{\gHt}{g_{H_1}}
\newcommand{\gHct}{g_{H^*_1}}

\begin{document}
 
\setcounter{page}0
\def\ppnumber{\vbox{\baselineskip14pt
%\hbox{hep-th/0000000}
}}
\def\ppdate{\footnotesize{}} \date{}

\author{Carlos Tamarit\\
[7mm]
{\normalsize  \it Perimeter Institute for Theoretical Physics}\\
{\normalsize \it Waterloo, ON, N2L 2Y5, Canada}\\
[3mm]
{\tt \footnotesize ctamarit at perimeterinstitute.ca}
}

\bigskip
\title{\bf Decoupling heavy sparticles in hierarchical SUSY scenarios: Two-loop Renormalization Group equations}\vskip 0.5cm
\maketitle

\begin{abstract} \normalsize
\noindent 
Two loop renormalization group equations for dimensionless as well as dimensionful parameters are obtained for the low energy theories that result from decoupling heavy scalar particles in Split SUSY and Effective SUSY scenarios, assuming that only a single Higgs field survives at low energy. For the Effective SUSY case two scenarios are considered: first, when the only light third generation scalars are the partners of the left-handed quark doublet and the right-handed top quark --which yields the minimal matter content compatible with naturalness-- and second, when all the scalars of the third generation are light. These  beta functions implementing decoupling will be useful to avoid the problems of perturbation theory in the MSSM in a mass-independent scheme such as $\overline{\rm DR}$ when large hierarchies in the spectrum are present.

\end{abstract}
\bigskip
\newpage
\tableofcontents
%%%%%%%%%%%%%%%%%%%%%%%%%%%%%%%%%%%%%%%%%%%%%%%%%%%%%%%%%%%%%%%%%%%%%%%%%%%%%%%%%%%%%%%%
%%%%%%%%%%%%%%%%%%%%%%%%%%%%%%%%%%%%%%%%%%%%%%%%%%%%%%%%%%%%%%%%%%%%%%%%%%%%%%%%%%%%%%%%
%%%%%%%%%%%%%%%%%%%%%%%%%%%%%%%%%%%%%%%%%%%%%%%%%%%%%%%%%%%%%%%%%%%%%%%%%%%%%%%%%%%%%%%%
%%%%%%%%%%%%%%%%%%%%%%%%%%%%%%%%%%%%%%%%%%%%%%%%%%%%%%%%%%%%%%%%%%%%%%%%%%%%%%%%%%%%%%%%
%%%%%%%%%%%%%%%%%%%%%%%%%%%%%%%%%%%%%%%%%%%%%%%%%%%%%%%%%%%%%%%%%%%%%%%%%%%%%%%%%%%%%%%%
%%%%%%%%%%%%%%%%%%%%%%%%%%%%%%%%%%%%%%%%%%%%%%%%%%%%%%%%%%%%%%%%%%%%%%%%%%%%%%%%%%%%%%%%

\section{Introduction}

Low energy Supersymmetry (SUSY) is mainly motivated by the hierarchy problem, the unification of gauge couplings and by the fact that, in the presence of R-parity, it provides stable dark matter candidates with the correct relic density.

Due to the large number of parameters of the Minimal Supersymmetric Standard Model (MSSM), there is a large variety of possible spectra for the new particles predicted by SUSY. The measurements being performed at the Large Hadron Collider (LHC) are already discarding sizable portions of constrained parameter spaces, obtained by assuming symmetries or degeneracies among the MSSM parameters at some high energy scale. Such is the case, among others, of the sugra-inspired constrained MSSM (CMSSM), or  models of gauge mediation (see for example \cite{Buchmueller:2011sw}, \cite{Kats:2011qh}).

If SUSY is still realized in Nature, the lack of evidence for supersymmetric particles may put into question its relevance as a solution of the hierarchy problem or encourage to look for less constrained spectra which still allow for naturalness. In these two cases, Split-SUSY \cite{ArkaniHamed:2004fb,Giudice:2004tc} and Effective SUSY \cite{Cohen:1996vb} may be singled out as the best motivated low energy scenarios. The spectrum of Split-SUSY at low energy includes, apart from the Standard Model particles, gauginos and higgsinos; it still allows for gauge coupling unification (or even improves it) and has dark matter candidates; however, it is heavily tuned. On the other hand, the Effective SUSY scenario at low energy adds to the previous set of particles some or all of the third generation scalar superpartners, which allows to retain naturalness up to scales of 10-20TeV; also, the flavor problem is solved by decoupling of the first two generations \cite{Dimopoulos:1995mi}. The minimal choice of light third 
generation scalars with a single low energy Higgs field includes only the left handed squark doublet and the right handed stop \cite{Brust:2011tb}. When all the third generation superpartners are present, more fine-tuning is expected due to the appearance of hypercharge D-term contributions in the beta function of the Higgs mass, which are absent in the minimal case. 

The two scenarios outlined above involve particle spectra with large hierarchies in the mass parameters. This immediately raises the question of whether perturbative calculations in a mass-independent scheme such as  $\overline{\text{DR}}$  (using for example the 2-loop MSSM beta functions obtained in ref.~\cite{Martin:1993zk}) are spoiled by the appearance of large terms involving logarithms of the masses of the heavy particles. Such  mass-independent schemes are not physical, and when using them the effect of the heavy particles does not decouple at low energies.  For example, in Effective SUSY scenarios it is known that the 2 loop RG equations can drive the third generation soft masses to negative values \cite{ArkaniHamed:1997ab}. However, finite quantum corrections can become quite significant due to the large logarithms mentioned before, and can drive the tachyonic masses to positive values. The fact that these quantum corrections become large raises doubts about the precision of the calculations, 
especially if one is interested in the cases in which third generation sparticles become light. This motivates the use of a physical mass-dependent scheme in which the heavy sparticles decouple. 

Such a scheme can be approximated by a stepwise running of the parameters of the theory, in which the heavy particles decouple at their thresholds. This requires to obtain the beta functions for all the parameters of the low energy theories that result after decoupling the heavy particles in the different scenarios. There are already results in the literature in the Split SUSY case:  the one-loop RG equations where obtained in ref.~\cite{Giudice:2004tc}, with two-loop contributions included for the gauge couplings. Other two-loop beta functions, ignoring flavor mixing effects, were obtained for dimensionless parameters in  refs.~\cite{Binger:2004nn}  and ref.~\cite{Giudice:2011cg}, with small discrepancies in the running of the quartic coupling. This article reports the computations of the full 2-loop beta functions in the $\overline{\text{MS}}$ scheme of all the dimensionless and dimensionful couplings appearing in the low energy theories (with heavy particles decoupled) describing the Split SUSY scenario 
as well as the two Effective SUSY realizations alluded to above: the minimal one, in which the only light scalars of the third generation are the left-handed quark doublet and the right handed stop, and the scenario in which all the third generations scalars are light. The RG equations were obtained from the results of ref.~\cite{Luo:2002ti}, which build upon the classic papers \cite{Machacek:1983tz}, \cite{Machacek:1983fi} and \cite{Machacek:1984zw} by properly dealing with complex fermion fields. The parameters of the low energy theories are all those consistent with the gauge symmetries and lepton number conservation, and the beta functions were computed including phases for Yukawa couplings and fermion masses and taking into account off-diagonal flavor contributions --however, for reasons of space most of the formulae displayed in the paper neglect them. The full formulae are available in the arXiv source material. In the Split-SUSY case, the results of refs.~\cite{Binger:2004nn} and~\cite{Giudice:2011cg}
 are reproduced safe for small differences: the RG equation for the scalar quartic coupling coincides with that of ref.~\cite{Binger:2004nn}, while some discrepancies arise with respect to ref.~\cite{Giudice:2011cg} in 2-loop terms involving the gauge coupling $g_2$. 

\begin{figure}[h]\centering
\begin{minipage}{0.5\textwidth}
 \includegraphics[scale=.9]{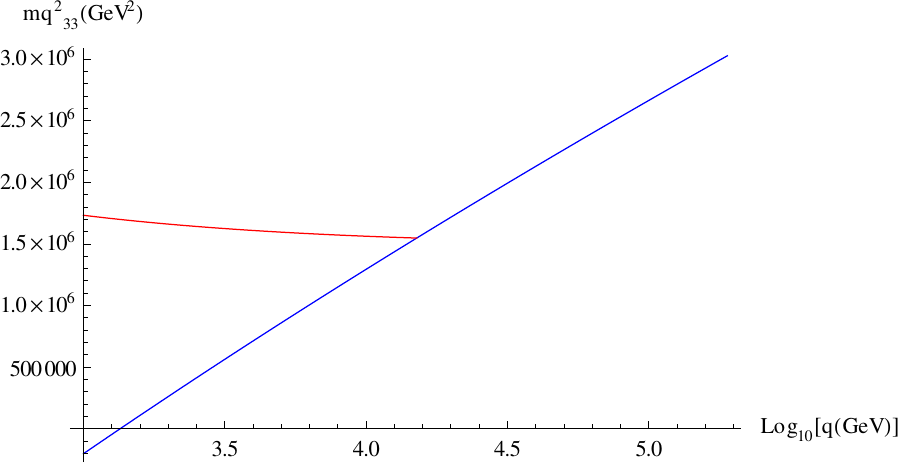}
\end{minipage}\begin{minipage}{0.5\textwidth}
 \includegraphics[scale=.9]{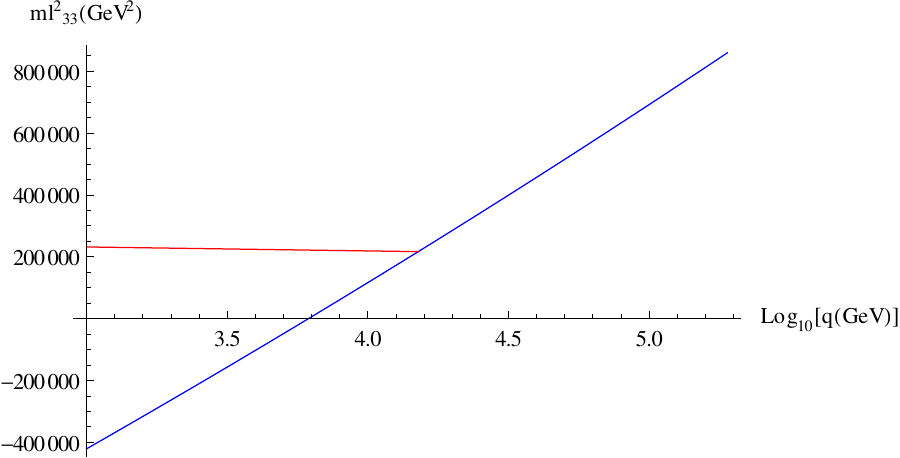}
\end{minipage}
\caption{\label{fig:1}2 loop running of ${m^2_Q}_{33}$ in a minimal Effective SUSY scenario (left), and  ${m^2_L}_{33}$ in a non-minimal one (right). The blue line represents the MSSM $\overline{\rm DR}$ running, while the red line represents the running in the the theory with heavy scalars decoupled at 15 TeV. Threshold effects at this scale were ignored.}
\end{figure}

To motivate the need for these new beta functions, Fig.~\ref{fig:1} shows an example of the 2 loop MSSM $\overline{\rm DR}$ running \cite{ Martin:1993zk} of the soft mass ${m^2_Q}_{33}$ in a minimal Effective SUSY scenario, as well as ${m^2_L}_{33}$ in a nonminimal one, compared with the running in the theories implementing decoupling. In both cases, first generation sparticles were given masses of 15 TeV, and boundary conditions inspired by gauge mediation were imposed for the light supersymmetric particles at scales $\Lambda_G=180$ TeV and $\Lambda_G=280$ TeV, respectively. 
\footnote{The boundary conditions considered are $m^2_{\rm light}=\frac{2\Lambda^2_G}{(16\pi^2)^2}\sum_ig_i^4 C_2(i)$ for the light squarks and leptons and $M_{i}=g^2_i \Lambda_g$ for the gauginos; trilinear couplings were set to zero. $\mu$ was taken as $M_1$ at the high scale and $m^2_{H_u}$, $m^2_{H_d}$ and $B_\mu$ were determined by demanding a correct electroweak symmetry breaking together with a consistent Higgs decoupling limit in the MSSM, yielding a Standard-Model like light Higgs and heavy Higgs states with masses of $15$ TeV. $\tan\beta$ was fixed at 10 and $\Lambda_g$ was set at 1 TeV. }. Clearly there are  large deviations from the MSSM $\overline{\rm DR}$ running, which drives the soft masses to tachyonic values. The difference between the 2 RG flows would correspond to finite corrections in the MSSM $\overline{\rm DR}$ scheme, which, as anticipated, become very large and put into question the precision of the perturbative calculation, which furthermore becomes problematic due to the tree-level 
tachyonic masses. These problems are avoided by using the decoupled RG flows. A more detailed phenomenological study of Effective SUSY scenarios using RG flows implementing decoupling will be given in a separate work \cite{Tamarit:2012ry}.

The paper is organized as follows. Section \ref{sec:Lagrangians} introduces the field content and Lagrangians of each of the three scenarios: Split SUSY in \S \ref{subsec:Larangians:SplitSUSY}, the minimal Effective SUSY scenario in \S \ref{subsec:Lagrangians:MinEffSUSY}, and the non-minimal one in \S \ref{subsec:Lagrangians:NonMinEffSUSY}. The beta functions are given in Section \ref{sec:Betas}, with the different scenarios being considered in \S\S \ref{subsec:Betas:SplitSUSY}, \ref{subsec:Betas:MinEffSUSY} and \ref{subsec:Betas:NonMinEffSUSY}, respectively. Final comments are given in \S \ref{sec:finalcomments}. For reasons of space, most of the formulae in the paper ignore complex phases. In the Split-SUSY case the formulae include off-diagonal flavor contributions, but these have been neglected in the expressions given for the beta functions in  Effective SUSY scenarios. Also, for the latter, two-loop formulae are provided for the beta functions of the gauge couplings, Standard-Model like Yukawa couplings and fermion and scalar mass terms, while the rest of 
the expressions are given at one-loop. The full 2 loop formulae including complex phases for Yukawa couplings and fermion masses as well as off-diagonal flavor contributions are available in the arXiv source material.

%%%%%%%%%%%%%%%%%%%%%%%%%%%%%%%%%%%%%%%%%%%%%%%%%%%%%%%%%%%%%%%%%%%%%%%%%%%%%%%%%
%%%%%%%%%%%%%%%%%%%%%%%%%%%%%%%%%%%%%%%%%%%%%%%%%%%%%%%%%%%%%%%%%%%%%%%%%%%%%%%%%
%%%%%%%%%%%%%%%%%%%%%%%%%%%%%%%%%%%%%%%%%%%%%%%%%%%%%%%%%%%%%%%%%%%%%%%%%%%%%%%%%
%%%%%%%%%%%%%%%%%%%%%%%%%%%%%%%%%%%%%%%%%%%%%%%%%%%%%%%%%%%%%%%%%%%%%%%%%%%%%%%%%
%%%%%%%%%%%%%%%%%%%%%%%%%%%%%%%%%%%%%%%%%%%%%%%%%%%%%%%%%%%%%%%%%%%%%%%%%%%%%%%%%

\section{Low energy Lagrangians\label{sec:Lagrangians}}

The field content and Lagrangians of the three different low energy scenarios outlined above are introduced next. Throughout this paper, fermion fields are denoted with lower case letters, and scalars with upper case ones.
%%%%%%%%%%%%%%%%%%%%%%%%%%%%%%%%%%%%%%%%%%%%%%%%%%%%%%%%%%%%%%%%%%%%%%%%%%%%%%%%%
%%%%%%%%%%%%%%%%%%%%%%%%%%%%%%%%%%%%%%%%%%%%%%%%%%%%%%%%%%%%%%%%%%%%%%%%%%%%%%%%%
%%%%%%%%%%%%%%%%%%%%%%%%%%%%%%%%%%%%%%%%%%%%%%%%%%%%%%%%%%%%%%%%%%%%%%%%%%%%%%%%%
%%%%%%%%%%%%%%%%%%%%%%%%%%%%%%%%%%%%%%%%%%%%%%%%%%%%%%%%%%%%%%%%%%%%%%%%%%%%%%%%%
%%%%%%%%%%%%%%%%%%%%%%%%%%%%%%%%%%%%%%%%%%%%%%%%%%%%%%%%%%%%%%%%%%%%%%%%%%%%%%%%%

\subsection{Split SUSY\label{subsec:Larangians:SplitSUSY}}

The low energy theory includes the Standard Model particles as well as gauginos and higgsinos. The notation for the particle fields as well as their
representations under the Standard Model gauge group are given in the next table. 
\begin{center}
\begin{tabular}{c|c|c|c}\label{table:SplitSUSY}
 \   & SU(3) & SU(2) & U(1) \\
\hline  
%%%%%%%%%%%%%%%%%%%%%%%%%%%%%%%%%%%%%%%%
$q_i,i=1\dots 3$ & $\square$ & $\square$ & 1/6 \\
\hline
%%%%%%%%%%%%%%%%%%%%%%%%%%%%%%%%%%%%%%%%
$u^c_{i},i=1\dots 3$ & $\overline\square$ & $\mathbb{I}$ & -2/3 \\
\hline
%%%%%%%%%%%%%%%%%%%%%%%%%%%%%%%%%%%%%%%%
$d^c_{i},i=1\dots 3$ & $\overline\square$ & $\mathbb{I}$ & 1/3 \\
\hline
%%%%%%%%%%%%%%%%%%%%%%%%%%%%%%%%%%%%%%%%
$l_{i},i=1\dots 3$ & $\mathbb{I}$ & $\square$ & -1/2 \\
\hline
%%%%%%%%%%%%%%%%%%%%%%%%%%%%%%%%%%%%%%%%
$e^c_{i},i=1\dots 3$ & $\mathbb{I}$ & $\mathbb{I}$ & 1\\
\hline
%%%%%%%%%%%%%%%%%%%%%%%%%%%%%%%%%%%%%%%%
$h_u $ & $\mathbb{I}$ & $\square$ & 1/2 \\
\hline
%%%%%%%%%%%%%%%%%%%%%%%%%%%%%%%%%%%%%%%%
$h_d $ & $\mathbb{I}$ & $\overline\square$ & -1/2 \\
\hline
%%%%%%%%%%%%%%%%%%%%%%%%%%%%%%%%%%%%%%%%
$\lambda_3$ & Ad & $\mathbb{I}$ & 0 \\
\hline
%%%%%%%%%%%%%%%%%%%%%%%%%%%%%%%%%%%%%%%%
$\lambda_2 $ & $\mathbb{I}$ & Ad & 0 \\
\hline
%%%%%%%%%%%%%%%%%%%%%%%%%%%%%%%%%%%%%%%%
$\lambda_1 $ & $\mathbb{I}$ & $\mathbb{I}$ & 0\\
\hline
%%%%%%%%%%%%%%%%%%%%%%%%%%%%%%%%%%%%%%%%
$H $ & $\mathbb{I}$ & $\square$ & 1/2\\
\hline
%%%%%%%%%%%%%%%%%%%%%%%%%%%%%%%%%%%%%%%%
\end{tabular} 
\end{center}\par

The Lagrangian is taken as follows,
\begin{align}
\nonumber
 {\cal L}=&{\cal L}_{SM}\!-\!\left\{\mu h_u h_d \!+\!\frac{1}{2}\sum_{k=1}^3\sum_{A=1}^{l(k)}M_k \lambda^A_k\lambda^A_k\!+\!\sum_{k=1}^2\sum_{A=1}^{l(k)}(g_{H_k}H^\dagger T_k^A\lambda^A h_u\!+\!g_{H^*_k}H T_k^A\lambda^A h_d)\!+\!\rm{c.c.}\!\right\},\\
%%%%
\,\label{eq:LagrangianSS}%\\\
%%%%
%&-m^2_H H^\dagger H-\frac{1}{2}\lambda (H^\dagger H)^2,
\end{align}
where ${\cal L}_{SM}$ is the Standard Model Lagrangian and $\{l(k)\}=\{1,3,8\}$ designate the dimension of the adjoint representation of the $k$th group, with the ordering $\{G_k\}=\{U(1),SU(2),SU(3)\}$. In order to fix the notation, the gauge couplings are denoted by $g_k$, the Yukawa matrices are ${y_u}_{ij},{y_d}_{ij},{y_l}_{ij},$ and the mass and quartic parameter of the Higgs potential are $m^2$ and $\lambda$; the Yukawa interactions and Higgs potential are then
\begin{align*}
 {\cal L}_{SM}\supset-(y_u)_{ji}q_{i}\epsilon H u^c_j-(y_d)_{ji}q_i H^\dagger d^c_j-(y_l)_{ji}l_i H^\dagger e^c_j-m^2_H H^\dagger H-\frac{1}{2}\lambda (H^\dagger H)^2,
\end{align*}
where $i,j$ are family indices, $ q_{i}\epsilon H =q_{i}^{a}\epsilon^{ab} H^b$,  $a,b$ being SU(2) indices and $\epsilon^{ab}$ the usual antisymmetric tensor with $\epsilon^{12}=1$.

The resulting beta functions of the couplings (with a GUT normalization convention for the gauge coupling $g_1$) are shown in \S\ref{subsec:Betas:SplitSUSY}.

%%%%%%%%%%%%%%%%%%%%%%%%%%%%%%%%%%%%%%%%%%%%%%%%%%%%%%%%%%%%%%%%%%%%%%%%%%%%%%%%%
%%%%%%%%%%%%%%%%%%%%%%%%%%%%%%%%%%%%%%%%%%%%%%%%%%%%%%%%%%%%%%%%%%%%%%%%%%%%%%%%%
%%%%%%%%%%%%%%%%%%%%%%%%%%%%%%%%%%%%%%%%%%%%%%%%%%%%%%%%%%%%%%%%%%%%%%%%%%%%%%%%%
%%%%%%%%%%%%%%%%%%%%%%%%%%%%%%%%%%%%%%%%%%%%%%%%%%%%%%%%%%%%%%%%%%%%%%%%%%%%%%%%%
%%%%%%%%%%%%%%%%%%%%%%%%%%%%%%%%%%%%%%%%%%%%%%%%%%%%%%%%%%%%%%%%%%%%%%%%%%%%%%%%%
\subsection{Minimal Effective SUSY\label{subsec:Lagrangians:MinEffSUSY}}

The low energy theory includes the fields of the previous section plus the third generation left-handed squark doublet and the right handed stop of the MSSM. The field content is summarized in the next table.

\begin{center}
\begin{tabular}{c|c|c|c}
 \   & SU(3) & SU(2) & U(1) \\
\hline  
%%%%%%%%%%%%%%%%%%%%%%%%%%%%%%%%%%%%%%%%
$q_i,i=1\dots 3$ & $\square$ & $\square$ & 1/6 \\
\hline
%%%%%%%%%%%%%%%%%%%%%%%%%%%%%%%%%%%%%%%%
$u^c_{i},i=1\dots 3$ & $\overline\square$ & $\mathbb{I}$ & -2/3 \\
\hline
%%%%%%%%%%%%%%%%%%%%%%%%%%%%%%%%%%%%%%%%
$d^c_{i},i=1\dots 3$ & $\overline\square$ & $\mathbb{I}$ & 1/3 \\
\hline
%%%%%%%%%%%%%%%%%%%%%%%%%%%%%%%%%%%%%%%%
$l_{i},i=1\dots 3$ & $\mathbb{I}$ & $\square$ & -1/2 \\
\hline
%%%%%%%%%%%%%%%%%%%%%%%%%%%%%%%%%%%%%%%%
$e^c_{i},i=1\dots 3$ & $\mathbb{I}$ & $\mathbb{I}$ & 1\\
\hline
%%%%%%%%%%%%%%%%%%%%%%%%%%%%%%%%%%%%%%%%
$h_u $ & $\mathbb{I}$ & $\square$ & 1/2 \\
\hline
%%%%%%%%%%%%%%%%%%%%%%%%%%%%%%%%%%%%%%%%
$h_d $ & $\mathbb{I}$ & $\overline\square$ & -1/2 \\
\hline
%%%%%%%%%%%%%%%%%%%%%%%%%%%%%%%%%%%%%%%%
$\lambda_3$ & Ad & $\mathbb{I}$ & 0 \\
\hline
%%%%%%%%%%%%%%%%%%%%%%%%%%%%%%%%%%%%%%%%
$\lambda_2 $ & $\mathbb{I}$ & Ad & 0 \\
\hline
%%%%%%%%%%%%%%%%%%%%%%%%%%%%%%%%%%%%%%%%
$\lambda_1 $ & $\mathbb{I}$ & $\mathbb{I}$ & 0\\
\hline
%%%%%%%%%%%%%%%%%%%%%%%%%%%%%%%%%%%%%%%%
$H $ & $\mathbb{I}$ & $\square$ & 1/2\\
\hline
%%%%%%%%%%%%%%%%%%%%%%%%%%%%%%%%%%%%%%%%
$Q$ & $\square$ & $\square$ & 1/6 \\
\hline
%%%%%%%%%%%%%%%%%%%%%%%%%%%%%%%%%%%%%%%%
$U^c$ & $\overline\square$ & $\mathbb{I}$ & -2/3 \\
\end{tabular} 
\end{center}\par

The most general renormalizable Lagrangian without lepton or baryon number violating terms is
\begin{align}
\nonumber {\cal L}=&{\cal L}_{SM}-\Big\{\mu h_u h_d +{z_q}_j U^c q_j\epsilon h_u+{z_u}_jQ\epsilon h_u u^c_j+{z_d}_jQ h_d d^c_j+\frac{1}{2}\sum_{k=1}^3\sum_{A=1}^{l(k)}M_k \lambda^A_k\lambda^A_k\\
%%%
\nonumber&.+\sum_{k=1}^3\sum_{A=1}^{l(k)}(g_{H_k}H^\dagger T_k^A\lambda^A h_u+g_{H^*_k}H T_k^A\lambda^A h_d+g_{Q_{j,k}} Q^\dagger T_k^A\lambda^A q_j+g_{U_{j,k}} {U^c}^\dagger T_k^A\lambda^A u^c_j)+{\rm{c.c.}}\Big\}\\
%%%%
&\nonumber-\frac{1}{2}\sum_{k=1}^3\sum_{A=1}^{l(k)}\overline\gamma_{k,S,S'}D^{k,A}_S D^{k,A}_{S'}-\frac{1}{2}\sum_S\lambda_S(S^\dagger S)^2-\sum_{S\neq S'}\lambda_{SS'}(S^\dagger S)({S'}^\dagger S')\\
%%%
\nonumber &-\lambda'_{QU}(QU^c)(Q^\dagger {U^c}^\dagger)-\lambda'_{HQ}(H\epsilon Q)({H^\dagger}\epsilon{Q^\dagger})-\lambda''_{HQ}(H Q^\dagger)({H^\dagger}Q) -m^2_Q Q^\dagger Q-m^2_U {U^c}^\dagger U^c\\
%%%%
&-(a_uQ\epsilon H U^c+{\rm{c.c.}}),\quad \quad
%%%%
D^{k,A}_S\equiv S^\dagger T_k^A S. \label{eq:LagrangianMES}
%%%%
%&-m^2_H H^\dagger H-\frac{1}{2}\lambda (H^\dagger H)^2,
\end{align}
In the expression above,  $S,S'$ denote the scalar fields in the theory, and $\overline\gamma_{k,S,S'}=\overline\gamma_{k,S',S}$; $j$ is summed over and runs over the three generations, and $a,b$ --which are also summed over-- are SU(2) indices of the fundamental representation. Given the transformation properties of the fields under the gauge groups, $g_{H_3}=g_{H_3^*}=g_{U_{j,2}}=\overline\gamma_{2,S,U}=\overline\gamma_{3,H,S}=0$. Also, several of the quartic couplings in eq.\eqref{eq:LagrangianMES} are redundant and can be ignored. This follows from the fact that after expanding the Lagrangian the quartic couplings $\overline\gamma_{k,S',S}$ and $\lambda$ end up appearing only in combinations that are equal to or proportional to the following:
\begin{align*}
\begin{array}{cc}
 3 \lambda _H+\frac{3}{4} \overline\gamma _{1,H,H}+\frac{3}{4} \overline\gamma _{2,H,H} , &
 \lambda''_{HQ}+\lambda _{HQ}+\frac{1}{12} \overline\gamma _{1,H,Q}+\frac{1}{4} \overline\gamma _{2,H,Q}, \\
  \lambda'_{HQ}+\lambda _{HQ}+\frac{1}{12} \overline\gamma _{1,H,Q}-\frac{1}{4} \overline\gamma _{2,H,Q},&
 \lambda _{HU}-\frac{1}{3} \overline\gamma _{1,H,U}, \\
 -\frac{\lambda'_{HQ}}{2}+\frac{\lambda''_{HQ}}{2}+\frac{1}{4} \overline\gamma _{2,H,Q},&
 3 \lambda _Q+\frac{1}{12} \overline\gamma _{1,Q,Q}+\frac{3}{4} \overline\gamma _{2,Q,Q}+\overline\gamma _{3,Q,Q}, \\
 \lambda _Q+\frac{1}{36} \overline\gamma _{1,Q,Q}-\frac{1}{4} \overline\gamma _{2,Q,Q}-\frac{1}{6} \overline\gamma _{3,Q,Q}, &
 \lambda'_{QU}+\lambda _{QU}-\frac{1}{9} \overline\gamma _{1,Q,U}-\frac{1}{3} \overline\gamma _{3,Q,U}, \\
 \lambda _{QU}-\frac{1}{9} \overline\gamma _{1,Q,U}+\frac{1}{6} \overline\gamma _{3,Q,U}, &
 \frac{1}{4} \overline\gamma _{2,Q,Q}+\frac{1}{4} \overline\gamma _{3,Q,Q}, \\
 \frac{\lambda'_{QU}}{2}-\frac{1}{4} \overline\gamma _{3,Q,U}, &
 3 \lambda _U+\frac{4}{3} \overline\gamma _{1,U,U}+\overline\gamma _{3,U,U}.
\end{array}
\end{align*}
This means that some of the $\lambda$ and $\overline\gamma_{k,S,S'}$ are redundant. Out of these twelve combinations, nine are independent, and they can be absorbed into new $\gamma_{k,SS'}$ couplings as follows
\begin{align}
\nonumber\gamma_{1,HH}&=\overline\gamma_{1,HH}+\overline\gamma_{2,HH}+4\lambda_H, & \gamma_{1,HQ}&=\overline\gamma_{1,HQ}+12\lambda_{HQ}+6\lambda'_{HQ}+6\lambda''_{HQ},\\
%%%%%
\nonumber\gamma_{1,HU}&=\overline\gamma_{1,HU}-3\lambda_{HU}, & \gamma_{1,QQ}&=\overline\gamma_{1,QQ}+36\lambda_Q+3\overline\gamma_{3,QQ},\\
%%%%%
\nonumber\gamma_{1,QU}&=\overline\gamma_{1,QU}-9\lambda_{QU}-3\lambda'_{QU}, & \gamma_{1,UU}&=\overline\gamma_{1,UU}+\frac{9}{4}\lambda_U+\frac{3}{4}\overline\gamma_{3,UU},\\
%%%%%
\nonumber\gamma_{2,HQ}&=\overline\gamma_{2,HQ}-2\lambda'_{HQ}+2\lambda''_{HQ}, & \gamma_{2,QQ}&=\overline\gamma_{2,QQ}+\overline\gamma_{3,QQ},\\
%%%
\gamma_{3,QU}&=\overline\gamma_{3,QU}-2\lambda'_{QU}.
\label{eq:MES:redefs}
\end{align}
Hence, the independent couplings of the Lagrangian can be taken as:
\begin{align}
\nonumber& g_i,\,\,{y_u}_{ij},\,\,{y_d}_{ij},\,\,{y_l}_{ij},\,\,z_{q_i},\,\,z_{u_i},\,\,z_{d_i},\,\,g_{Q_{i,k}},\,\,g_{U_{i,1}},\,\,g_{U_{i,3}}, M_i,\,\, i,j,k=1,2,3,\\
%%%%
\nonumber&g_{H_1},\,\,g_{H_1^*},\,\,g_{H_2},\,\,g_{H_2^*},\gamma_{1,H,H},\gamma_{1,H,Q},\gamma_{1,H,U},\gamma_{1,Q,Q},\gamma_{1,Q,U},\gamma_{1,U,U},\gamma_{2,H,Q},\gamma_{2,QQ,},\gamma_{3,Q,U},\\
%%%%
&\mu,a_u,m^2_H,m^2_Q,m^2_U.\label{eq:couplingsMES}
\end{align}
Eqs.~\eqref{eq:MES:redefs} will be needed when matching the MSSM quartic couplings to the independent couplings of eq.~\eqref{eq:couplingsMES}. Note that several terms in the Lagrangian corresponding to interactions beyond the Standard Model are not flavor diagonal, while flavor mixing contributions due to new Physics are expected to be small. Even though the beta functions in the full flavor mixed case were computed, for reasons of space and given the fact that off-diagonal flavor contributions are expected to be small, the formulae in \S \ref{subsec:Betas:MinEffSUSY} will neglect these, except in the case of the gauge couplings. To ease the notation, the following definitions will be used:
\begin{align}
\begin{array}{cccccc}
 g_{H_1}\equiv \bar g_{1}, & g_{H_1^*}\equiv \bar g_{2},& g_{H_2}\equiv\bar{g}_3, & g_{H_2^*}\equiv\bar{g}_4,&   g_{Q_{3,1}}\equiv\bar{g}_5, &  g_{Q_{3,2}}\equiv\bar{g}_6,\\
%%%%%%%%%
 g_{Q_{3,3}}\equiv\bar{g}_7, & g_{U_{3,1}}\equiv\bar{g}_8, & g_{U_{3,3}}\equiv\bar{g}_9, &  \gamma_{1,H,H}\equiv\hat{\gamma}_1, &  \gamma_{1,H,Q}\equiv\hat{\gamma}_2,&  \gamma_{1,H,U}\equiv\hat{\gamma}_3,\\
%%%%%%
  \gamma_{1,Q,Q}\equiv\hat{\gamma}_4,&  \gamma_{1,Q,U}\equiv\hat{\gamma}_5,&  \gamma_{1,U,U}\equiv\hat{\gamma}_6, &  \gamma_{2,H,Q}\equiv\hat{\gamma}_7, &  \gamma_{2,Q,Q}\equiv\hat{\gamma}_8,&  \gamma_{3,Q,U}\equiv\hat{\gamma}_9,
\end{array}\label{eq:couplingsMES:notation}
\end{align}

%%%%%%%%%%%%%%%%%%%%%%%%%%%%%%%%%%%%%%%%%%%%%%%%%%%%%%%%%%%%%%%%%%%%%%%%%%%%%%%%%
%%%%%%%%%%%%%%%%%%%%%%%%%%%%%%%%%%%%%%%%%%%%%%%%%%%%%%%%%%%%%%%%%%%%%%%%%%%%%%%%%
%%%%%%%%%%%%%%%%%%%%%%%%%%%%%%%%%%%%%%%%%%%%%%%%%%%%%%%%%%%%%%%%%%%%%%%%%%%%%%%%%
%%%%%%%%%%%%%%%%%%%%%%%%%%%%%%%%%%%%%%%%%%%%%%%%%%%%%%%%%%%%%%%%%%%%%%%%%%%%%%%%%
%%%%%%%%%%%%%%%%%%%%%%%%%%%%%%%%%%%%%%%%%%%%%%%%%%%%%%%%%%%%%%%%%%%%%%%%%%%%%%%%%
\subsection{Effective SUSY with a full generation of light scalars\label{subsec:Lagrangians:NonMinEffSUSY}}

Lastly, the low energy theory in this case involves all the fields in the previous sections plus a right-handed sbottom, a left-handed third-generation slepton doublet and a right handed stau, as summarized in the next table.
\begin{center}
\begin{tabular}{c|c|c|c}
 \   & SU(3) & SU(2) & U(1) \\
\hline  
%%%%%%%%%%%%%%%%%%%%%%%%%%%%%%%%%%%%%%%%
$q_i,i=1\dots 3$ & $\square$ & $\square$ & 1/6 \\
\hline
%%%%%%%%%%%%%%%%%%%%%%%%%%%%%%%%%%%%%%%%
$u^c_{i},i=1\dots 3$ & $\overline\square$ & $\mathbb{I}$ & -2/3 \\
\hline
%%%%%%%%%%%%%%%%%%%%%%%%%%%%%%%%%%%%%%%%
$d^c_{i},i=1\dots 3$ & $\overline\square$ & $\mathbb{I}$ & 1/3 \\
\hline
%%%%%%%%%%%%%%%%%%%%%%%%%%%%%%%%%%%%%%%%
$l_{i},i=1\dots 3$ & $\mathbb{I}$ & $\square$ & -1/2 \\
\hline
%%%%%%%%%%%%%%%%%%%%%%%%%%%%%%%%%%%%%%%%
$e^c_{i},i=1\dots 3$ & $\mathbb{I}$ & $\mathbb{I}$ & 1\\
\hline
%%%%%%%%%%%%%%%%%%%%%%%%%%%%%%%%%%%%%%%%
$h_u $ & $\mathbb{I}$ & $\square$ & 1/2 \\
\hline
%%%%%%%%%%%%%%%%%%%%%%%%%%%%%%%%%%%%%%%%
$h_d $ & $\mathbb{I}$ & $\overline\square$ & -1/2 \\
\hline
%%%%%%%%%%%%%%%%%%%%%%%%%%%%%%%%%%%%%%%%
$\lambda_3$ & Ad & $\mathbb{I}$ & 0 \\
\hline
%%%%%%%%%%%%%%%%%%%%%%%%%%%%%%%%%%%%%%%%
$\lambda_2 $ & $\mathbb{I}$ & Ad & 0 \\
\hline
%%%%%%%%%%%%%%%%%%%%%%%%%%%%%%%%%%%%%%%%
$\lambda_1 $ & $\mathbb{I}$ & $\mathbb{I}$ & 0\\
\hline
%%%%%%%%%%%%%%%%%%%%%%%%%%%%%%%%%%%%%%%%
$H $ & $\mathbb{I}$ & $\square$ & 1/2\\
\hline
%%%%%%%%%%%%%%%%%%%%%%%%%%%%%%%%%%%%%%%%
$Q$ & $\square$ & $\square$ & 1/6 \\
\hline
%%%%%%%%%%%%%%%%%%%%%%%%%%%%%%%%%%%%%%%%
$U^c$ & $\overline\square$ & $\mathbb{I}$ & -2/3 \\
%%%%%%%%%%%%%%%%%%%%%%%%%%%%%%%%%%%%%%%%
$D^c$ & $\overline\square$ & $\mathbb{I}$ & 1/3 \\
\hline
%%%%%%%%%%%%%%%%%%%%%%%%%%%%%%%%%%%%%%%%
$L$ & $\mathbb{I}$ & $\square$ & -1/2 \\
\hline
%%%%%%%%%%%%%%%%%%%%%%%%%%%%%%%%%%%%%%%%
$E^c$ & $\mathbb{I}$ & $\mathbb{I}$ & 1
\end{tabular} 
\end{center}\par
The most general renormalizable Lagrangian without lepton and baryon number violating terms is
\begin{align}
\nonumber {\cal L}=&{\cal L}_{SM}\!-\left\{\!\mu h_u h_d +{z_q}_j U^c q_j\epsilon h_u+{z_u}_jQ \epsilon h_u u^c_j+{z_d}_jQ h_d d^c_j+{z_{q^*}}_jD^c q_jh_d+{z_{l}}_jE^c l_jh_d\right.\\
%%%
\nonumber&\left.+{z_{e}}_jL e^c_jh_d+\frac{1}{2}\sum_{k=1}^3\sum_{A=1}^{l(k)}M_k \lambda^A_k\lambda^A_k+\sum_{k=1}^3\sum_{A=1}^{l(k)}(g_{H_k}H^\dagger T_k^A\lambda^A h_u+g_{H^*_k}H T_k^A\lambda^A h_d\right.\\
%%%
& \label{eq:LagrangianNMES}+g_{Q_{j,k}} Q^\dagger T_k^A\lambda^A q_j+g_{U_{j,k}} {U^c}^\dagger T_k^A\lambda^A u^c_j+g_{D_{j,k}} {D^c}^\dagger T_k^A\lambda^A d^c_j+g_{L_{j,k}} {L}^\dagger T_k^A\lambda^A l_j\\
%%%
\nonumber&\left.+g_{E_{j,k}} {E^c}^\dagger T_k^A\lambda^A e^c_j)+\rm{c.c.}\right\}-\frac{1}{2}\sum_{k=1}^3\sum_{A=1}^{l(k)}\overline\gamma_{k,S,S'}D^{k,A}_S D^{k,A}_{S'}-\frac{1}{2}\sum_S\lambda_S(S^\dagger S)^2\\
%%%%
\nonumber&-\sum_{S\neq S'}\lambda_{SS'}(S^\dagger S)({S'}^\dagger S')-\lambda'_{QU}(QU^c)(Q^\dagger {U^c}^\dagger)-\lambda'_{HQ}(H\epsilon Q)({H^\dagger}\epsilon{Q^\dagger})-\lambda''_{HQ}(H Q^\dagger)({H^\dagger}Q)\\
%%%
%\end{align*}\begin{align}
%%%
\nonumber&-\lambda'_{HL}(H\epsilon L)({H^\dagger}\epsilon{L^\dagger})-\lambda''_{HL}(H L^\dagger)({H^\dagger}L)-\lambda'_{QL}(Q\epsilon L)({Q^\dagger}\epsilon{L^\dagger})-\lambda''_{QL}(Q L^\dagger)({Q^\dagger}L)\\
%%%%
\nonumber&-\lambda'_{QD}(Q D^c)(Q^\dagger {D^c}^\dagger)\!-\!\lambda'_{UD}({U^c}^\dagger D^c)(U^c {D^c}^\dagger)\!-\!m^2_Q Q^\dagger Q\!-m^2_U {U^c}^\dagger U^c\!-m^2_D {D^c}^\dagger D^c\!-m^2_L L^\dagger L\end{align}\begin{align}
%%%%
&\nonumber-m^2_E {E^c}^\dagger E^c\!-(a_u Q\epsilon H U^c\!+c_d Q H^\dagger D^c+c_l Q H^\dagger E^c+\lambda'_{E}(Q L^\dagger)({E^c}^\dagger D^c)+\lambda''_{E}(Q\epsilon L)(E^cU^c)\\
%%%%
&\nonumber+{\rm c.c.})
%%%%
%&-m^2_H H^\dagger H-\frac{1}{2}\lambda (H^\dagger H)^2,
\end{align}
In this case, given the properties of the fields under gauge transformations, $g_{H_3}=g_{H_3^*}=g_{U_{j,2}}=g_{D_{j,2}}=g_{L_{j,3}}=g_{E_{j,2}}=g_{E_{j,3}}=\overline\gamma_{2,U/D/E,S}=\overline\gamma_{3,H/L/E,S}=0$; the notation and summing conventions are the same as in eqs.~\eqref{eq:LagrangianSS} and \eqref{eq:LagrangianMES}.

Again, some of the $\gamma_{k,S,S'}$ are redundant, as follows from the fact that the quartic couplings $\lambda_S,\lambda_{S,S'},\overline\gamma_{k,S,S'}$ only appear in combinations proportional to the following:
\begin{align*}
\begin{array}{cc}
 3 \lambda_H+\frac{3}{4} \overline\gamma _{1,H,H}+\frac{3}{4} \overline\gamma _{2,H,H},&\lambda''_{HQ}+\lambda_{HQ}+\frac{1}{12} \overline\gamma _{1,H,Q}+\frac{1}{4} \overline\gamma _{2,H,Q},\\
\lambda'_{HQ}+\lambda_{HQ}+\frac{1}{12} \overline\gamma _{1,H,Q}-\frac{1}{4} \overline\gamma _{2,H,Q},&\lambda_{HU}-\frac{1}{3} \overline\gamma _{1,H,U},\\
%%%
\lambda_{HD}+\frac{1}{6} \overline\gamma _{1,H,D},&\lambda''_{HL}+\lambda_{HL}-\frac{1}{4} \overline\gamma _{1,H,L}+\frac{1}{4} \overline\gamma _{2,H,L},\\
%%%%
\lambda'_{HL}+\lambda_{HL}-\frac{1}{4} \overline\gamma _{1,H,L}-\frac{1}{4} \overline\gamma _{2,H,L},&\lambda_{HE}+\frac{1}{2} \overline\gamma _{1,H,E},\\
%%%%
-\frac{1}{2}\lambda'_{HQ}+\frac{1}{2}\lambda''_{HQ}+\frac{1}{4} \overline\gamma _{2,H,Q},&-\frac{1}{2}\lambda'_{HL}+\frac{1}{2}\lambda''_{HL}+\frac{1}{4} \overline\gamma _{2,H,L},\\
%%%%
3 \lambda_Q+\frac{1}{12} \overline\gamma _{1,Q,Q}+\frac{3}{4} \overline\gamma _{2,Q,Q}+\overline\gamma _{3,Q,Q},& \lambda_Q+\frac{1}{36} \overline\gamma _{1,Q,Q}-\frac{1}{4} \overline\gamma _{2,Q,Q}-\frac{1}{6} \overline\gamma _{3,Q,Q}\\
%%%%
\lambda'_{QU}+\lambda_{QU}-\frac{1}{9} \overline\gamma _{1,Q,U}-\frac{1}{3} \overline\gamma _{3,Q,U}, & 3 \lambda_E+3 \overline\gamma _{1,E,E}\\
%%%%
\lambda_{QU}-\frac{1}{9} \overline\gamma _{1,Q,U}+\frac{1}{6} \overline\gamma _{3,Q,U},&\lambda'_{QD}+\lambda_{QD}+\frac{1}{18} \overline\gamma _{1,Q,D}-\frac{1}{3} \overline\gamma _{3,Q,D},\\
\lambda_{QD}+\frac{1}{18} \overline\gamma _{1,Q,D}+\frac{1}{6} \overline\gamma _{3,Q,D},&\lambda''_{QL}+\lambda_{QL}-\frac{1}{12} \overline\gamma _{1,Q,L}+\frac{1}{4} \overline\gamma _{2,Q,L},\\
%%%%
\lambda'_{QL}+\lambda_{QL}-\frac{1}{12} \overline\gamma _{1,Q,L}-\frac{1}{4} \overline\gamma _{2,Q,L},&\lambda_{QE}+\frac{1}{6} \overline\gamma _{1,Q,E},\\
%%%%
\frac{1}{4} \overline\gamma _{2,Q,Q}+\frac{1}{4} \overline\gamma _{3,Q,Q},&\frac{1}{2}\lambda'_{QU}-\frac{1}{4} \overline\gamma _{3,Q,U},\\
%%%%
\frac{1}{2}\lambda'_{QD}-\frac{1}{4} \overline\gamma _{3,Q,D},&-\frac{1}{2}\lambda'_{QL}+\frac{1}{2}\lambda''_{QL}+\frac{1}{4} \overline\gamma _{2,Q,L},\\
%%%%
3 \lambda_U+\frac{4}{3} \overline\gamma _{1,U,U}+\overline\gamma _{3,U,U},&\lambda'_{UD}+\lambda_{UD}-\frac{2}{9} \overline\gamma _{1,U,D}+\frac{1}{3} \overline\gamma _{3,U,D},\\
%%%%
\lambda_{UD}-\frac{2}{9} \overline\gamma _{1,U,D}-\frac{1}{6} \overline\gamma _{3,U,D},&\lambda_{UL}+\frac{1}{3} \overline\gamma _{1,U,L},\\
%%%%
\lambda_{UE}-\frac{2}{3} \overline\gamma _{1,U,E},&\frac{1}{2}\lambda'_{UD}+\frac{1}{4} \overline\gamma _{3,U,D},\\
%%%%
3 \lambda_D+\frac{1}{3} \overline\gamma _{1,D,D}+\overline\gamma _{3,D,D},&\lambda_{DL}-\frac{1}{6} \overline\gamma _{1,D,L},\\
%%%%
\lambda_{DE}+\frac{1}{3} \overline\gamma _{1,D,E},& 3 \lambda_L+\frac{3}{4} \overline\gamma _{1,L,L}+\frac{3}{4} \overline\gamma _{2,L,L},\\
%%%
\lambda_{LE}-\frac{1}{2} \overline\gamma _{1,L,E},& \lambda'_E,\quad\lambda''_E.\\
%%%%
\end{array}
\end{align*}
Only 30 linear combinations of the quartic couplings give independent contributions to the Lagrangian; these are
\begin{align*}
\nonumber\gamma_{1,HH}&=\overline\gamma_{1,HH}+\overline\gamma_{2,HH}+4\lambda_H, & \gamma_{1,HQ}&=\overline\gamma_{1,HQ}+12\lambda_{HQ}+6\lambda'_{HQ}+6\lambda''_{HQ},\\
%%%%%
\nonumber\gamma_{1,HU}&=\overline\gamma_{1,HU}-3\lambda_{HU}, &\nonumber\gamma_{1,HD}&=\overline\gamma_{1,HD}+6\lambda_{HD},\\
%%%%
\nonumber\gamma_{1,HL}&=\overline\gamma_{1,HL}-4\lambda_{HL}-2\lambda'_{HL}-2\lambda''_{HL}, &\nonumber\gamma_{1,HE}&=\overline\gamma_{1,HE}+2\lambda_{HE},\\
%%%
 \gamma_{1,QQ}&=\overline\gamma_{1,QQ}+36\lambda_Q+3\overline\gamma_{3,QQ}, &\gamma_{1,QU}&=\overline\gamma_{1,QU}-9\lambda_{QU}-3\lambda'_{QU},\\
%%%
\gamma_{1,QD}&=\overline\gamma_{1,QD}+18\lambda_{QD}+6\lambda'_{QD}, & \gamma_{1,QL}&=\overline\gamma_{1,QL}-12\lambda_{QL}-6\lambda'_{QL}-6\lambda''_{QL},\\
%%%
\gamma_{1,QE}&=\overline\gamma_{1,QE}+6\lambda_{QE},& \gamma_{1,UU}&=\overline\gamma_{1,UU}+\frac{9}{4}\lambda_U+\frac{3}{4}\overline\gamma_{3,UU},
\\
%%%%%
%%%
\gamma_{1,UD}&=\overline\gamma_{1,UD}-\frac{9}{2}\lambda_{UD}-\frac{3}{2}\lambda'_{UD}, &\gamma_{1,UL}&=\overline\gamma_{1,UL}+3\lambda_{UL},\\
%%%
%%%%%
\gamma_{1,UE}&=\overline\gamma_{1,UE}-\frac{3}{2}\lambda_{UE}, &\gamma_{1,DD}&=\overline\gamma_{1,DD}+3\overline\gamma_{3,DD}+9\lambda_{D},\end{align*}\begin{align*}
%%%%
\gamma_{1,DL}&=\overline\gamma_{1,DL}-6\lambda_{DL}, & \gamma_{1,DE}&=\overline\gamma_{1,DE}+3\lambda_{DE},\\
%%%%%%
\gamma_{1,LL}&=\overline\gamma_{1,LL}+\overline\gamma_{2,LL}+4\lambda_{L}, &\gamma_{1,LE}&=\overline\gamma_{1,LE}-2\lambda_{LE},\\
%%%%%
\gamma_{1,EE}&=\overline\gamma_{1,EE}+\lambda_{E}, &\gamma_{2,HQ}&=\overline\gamma_{2,HQ}-2\lambda'_{HQ}+2\lambda''_{HQ},\\
%%%%%
\gamma_{2,H,L}&=\overline\gamma_{2,H,L}-2\lambda'_{HL}+2\lambda''_{HL}, &   \gamma_{2,QQ}&=\overline\gamma_{2,QQ}+\overline\gamma_{3,QQ},\\
%%%
\nonumber\gamma_{2,Q,L}&=\overline\gamma_{2,Q,L}-2\lambda'_{QL}+2\lambda''_{QL},& \gamma_{3,QU}&=\overline\gamma_{3,QU}-2\lambda'_{QU},\\
%%%%%
\nonumber\gamma_{3,QD}&=\overline\gamma_{3,QD}-2\lambda'_{QD}, & \gamma_{3,UD}&=\overline\gamma_{3,UD}+2\lambda'_{UD},\\
%%%%
\lambda'_E&, & \lambda''_E&.
%\label{eq:ES:redefs}
\end{align*}
The choice of independent couplings for the Lagrangian is given then by:
\begin{align}
\nonumber& g_i,\,\,{y_u}_{ij},\,\,{y_d}_{ij},\,\,{y_l}_{ij},\,\,z_{q_i},\,\,z_{u_i},\,\,z_{d_i},\,\,z_{q^*_i},\,\,z_{l_i},\,\,z_{e_i},\,g_{Q_{i,k}},\,g_{{U/D}_{i,1/3}},\,g_{{L}_{i,1/2}},\,g_{{E}_{i,1}}, M_i, i,j,k=1,2,3,\\
%%%%
\nonumber&g_{H_{1/2}},\,g_{H_{1/2}^*},\,\gamma_{k,S,S'},\,\{k,S,S'\}\notin\{\{2,U/D/E,S\},\{3,H/L/E,S\},\{3,Q,Q\},\{3,U,U\},\\
%%%%
&\{3,D,D\},\{2,H,H\},\{2,L,L\}\},\,\,\lambda'_E,\lambda''_E,\,\mu,a_u,c_d,c_l,m^2_H,m^2_Q,m^2_U,m^2_D,m^2_L,m^2_E.
\label{eq:couplingsNMES}
\end{align}
As in the case of the minimal realization of low-energy Effective Supersymmetry, the formulae for the beta functions given in \S\ref{subsec:Betas:NonMinEffSUSY} neglect off-diagonal flavor contributions as well as phases in Yukawa couplings and fermion mass parameters. To simplify the notation, the following definitions are used
\begin{align}
\nonumber g_{H_1}&\equiv \bar g_{1}, & g_{H_1^*}&\equiv \bar g_{2},& g_{H_2}&\equiv\bar{g}_3, & g_{H_2^*}&\equiv\bar{g}_4,&   g_{Q_{3,1}}&\equiv\bar{g}_5, &  g_{Q_{3,2}}&\equiv\bar{g}_6,\\
%%%%%%%%%
\nonumber g_{Q_{3,3}}&\equiv\bar{g}_7, & g_{U_{3,1}}&\equiv\bar{g}_8, & g_{U_{3,3}}&\equiv\bar{g}_9, & g_{D_{3,1}}&\equiv\bar{g}_{10}, & g_{D_{3,3}}&\equiv\bar{g}_{11},  & g_{L_{3,1}}&\equiv\bar{g}_{12},\\
%%%%%%%%%%
\nonumber g_{L_{3,2}}&\equiv\bar{g}_{13}, & g_{E_{3,1}}&\equiv\bar{g}_{14}, &\gamma_{1,H,H}&\equiv\tilde{\gamma}_1, &\gamma_{1,H,Q}&\equiv\tilde{\gamma}_2 , & \gamma_{1,H,U}&\equiv\tilde{\gamma}_3, & \gamma_{1,H,D}&\equiv\tilde{\gamma}_4\\
%%%%%%
 \nonumber\gamma_{1,H,L}&\equiv\tilde{\gamma}_5, &  \gamma_{1,H,E}&\equiv\tilde{\gamma}_6, & \gamma_{1,Q,Q}&\equiv\tilde{\gamma}_7, &  \gamma_{1,Q,U}&\equiv\tilde{\gamma}_8, & \gamma_{1,Q,D}&\equiv\tilde{\gamma}_9, & \gamma_{1,Q,L}&\equiv\tilde{\gamma}_{10}\\
%%%%%%
\nonumber\gamma_{1,Q,E}&\equiv\tilde{\gamma}_{11}, &\gamma_{1,U,U}&\equiv\tilde{\gamma}_{12} , &\gamma_{1,U,D}&\equiv\tilde{\gamma}_{13}, & \gamma_{1,U,L}&\equiv\tilde{\gamma}_{14} , &\gamma_{1,U,E}&\equiv\tilde{\gamma}_{15}, &\gamma_{1,D,D}&\equiv\tilde{\gamma}_{16}\\
%%%%%%
\nonumber\gamma_{1,D,L}&\equiv\tilde{\gamma}_{17}, &\gamma_{1,D,E}&\equiv\tilde{\gamma}_{18} , &\gamma_{1,L,L}&\equiv\tilde{\gamma}_{19}, & \gamma_{1,L,E}&\equiv\tilde{\gamma}_{20} , &\gamma_{1,E,E}&\equiv\tilde{\gamma}_{21}, &\gamma_{2,H,Q}&\equiv\tilde{\gamma}_{22}\\
%%%%
\nonumber\gamma_{2,H,L}&\equiv\tilde{\gamma}_{23}, &\gamma_{2,Q,Q}&\equiv\tilde{\gamma}_{24} , &\gamma_{2,Q,L}&\equiv\tilde{\gamma}_{25}, & \gamma_{3,Q,U}&\equiv\tilde{\gamma}_{26} , &\gamma_{3,Q,D}&\equiv\tilde{\gamma}_{27}, &\gamma_{3,U,D}&\equiv\tilde{\gamma}_{28}\\
%%%%%
\lambda'_E&\equiv\tilde{\gamma}_{29}, & \lambda''_E&\equiv\tilde{\gamma}_{30}.
\label{eq:couplingsNMES:notation}
\end{align}

To simplify the formulae, some beta functions are given in comparison with those in the Minimal Effective Susy scenario, which are denoted with the superscript ``MES''. In this respect, note that the parameters $\hat\gamma$ in the Minimal Effective Susy case can be expressed in terms of the parameters $\bar\gamma$ of the nonminimal scenario with the aid of eqs.~\eqref{eq:couplingsMES:notation} and \eqref{eq:couplingsNMES:notation}.

%%%%%%%%%%%%%%%%%%%%%%%%%%%%%%%%%%%%%%%%%%%%%%%%%%%%%%%%%%%%%%%%%%%%%%%%%%%%%%%%%
%%%%%%%%%%%%%%%%%%%%%%%%%%%%%%%%%%%%%%%%%%%%%%%%%%%%%%%%%%%%%%%%%%%%%%%%%%%%%%%%%
%%%%%%%%%%%%%%%%%%%%%%%%%%%%%%%%%%%%%%%%%%%%%%%%%%%%%%%%%%%%%%%%%%%%%%%%%%%%%%%%%
%%%%%%%%%%%%%%%%%%%%%%%%%%%%%%%%%%%%%%%%%%%%%%%%%%%%%%%%%%%%%%%%%%%%%%%%%%%%%%%%%
%%%%%%%%%%%%%%%%%%%%%%%%%%%%%%%%%%%%%%%%%%%%%%%%%%%%%%%%%%%%%%%%%%%%%%%%%%%%%%%%%
%%%%%%%%%%%%%%%%%%%%%%%%%%%%%%%%%%%%%%%%%%%%%%%%%%%%%%%%%%%%%%%%%%%%%%%%%%%%%%%%%
%%%%%%%%%%%%%%%%%%%%%%%%%%%%%%%%%%%%%%%%%%%%%%%%%%%%%%%%%%%%%%%%%%%%%%%%%%%%%%%%%
%%%%%%%%%%%%%%%%%%%%%%%%%%%%%%%%%%%%%%%%%%%%%%%%%%%%%%%%%%%%%%%%%%%%%%%%%%%%%%%%%
%%%%%%%%%%%%%%%%%%%%%%%%%%%%%%%%%%%%%%%%%%%%%%%%%%%%%%%%%%%%%%%%%%%%%%%%%%%%%%%%%
%%%%%%%%%%%%%%%%%%%%%%%%%%%%%%%%%%%%%%%%%%%%%%%%%%%%%%%%%%%%%%%%%%%%%%%%%%%%%%%%%
%%%%%%%%%%%%%%%%%%%%%%%%%%%%%%%%%%%%%%%%%%%%%%%%%%%%%%%%%%%%%%%%%%%%%%%%%%%%%%%%%
%%%%%%%%%%%%%%%%%%%%%%%%%%%%%%%%%%%%%%%%%%%%%%%%%%%%%%%%%%%%%%%%%%%%%%%%%%%%%%%%%
%%%%%%%%%%%%%%%%%%%%%%%%%%%%%%%%%%%%%%%%%%%%%%%%%%%%%%%%%%%%%%%%%%%%%%%%%%%%%%%%%
%%%%%%%%%%%%%%%%%%%%%%%%%%%%%%%%%%%%%%%%%%%%%%%%%%%%%%%%%%%%%%%%%%%%%%%%%%%%%%%%%
%%%%%%%%%%%%%%%%%%%%%%%%%%%%%%%%%%%%%%%%%%%%%%%%%%%%%%%%%%%%%%%%%%%%%%%%%%%%%%%%%
\newpage
\section{Beta functions\label{sec:Betas}}

This section provides the beta functions in the ${\overline{\text{MS}}}$ scheme for the parameters of the theories described in \S\ref{sec:Lagrangians}. As usual, for a coupling $\alpha$, $\beta_\alpha=\mu \frac{d\alpha}{d\mu}$. The $g_1$ hypercharge gauge coupling is taken in the GUT normalization, and $\xi$ denotes the standard gauge parameter. For reasons of space, complex phases are in general ignored, while off-diagonal flavor couplings are only taken into account in the Split SUSY case and in the beta functions for the gauge couplings in Effective SUSY scenarios. For the latter, two-loop contributions are given only for the gauge couplings, Standard Model-like Yukawa couplings, and fermion and scalar mass parameters. The full expressions, including complex phases for Yukawa couplings and fermion masses, as well as off-diagonal flavor contributions, can be found online in the arXiv source material.
%%%%%%%%%%%%%%%%%%%%%%%%%%%%%%%%%%%%%%%%%%%%%%%%%%%%%%%%%%%%%%%%%%%%%%%%%%%%%%%%%
%%%%%%%%%%%%%%%%%%%%%%%%%%%%%%%%%%%%%%%%%%%%%%%%%%%%%%%%%%%%%%%%%%%%%%%%%%%%%%%%%
%%%%%%%%%%%%%%%%%%%%%%%%%%%%%%%%%%%%%%%%%%%%%%%%%%%%%%%%%%%%%%%%%%%%%%%%%%%%%%%%%
%%%%%%%%%%%%%%%%%%%%%%%%%%%%%%%%%%%%%%%%%%%%%%%%%%%%%%%%%%%%%%%%%%%%%%%%%%%%%%%%%
%%%%%%%%%%%%%%%%%%%%%%%%%%%%%%%%%%%%%%%%%%%%%%%%%%%%%%%%%%%%%%%%%%%%%%%%%%%%%%%%%
\subsection{Split SUSY\label{subsec:Betas:SplitSUSY}}
This section presents the beta functions for the couplings in the Lagrangian of eq.~\eqref{eq:LagrangianSS}.
%%%%%%%%%%%%%%%%%%%%%%%%%%%%%%%%%%%%%%%%%%%%%%%%%%%%%%%%%%%%%%%%%%%%%%%%%%%%%%%%%
%%%%%%%%%%%%%%%%%%%%%%%%%%%%%%%%%%%%%%%%%%%%%%%%%%%%%%%%%%%%%%%%%%%%%%%%%%%%%%%%%
%%%%%%%%%%%%%%%%%%%%%%%%%%%%%%%%%%%%%%%%%%%%%%%%%%%%%%%%%%%%%%%%%%%%%%%%%%%%%%%%%
\subsubsection{Gauge couplings}
%%%%%%%%%%%%%%%%%%%%%%%%%%%%%%%%%%%%%%%%%%%%%%%%%%%%%%%%%%%%%%%%%%%%%%%%%%%%%%%%%
%%%%%%%%%%%%%%%%%%%%%%%%%%%%%%%%%%%%%%%%%%%%%%%%%%%%%%%%%%%%%%%%%%%%%%%%%%%%%%%%%
%%%%%%%%%%%%%%%%%%%%%%%%%%%%%%%%%%%%%%%%%%%%%%%%%%%%%%%%%%%%%%%%%%%%%%%%%%%%%%%%%
\begin{align*}
\beta_{g_1}=&\frac{9}{32\pi^2}g_1^3+\frac{1}{(16\pi^2)^2}\Big\{\frac{104 }{25}g_1^5+\frac{18}{5} {g_1}^3 {g_2}^2+\frac{44}{5} {g_1}^3 {g_3}^2-\frac{9}{40} {g_1}^3 (g_{H_2}^2+g_{H^*_2}^2)\\
%%%%%
&-\frac{3}{40} {g_1}^3(g_{H_1}^2+g_{H^*_1}^2)-g_1^3\big(\frac{1}{2}\Tr\, y_d^\dagger y_d+\frac{3}{2}\Tr\, y_l^\dagger y_l+\frac{17}{10}\Tr\, y_u^\dagger y_u\big)\Big\},\\
%%%%%
\beta_{g_2}=&-\frac{7}{96\pi^2}g_2^3+\frac{1}{(16\pi^2)^2}\Big\{\frac{6}{5} {g_1}^2 {g_2}^3+\frac{106 {g_2}^5}{3}+12 {g_2}^3 {g_3}^2-\frac{11}{8} {g_2}^3 (g_{H_2}^2+g_{H^*_2}^2)\\
%%%%
&-\frac{1}{8} {g_2}^3 (g_{H_1}^2+g_{H^*_1}^2)-g_2^3\big(\frac{3}{2}\Tr\, y_d^\dagger y_d+\frac{1}{2}\Tr\, y_l^\dagger y_l+\frac{3}{2}\Tr\, y_u^\dagger y_u\big)\Big\},\\
%%%%%%
\beta_{g_3}=&-\frac{5}{16\pi^2}g_3^3+\frac{1}{(16\pi^2)^2}\Big\{\frac{11}{10} g_1^2 g_3^3 + \frac{9}{2} g_2^2 g_3^3 + 22 g_3^5-2g_3^3(\Tr\, y_d^\dagger y_d+\Tr\, y_u^\dagger y_u)\Big\}.
\end{align*}
%%%%%%%%%%%%%%%%%%%%%%%%%%%%%%%%%%%%%%%%%%%%%%%%%%%%%%%%%%%%%%%%%%%%%%%%%%%%%%%%%
%%%%%%%%%%%%%%%%%%%%%%%%%%%%%%%%%%%%%%%%%%%%%%%%%%%%%%%%%%%%%%%%%%%%%%%%%%%%%%%%%
%%%%%%%%%%%%%%%%%%%%%%%%%%%%%%%%%%%%%%%%%%%%%%%%%%%%%%%%%%%%%%%%%%%%%%%%%%%%%%%%%
\subsubsection{Yukawas}
%%%%%%%%%%%%%%%%%%%%%%%%%%%%%%%%%%%%%%%%%%%%%%%%%%%%%%%%%%%%%%%%%%%%%%%%%%%%%%%%%
%%%%%%%%%%%%%%%%%%%%%%%%%%%%%%%%%%%%%%%%%%%%%%%%%%%%%%%%%%%%%%%%%%%%%%%%%%%%%%%%%
%%%%%%%%%%%%%%%%%%%%%%%%%%%%%%%%%%%%%%%%%%%%%%%%%%%%%%%%%%%%%%%%%%%%%%%%%%%%%%%%%
\begin{align*}
 \beta_{y_u}=& {\beta_{y_u}}^{SM}+
\frac{1}{16\pi^2}\Big(\frac{3}{4} (g_{H_2}^2+g_{H^*_2}^2) +\frac{1}{4} (g_{H_1}^2+g_{H^*_1}^2) \Big)y_u\\
%%%
& \!+ \frac{1}{(16\pi^2)^2} y_u\Big\{(g_{H_2}^2+g_{H^*_2}^2)\Big(\frac{15}{16}y^\dagger_d y_d\!-\!\frac{27}{16}y^\dagger_u y_u+\frac{165}{32}g_2^2+\frac{9}{32}g_1^2\Big)-\frac{9}{32}\gHd^2\gHt^2-\frac{3}{16}\gHd^2\gHcd^2\\
%%%%
&-\frac{9}{32}\gHcd^2\gHct^2+(g_{H_1}^2+g_{H^*_1}^2)\Big(\frac{5}{16}y^\dagger_d y_d\!-\!\frac{9}{16}y^\dagger_u y_u+\frac{15}{32}g_2^2+\frac{3}{32}g_1^2\Big)-\frac{5}{16}\gHt^2\gHct^2+\frac{40}{3}g_3^4\\
%%%%
&+\frac{3}{2}g_2^4+\frac{29}{150}g_1^4-\frac{45}{64}(\gHd^4+\gHcd^4)-\frac{9}{64}(\gHt^4+\gHct^4)
-\frac{3}{4}\gHd\gHcd\gHt\gHct\Big\},
\end{align*}
\begin{align*}
\beta_{y_d}=& {\beta_{y_d}}^{SM}+
\frac{1}{16\pi^2}\Big(\frac{3}{4} (g_{H_2}^2+g_{H^*_2}^2) +\frac{1}{4} (g_{H_1}^2+g_{H^*_1}^2) \Big)y_d\\
%%%
& \!+ \frac{1}{(16\pi^2)^2} y_d\Big\{(g_{H_2}^2+g_{H^*_2}^2)\Big(\frac{15}{16}\yud\yu\!-\!\frac{27}{16}y^\dagger_d y_d+\frac{165}{32}g_2^2+\frac{9}{32}g_1^2\Big)-\frac{9}{32}\gHd^2\gHt^2-\frac{3}{16}\gHd^2\gHcd^2\\
%%%%
&-\frac{9}{32}\gHcd^2\gHct^2+(g_{H_1}^2+g_{H^*_1}^2)\Big(\frac{5}{16}y^\dagger_u y_u\!-\!\frac{9}{16}y^\dagger_d y_d+\frac{15}{32}g_2^2+\frac{3}{32}g_1^2\Big)-\frac{5}{16}\gHt^2\gHct^2+\frac{40}{3}g_3^4\\
%%%%
&+\frac{3}{2}g_2^4-\frac{1}{150}g_1^4-\frac{45}{64}(\gHd^4+\gHcd^4)-\frac{9}{64}(\gHt^4+\gHct^4)-\frac{3}{4}\gHd\gHcd\gHt\gHct\Big\},
\end{align*}
\begin{align*}
\hskip-15pt\beta_{y_l}=& {\beta_{y_l}}^{SM}+
\frac{1}{16\pi^2}\Big( \frac{3}{4} (g_{H_2}^2+g_{H^*_2}^2) +\frac{1}{4} (g_{H_1}^2+g_{H^*_1}^2) \Big)y_l\\
%%%
& \!+ \frac{1}{(16\pi^2)^2} y_l\Big\{(g_{H_2}^2+g_{H^*_2}^2)\Big(-\frac{27}{16}\yld\yl+\frac{165}{32}g_2^2+\frac{9}{32}g_1^2\Big)-\frac{9}{32}\gHd^2\gHt^2-\frac{3}{16}\gHd^2\gHcd^2\\
%%%%
&-\frac{9}{32}\gHcd^2\gHct^2+(g_{H_1}^2+g_{H^*_1}^2)\Big(-\frac{9}{16}y^\dagger_l y_l+\frac{15}{32}g_2^2+\frac{3}{32}g_1^2\Big)-\frac{5}{16}\gHt^2\gHct^2+\frac{3}{2}g_2^4+\frac{33}{50}g_1^4\\
%%%%
&-\frac{45}{64}(\gHd^4+\gHcd^4)-\frac{9}{64}(\gHt^4+\gHct^4)-\frac{3}{4}\gHd\gHcd\gHt\gHct\Big\},
\end{align*}
\begin{align*}
 \beta_{\gHd}=&\frac{1}{16\pi^2}\Big\{\frac{11 g_{H_2}^3}{8}+\frac{1}{2} g_{H_1} g_{H_1^*} g_{H_2^*}+\gHd\Big(-\frac{9 g_1^2}{20}-\frac{33 g_2^2}{4}+\frac{3 g_{H_1}^2}{8}+\frac{g_{H^*_1}^2}{4}+\frac{g_{H^*_2}^2}{2}\\
%%%
&+\Tr[3\ydd\yd+3\yud\yu+\yld\yl]\Big)\Big\}+\frac{1}{(16\pi^2)^2}\Big\{g_{H_2} \left(\frac{3 \lambda ^2}{2}+\frac{117 g_1^4}{200}+\frac{9}{20} g_1^2 g_2^2-\frac{409 g_2^4}{12}\right.\\
&\left.-\frac{1}{2} \lambda  g_{H_1}^2+\frac{63}{320} g_1^2 g_{H_1}^2+\frac{111}{64} g_2^2 g_{H_1}^2-\frac{5 g_{H_1}^4}{64}+\frac{3}{32} g_1^2 g_{H_1^*}^2+\frac{15}{32} g_2^2 g_{H_1^*}^2-\frac{3}{8} g_{H_1}^2 g_{H_1^*}^2-\frac{9 g_{H_1^*}^4}{64}\right.\\
%%%%
&\left.-\frac{1}{2} \lambda  g_{H_2^*}^2+\frac{3}{40} g_1^2 g_{H_2^*}^2+\frac{17}{8} g_2^2 g_{H_2^*}^2-\frac{31}{64} g_{H_1}^2 g_{H_2^*}^2-\frac{13}{64} g_{H_1^*}^2 g_{H_2^*}^2-\frac{11 g_{H_2^*}^4}{32}\right.  \\
%%%%
&+\left(\frac{5 g_1^2}{8}+\frac{45 g_2^2}{8}+20 g_3^2-\frac{9 g_{H_1}^2}{16}+\frac{3 g_{H_2^*}^2}{8}\right) \text{Tr}[\ydd\yd]+\left(\frac{15 g_1^2}{8}+\frac{15 g_2^2}{8}-\frac{3 g_{H_1}^2}{16}\right.\\
%%%
&\left.+\frac{g_{H_2^*}^2}{8}\right) \text{Tr}[\yld\yl]+\left(\frac{17 g_1^2}{8}+\frac{45 g_2^2}{8}+20 g_3^2-\frac{9 g_{H_1}^2}{16}+\frac{3 g_{H_2^*}^2}{8}\right) \text{Tr}[\yud\yu]-\frac{27}{4} \text{Tr}[\ydd\yd\ydd\yd]\\
%%%%
&\left.+\frac{3}{2} \text{Tr}[\ydd\yd\yud\yu]-\frac{9}{4} \text{Tr}[\yld\yl\yld\yl]-\frac{27}{4} \text{Tr}[\yud\yu\yud\yu]\right)-\frac{5}{2} \lambda  g_{H_2}^3+\frac{87}{64} g_1^2 g_{H_2}^3+\frac{875}{64} g_2^2 g_{H_2}^3\\
%%%%
&-\frac{59}{64} g_{H_1}^2 g_{H_2}^3-\frac{15}{64} g_{H_1^*}^2 g_{H_2}^3-\frac{7 g_{H_2}^5}{8}-\frac{1}{2} \lambda  g_{H_1} g_{H_1^*} g_{H_2^*}+\frac{3}{40} g_1^2 g_{H_1} g_{H_1^*} g_{H_2^*}+\frac{9}{8} g_2^2 g_{H_1} g_{H_1^*} g_{H_2^*}\\
%%%
&-\frac{3}{8} g_{H_1}^3 g_{H_1^*} g_{H_2^*}-\frac{5}{16} g_{H_1} g_{H_1^*}^3 g_{H_2^*}-g_{H_1} g_{H_1^*} g_{H_2}^2 g_{H_2^*}-\frac{27}{32} g_{H_2}^3 g_{H_2^*}^2-\frac{9}{16} g_{H_1} g_{H_1^*} g_{H_2^*}^3\\
%%%
&-\frac{45}{16} g_{H_2}^3 \left(\text{Tr}[\ydd\yd]+\frac{1}{3} \text{Tr}[\yld\yl]+\text{Tr}[\yud\yu]\right)-\frac{3}{2} g_{H_1} g_{H_1^*} g_{H_2^*} \left(\text{Tr}[\ydd\yd]+\frac{1}{3} \text{Tr}[\yld\yl]\right.\\
%%%
&\left.+\text{Tr}[\yud\yu]\right),
\end{align*}
\begin{align*}
 \beta_{\gHt}=&\frac{1}{16\pi^2}\Big\{\frac{3 g_{H_1^*} g_{H_2} g_{H_2^*}}{2 }+\gHt\left(-\frac{9 g_1^2}{20}-\frac{9 g_2^2}{4}+\frac{5 g_{H_1}^2}{8}+g_{H_1^*}^2+\frac{9 g_{H_2}^2}{8}+\frac{3 g_{H_2^*}^2}{4}\right.\\
%%%
&\left.+\text{Tr}[3 \ydd\yd+\yld\yl+3 \yud\yu]\right)\Big\}\!+\frac{1}{(16\pi^2)^2}\Big\{g_{H_1} \left(\frac{3 \lambda ^2}{2}+\frac{117 g_1^4}{200}\!-\!\frac{27}{20} g_1^2 g_2^2\!-\!\frac{17 g_2^4}{4}\!-\!\frac{3}{2} \lambda  g_{H_1^*}^2\right.\\
%%%%
&\left.+\frac{3}{80} g_1^2 g_{H_1^*}^2+\frac{39}{16} g_2^2 g_{H_1^*}^2-\frac{9 g_{H_1^*}^4}{16}-\frac{3}{2} \lambda  g_{H_2}^2+\frac{189}{320} g_1^2 g_{H_2}^2+\frac{549}{64} g_2^2 g_{H_2}^2-\frac{75}{64} g_{H_1^*}^2 g_{H_2}^2\!-\!\frac{99 g_{H_2}^4}{64}\right.\\
%%%%
&+\frac{9}{32} g_1^2 g_{H_2^*}^2+\frac{165}{32} g_2^2 g_{H_2^*}^2-\frac{75}{64} g_{H_1^*}^2 g_{H_2^*}^2-\frac{21}{32} g_{H_2}^2 g_{H_2^*}^2-\frac{45 g_{H_2^*}^4}{64}+\left(\frac{5 g_1^2}{8}+\frac{45 g_2^2}{8}+20 g_3^2\right.\\
%%%
&\left.-\frac{21 g_{H_1^*}^2}{8}-\frac{27 g_{H_2}^2}{16}\right) \text{Tr}[\ydd\yd]+\left(\frac{15 g_1^2}{8}+\frac{15 g_2^2}{8}-\frac{7 g_{H_1^*}^2}{8}-\frac{9 g_{H_2}^2}{16}\right) \text{Tr}[\yld\yl]+\left(\frac{17 g_1^2}{8}\right.\\
%%%
&\left.+\frac{45 g_2^2}{8}+20 g_3^2-\frac{21 g_{H_1^*}^2}{8}-\frac{27 g_{H_2}^2}{16}\right) \text{Tr}[\yud\yu]-\frac{27}{4} \text{Tr}[\ydd\yd\ydd\yd]+\frac{3}{2} \text{Tr}[\ydd\yd\yud\yu]\\
%%%
&\left.-\frac{9}{4} \text{Tr}[\yld\yl\yld\yl]-\frac{27}{4} \text{Tr}[\yud\yu\yud\yu]\right)-\frac{3 g_{H_1}^5}{16}+g_{H_1^*} g_{H_2} g_{H_2^*}\left(-\frac{3}{2} \lambda+\frac{9}{40} g_1^2 +\frac{51}{8} g_2^2\right.  \\
%%%%
&\left. -\frac{3}{2} g_{H_1}^2 \right)-\frac{9}{16} g_{H_1^*}^3 g_{H_2} g_{H_2^*}-\frac{9}{8} g_{H_1^*} g_{H_2}^3 g_{H_2^*}-\frac{33}{16} g_{H_1^*} g_{H_2} g_{H_2^*}^3-\frac{9}{2} g_{H_1^*} g_{H_2} g_{H_2^*} \left(\text{Tr}[\ydd \yd]\right.\\
%%%%
&\left.+\frac{1}{3} \text{Tr}[\yld \yl]+\text{Tr}[\yud \yu]\right)+g_{H_1}^3 \left(-\frac{3 \lambda }{2}+\frac{309 g_1^2}{320}+\frac{165 g_2^2}{64}-\frac{15 g_{H_1^*}^2}{16}-\frac{9 g_{H_2}^2}{64}-\frac{27 g_{H_2^*}^2}{64}\right.\\
%%%
&\left.-\frac{27}{16} \left(\text{Tr}[\ydd\yd]+\frac{1}{3} \text{Tr}[\yld\yl]+\text{Tr}[\yud\yu]\right)\right).
\end{align*}
In the above equations, $\beta^{SM}_{y_t},\beta^{SM}_{y_b}$ and $\beta^{SM}_{y_\tau}$ denote the Standard Model beta functions, which are given for example in ref.~\cite{Luo:2002ey} and reproduced next,
\begin{align*}
 \beta_{y_u}^{SM} = &\frac{1}{16 \pi ^2}\yu \left\{-\frac{17}{20}g_1^2-\frac{9}{4}g_2^2-8 g_3^2+Y_2+\frac{3}{2} (\yud \yu-\ydd \yd)\right\}\!+\!\frac{1}{(16\pi^2)^2}y_u\left\{\frac{1187}{600}g_1^4 
           \!- \frac{9}{20}g_1^2g_2^2 \right. \\
%%%%
&+ \frac{19}{15}g_1^2g_3^2 
      - \frac{23}{4}g_2^4 
        + 9 g_2^2g_3^2 - 108 g_3^4+
 \left(\frac{5 \ydd \yd}{4}-\frac{9 \yud \yu}{4}\right) Y_2\!-\left(16 {g_3}^2-\frac{9 {g_2}^2}{16}\right.\\
%%
%%%%
&\left.+\frac{43 {g_1}^2}{80}\right) \ydd \yd+\left(16 {g_3}^2+\frac{135 {g_2}^2}{16}+\frac{223 {g_1}^2}{80}\right) \yud \yu+\frac{3 \lambda^2}{2}-6 \lambda \yud \yu-\chi_4+\frac{5 Y_4}{2}\\
%%%
&\left.-\frac{1}{4} \ydd \yd \yud \yu-\yud \yu \ydd \yd+\frac{11}{4} \ydd \yd \ydd \yd+\frac{3}{2} \yud \yu \yud \yu\right\},\\
\end{align*}
\begin{align*}
 \beta_{y_d}^{SM} = &\frac{1}{16 \pi ^2}\yd \left\{-\frac{1}{4}g_1^2-\frac{9}{4}g_2^2-8g_3^2 
+Y_2+\frac{3}{2} (\ydd \yd-\yud \yu)\right\}+\frac{1}{(16\pi^2)^2}y_d\left\{
-\frac{127}{600}g_1^4 \right.\\
%%%
&
      - \frac{27}{20}g_1^2g_2^2\! +\! \frac{31}{15}g_1^2g_3^2 \!- \frac{23}{4} g_2^4 
        + 9 g_2^2g_3^2 \!- 108 g_3^4\!+\!\left(\frac{5 \yud\yu}{4}\!-\frac{9 \ydd\yd}{4}\right) Y_2+\left(16 {g_3}^2+\frac{135 {g_2}^2}{16}\right.\\
%%%%
&+\left.\frac{187 {g_1}^2}{80}\right) \ydd\yd\!-\left(16 {g_3}^2\!-\frac{9 {g_2}^2}{16}\!+\!\frac{79 {g_1}^2}{80}\right) \yud\yu\!+\!\frac{3 \lambda^2 }{2}-6 \lambda \ydd\yd-\ydd\yd\yud\yu\\
%%%
&\left.+\frac{3}{2} \ydd\yd\ydd\yd-\chi_4 +\frac{5Y_4 }{2}-\frac{1}{4} \yud\yu\ydd\yd+\frac{11}{4} \yud\yu\yud\yu\right\},\\
\end{align*}
\begin{align*}
 \beta_{y_l}^{SM} = &\frac{1}{16 \pi ^2}\yl \left\{
-\frac{9}{4}(g_1^2\!+\!g_2^2)+Y_2\!+\!\frac{3}{2} \yld \yl\right\}+\frac{1}{(16\pi^2)^2}y_l\left\{\frac{1371}{200}g_1^4 + \frac{27}{20}g_1^2g_2^2 - \frac{23}{4} g_2^4-\frac{9}{4} \yld\yl Y_2\right.\\
%%%
&\left.+\left(\frac{135 g_2^2}{16}+\frac{387 g_1^2}{80}\right) \yld\yl+\frac{3 \lambda^2 }{2}-6 \lambda \yld\yl+\frac{3}{2} \yld\yl\yld\yl-\chi_4 +\frac{5 Y_4}{2}
\right\},
\end{align*}
where
\begin{align*}
Y_2=&\Tr[3\yud\yu+\yld\yl+3\ydd\yd],\\
%%%
\chi_4=&\frac{9}{4}{\rm Tr} \left[ 3(\yud\yu)^2 + 3(\ydd\yd)^2
         + (\yld\yl)^2 \right. 
       - \left. \frac{1}{3} \left\{ \yud\yu, \ydd\yd \right\} \right],\\
%%%
Y_4=&\left( \frac{17}{20}g_1^2+\frac{9}{4}g_2^2+8g_3^2 \right) {\rm Tr}(\yud\yu) 
     +  \left( \frac{1}{4}g_1^2+\frac{9}{4}g_2^2+8g_3^2 \right) {\rm Tr}(\ydd\yd ) 
     +  \frac{3}{4} \left( g_1^2+g_2^2 \right) {\rm Tr}(\yld\yl).
\end{align*}
The beta functions for $\gHcd$ and $\gHct$ are obtained from those of $\gHd$ and $\gHt$ by making the substitutions
\begin{align*}
 \gHd\leftrightarrow\gHcd, \quad \gHt\leftrightarrow\gHct.
\end{align*}
%%%%%%%%%%%%%%%%%%%%%%%%%%%%%%%%%%%%%%%%%%%%%%%%%%%%%%%%%%%%%%%%%%%%%%%%%%%%%%%%%
%%%%%%%%%%%%%%%%%%%%%%%%%%%%%%%%%%%%%%%%%%%%%%%%%%%%%%%%%%%%%%%%%%%%%%%%%%%%%%%%%
%%%%%%%%%%%%%%%%%%%%%%%%%%%%%%%%%%%%%%%%%%%%%%%%%%%%%%%%%%%%%%%%%%%%%%%%%%%%%%%%%
\subsubsection{Quartic coupling}
%%%%%%%%%%%%%%%%%%%%%%%%%%%%%%%%%%%%%%%%%%%%%%%%%%%%%%%%%%%%%%%%%%%%%%%%%%%%%%%%%
%%%%%%%%%%%%%%%%%%%%%%%%%%%%%%%%%%%%%%%%%%%%%%%%%%%%%%%%%%%%%%%%%%%%%%%%%%%%%%%%%
%%%%%%%%%%%%%%%%%%%%%%%%%%%%%%%%%%%%%%%%%%%%%%%%%%%%%%%%%%%%%%%%%%%%%%%%%%%%%%%%%
\begin{align*}
 \beta_\lambda=&\frac{1}{16\pi^2}\left\{12 \lambda ^2-\frac{9 \lambda  g_1^2}{5}+\frac{27 g_1^4}{100}-9 \lambda  g_2^2+\frac{9}{10} g_1^2 g_2^2+\frac{9 g_2^4}{4}+\lambda  g_{H_1}^2-\frac{g_{H_1}^4}{4}+\lambda  g_{H_1^*}^2-\frac{1}{2} g_{H_1}^2 g_{H_1^*}^2\right.\\
%%%
&\!-\!\frac{g_{H_1^*}^4}{4}+3 \lambda  g_{H_2}^2\!-\!\frac{1}{2} g_{H_1}^2 g_{H_2}^2\!-\!\frac{5 g_{H_2}^4}{4}\!-\!g_{H_1} g_{H_1^*} g_{H_2} g_{H_2^*}+3 \lambda  g_{H_2^*}^2-\frac{1}{2} g_{H_1^*}^2 g_{H_2^*}^2\!-\!\frac{1}{2} g_{H_2}^2 g_{H_2^*}^2\!-\frac{5 g_{H_2^*}^4}{4}\\
%%%
&\left.+12 \lambda  \text{Tr}\left[\ydd\yd+\frac{\yld\yl}{3}+\yud\yu\right]-12 \text{Tr}[\ydd\yd\ydd\yd]-4 \text{Tr}[\yld\yl\yld\yl]-12 \text{Tr}[\yud\yu\yud\yu]\right\}\\
%%%
&+\frac{1}{(16\pi^2)^2}\left\{-78 \lambda ^3+\frac{54}{5} \lambda ^2 g_1^2+\frac{2007 \lambda  g_1^4}{200}-\frac{3699 g_1^6}{1000}+54 \lambda ^2 g_2^2+\frac{117}{20} \lambda  g_1^2 g_2^2-\frac{1773}{200} g_1^4 g_2^2\right.\\
%%%
&+\frac{47 \lambda  g_2^4}{8}-\frac{77}{8} g_1^2 g_2^4+\frac{209 g_2^6}{8}-6 \lambda ^2 (g_{H_1}^2+g_{H_1^*}^2)+\frac{3}{8} \lambda  g_1^2 (g_{H_1}^2+g_{H_1^*}^2)-\frac{9}{200} g_1^4 (g_{H_1}^2+g_{H_1^*}^2)
%%%%
\end{align*}
\begin{align*}
&+\frac{15}{8} \lambda  g_2^2 (g_{H_1}^2+g_{H_1^*}^2)-\frac{3}{20} g_1^2 g_2^2 (g_{H_1}^2+g_{H_1^*}^2)-\frac{3}{8} g_2^4 (g_{H_1}^2+g_{H_1^*}^2)-\frac{1}{16} \lambda ( g_{H_1}^4+g_{H_1^*}^4)\\
%%%
\phantom{\beta_\lambda=}&+\frac{5}{16}(g_{H_1}^6+g_{H^c_1}^6)+\frac{3}{4} \lambda  g_{H_1}^2 g_{H_1^*}^2+\frac{17}{16} (g_{H_1}^4 g_{H_1^*}^2+ g_{H_1}^2 g_{H_1^*}^4)-18 \lambda ^2 (g_{H_2}^2+g_{H_2^*}^2)\\
%%%
%%%
&+\frac{9}{8} \lambda  g_1^2 (g_{H_2}^2+g_{H_2^*}^2)-\frac{27}{200} g_1^4 (g_{H_2}^2+g_{H_2^*}^2)+\frac{165}{8} \lambda  g_2^2 (g_{H_2}^2+g_{H_2^*}^2)
+\frac{63}{20} g_1^2 g_2^2 (g_{H_2}^2+g_{H^c_2}^2)\\
%%%%
&
-\frac{153}{8} g_2^4 (g_{H_2}^2+g_{H^c_2}^2)
-\frac{1}{8} \lambda ( g_{H_1}^2 g_{H_2}^2+g_{H_1^*}^2 g_{H_2^*}^2)
-g_2^2 (g_{H_1}^2 g_{H_2}^2+g_{H_1^*}^2 g_{H_2^*}^2)\\
%%%
&
+\frac{17}{16} (g_{H_1}^4 g_{H_2}^2+g_{H_1^*}^4 g_{H_2^*}^2)+\frac{19}{16} g_{H_1}^2 g_{H_1^*}^2 (g_{H_2}^2+ g_{H_2^*}^2)-\frac{5}{16} \lambda  (g_{H_2}^4+g_{H_2^*}^4)-5 g_2^2 (g_{H_2}^4+g_{H_2^*}^4)\\
%%%
&
+\frac{11}{16} (g_{H_1}^2 g_{H_2}^4+g_{H_1^*}^2 g_{H_2^*}^4)+\frac{47}{16}(g_{H_2}^6+g_{H_2^*}^6)
+5 \lambda  g_{H_1} g_{H_1^*} g_{H_2} g_{H_2^*}-2 g_2^2 g_{H_1} g_{H_1^*} g_{H_2} g_{H_2^*}\\
%%%%%
&+\frac{21}{8} g_{H_2} g_{H_2^*}( g_{H_1}^3 g_{H_1^*}+ g_{H_1} g_{H_1^*}^3 )+\frac{19}{8} g_{H_1} g_{H_1^*} (g_{H_2}^3 g_{H_2^*}+g_{H_2} g_{H_2^*}^3)- g_{H_2}^2 g_{H_2^*}^2\left(\frac{11}{4} \lambda +2 g_2^2 \right)\\
%%%%
&+\frac{21}{16}  g_{H_2}^2 g_{H_2^*}^2(g_{H_1}^2+g_{H_1^*}^2)+\frac{7}{16}( g_{H_2}^4 g_{H_2^*}^2+g_{H_2}^2 g_{H_2^*}^4)\\
%%%
&+\left(-72 \lambda ^2+\frac{5 \lambda  g_1^2}{2}+\frac{9 g_1^4}{10}+\frac{45 \lambda  g_2^2}{2}+\frac{27}{5} g_1^2 g_2^2-\frac{9 g_2^4}{2}+80 \lambda  g_3^2\right) \text{Tr}[\ydd\yd]\\
%%%%
&+\left(-24 \lambda ^2+\frac{15 \lambda  g_1^2}{2}-\frac{9 g_1^4}{2}+\frac{15 \lambda  g_2^2}{2}+\frac{33}{5} g_1^2 g_2^2-\frac{3 g_2^4}{2}\right) \text{Tr}[\yld\yl]\\
%%%%
&+\left(-72 \lambda ^2+\frac{17 \lambda  g_1^2}{2}-\frac{171 g_1^4}{50}+\frac{45 \lambda  g_2^2}{2}+\frac{63}{5} g_1^2 g_2^2-\frac{9 g_2^4}{2}+80 \lambda  g_3^2\right) \text{Tr}[\yud\yu]\\
%%%
&+\left(-3 \lambda +\frac{8 g_1^2}{5}-64 g_3^2\right) \text{Tr}[\ydd\yd\ydd\yd]+\left(-\lambda -\frac{24 g_1^2}{5}\right) \text{Tr}[\yld\yl\yld\yl]-42 \lambda  \text{Tr}[\yud\yu\ydd\yd]\\
%%%%%
%%%%%
&+\left(-3 \lambda -\frac{16 g_1^2}{5}-64 g_3^2\right) \text{Tr}[\yud\yu\yud\yu]+60 \text{Tr}[\ydd\yd\ydd\yd\ydd\yd]-12 \text{Tr}[\ydd\yd\ydd\yd\yud\yu]\\
%%%%
&\left.-12 \text{Tr}[\ydd\yd\yud\yu\yud\yu]+20 \text{Tr}[\yld\yl\yld\yl\yld\yl]+60 \text{Tr}[\yud\yu\yud\yu\yud\yu]\right\}.
\end{align*}
%%%%%%%%%%%%%%%%%%%%%%%%%%%%%%%%%%%%%%%%%%%%%%%%%%%%%%%%%%%%%%%%%%%%%%%%%%%%%%%%%
%%%%%%%%%%%%%%%%%%%%%%%%%%%%%%%%%%%%%%%%%%%%%%%%%%%%%%%%%%%%%%%%%%%%%%%%%%%%%%%%%
%%%%%%%%%%%%%%%%%%%%%%%%%%%%%%%%%%%%%%%%%%%%%%%%%%%%%%%%%%%%%%%%%%%%%%%%%%%%%%%%%
\subsubsection{Fermion masses}
%%%%%%%%%%%%%%%%%%%%%%%%%%%%%%%%%%%%%%%%%%%%%%%%%%%%%%%%%%%%%%%%%%%%%%%%%%%%%%%%%
%%%%%%%%%%%%%%%%%%%%%%%%%%%%%%%%%%%%%%%%%%%%%%%%%%%%%%%%%%%%%%%%%%%%%%%%%%%%%%%%%
%%%%%%%%%%%%%%%%%%%%%%%%%%%%%%%%%%%%%%%%%%%%%%%%%%%%%%%%%%%%%%%%%%%%%%%%%%%%%%%%%
\begin{align*}
 \beta_\mu=&\frac{1}{16\pi^2}\left\{\mu\left(\!-\frac{9 g_1^2}{10}-\frac{9 g_2^2}{2}+\frac{1}{8}(g_{H_1}^2+g_{H_1^*}^2)+\frac{3 }{8}(g_{H_2}^2+g_{H_2^*}^2)\right)-\frac{1}{2} g_{H_1} g_{H_1^*} M_1-\frac{3}{2} g_{H_2} g_{H_2^*} M_2\right\}\\
%%%%
&+\frac{1}{(16\pi^2)^2}\left\{\mu  \left(g_{H_1}^2 \left(-\frac{g_{H_1^*}^2}{2}-\frac{9 g_{H_2^*}^2}{32}-\frac{9 g_{H_2}^2}{32}+\frac{33 g_1^2}{320}+\frac{33 g_2^2}{64}\right)+\left(-\frac{9 g_{H_2^*}^2}{32}-\frac{9 g_{H_2}^2}{32}\right.\right.\right.\\
%%%%
&\left.+\frac{33 g_1^2}{320}+\frac{33 g_2^2}{64}\right) g_{H_1^*}^2+g_{H_2}^2 \left(-\frac{45 g_{H_2^*}^2}{16}+\frac{99 g_1^2}{320}+\frac{363 g_2^2}{64}\right)+\left(\frac{99 g_1^2}{320}+\frac{363 g_2^2}{64}\right) g_{H_2^*}^2\end{align*}\begin{align*}
%%%%
&\left.+\frac{3}{4} g_{H_1} g_{H_2} g_{H_1^*} g_{H_2^*}-\frac{g_{H_1^*}^4}{16}-\frac{15 g_{H_2^*}^4}{32}-\frac{g_{H_1}^4}{16}-\frac{15 g_{H_2}^4}{32}+\frac{1359 g_1^4}{400}-\frac{27}{40} g_2^2 g_1^2-\frac{421 g_2^4}{16}\right)\\
%%%%
&+M_1 \left(\frac{1}{4} g_{H_1}^3 g_{H_1^*}+g_{H_1} \left(\frac{g_{H_1^*}^3}{4}+\left(-\frac{ 9}{20} g_1^2-\frac{9 g_2^2}{4}\right) g_{H_1^*}\right)\right)+M_2 \left(\frac{3}{4} g_{H_2}^3 g_{H_2^*}\right.\\
%%%
&\left.\left.+g_{H_2} \left(\frac{3 g_{H_2^*}^3}{4}+\left(-\frac{27}{20}  g_1^2-\frac{87 g_2^2}{4}\right) g_{H_2^*}\right)\right)\right\},
\end{align*}
\begin{align*}
\beta_{M_1}=&\frac{1}{16\pi^2}\left\{-2\mu \gHt\gHct+\frac{1}{2}M_1(\gHt^2+\gHct^2)\right\}+\frac{1}{(16\pi^2)^2}\left\{
\mu  \left(\frac{1}{4} g_{H_1}^3 g_{H_1^*}+g_{H_1} \left(\frac{g_{H_1^*}^3}{4}\right.\right.\right.\\
%%%%%
&\left.\left.+\left(\frac{3 g_{H_2^*}^2}{4}+\frac{3 g_{H_2}^2}{4}-\frac{12}{5}  g_1^2-12 g_2^2\right) g_{H_1^*}\right)\right)+M_1 \left(g_{H_1}^2 \left(-\frac{7 g_{H_1^*}^2}{8}-\frac{9 g_{H_2^*}^2}{16}-\frac{21 g_{H_2}^2}{32}\right.\right.\\
%%%%
&\left.\left.+\frac{51 g_1^2}{80}+\frac{51 g_2^2}{16}\right)+\frac{g_{H_1^*}^4}{32}+\left(-\frac{21 g_{H_2^*}^2}{32}-\frac{9 g_{H_2}^2}{16}+\frac{51 g_1^2}{80}+\frac{51 g_2^2}{16}\right) g_{H_1^*}^2+\frac{g_{H_1}^4}{32}\right)\\
%%%%
&+M_2 \left(-\frac{3}{2} g_{H_1} g_{H_2} g_{H_1^*} g_{H_2^*}+\frac{3}{4} g_{H_1^*}^2 g_{H_2^*}^2+\frac{3}{4} g_{H_1}^2 g_{H_2}^2\right)
-\frac{9}{4} \left(g_{H_1}^2+g_{H_1^*}^2\right)M_1 \text{Tr}\left[\ydd\yd\right.\\
%%%
&\left.\left.+\frac{\yld\yl}{3}+\yud\yd\right]\right\},
\end{align*}
\begin{align*}
\hskip-20pt\beta_{M_2}=&\frac{1}{16\pi^2}\left\{\left(-12 g_2^2+\frac{g_{H_2}^2}{2}+\frac{g_{H_2^*}^2}{2}\right) M_2-2 \mu  g_{H_2} g_{H_2^*}\right\}+\frac{1}{(16\pi^2)^2}\left\{\mu  \left(\frac{3}{4} g_{H_2}^3 g_{H_2^*}+g_{H_2} \left(\frac{3 g_{H_2^*}^3}{4}\right.\right.\right.\\
%%%%
&\left.\left.+\left(-\frac{12}{5}  g_1^2-24 g_2^2\right) g_{H_2^*}\right)+\frac{1}{4} g_{H_1}^2 g_{H_2} g_{H_2^*}+\frac{1}{4} g_{H_2} g_{H_1^*}^2 g_{H_2^*}\right)+M_1 \left(-\frac{1}{2} g_{H_1} g_{H_2} g_{H_1^*} g_{H_2^*}\right.\\
%%%%
&\left.+\frac{1}{4} g_{H_1^*}^2 g_{H_2^*}^2+\frac{1}{4} g_{H_1}^2 g_{H_2}^2\right)+M_2 \left(g_{H_1}^2 \left(-\frac{3 g_{H_2^*}^2}{16}-\frac{7}{32}  g_{H_2}^2\right)+\left(-\frac{7 g_{H_2^*}^2}{32}-\frac{3}{16}  g_{H_2}^2\right) g_{H_1^*}^2\right.\\
%%%%
&\left.+g_{H_2}^2 \left(-\frac{21 g_{H_2^*}^2}{8}+\frac{51 g_1^2}{80}+\frac{91 g_2^2}{16}\right)+\left(\frac{51 g_1^2}{80}+\frac{91 g_2^2}{16}\right) g_{H_2^*}^2-\frac{29 g_{H_2^*}^4}{32}-\frac{29 g_{H_2}^4}{32}-\frac{233}{3}  g_2^4\right)\\
%%%
&\left.-\frac{9}{4} \left(g_{H_2}^2+g_{H_2^*}^2\right) M_2\text{Tr}\left[\ydd\yd+\frac{\yld\yl}{3}+\yud\yu\right]\right\},
\end{align*}
\begin{align*}
\hskip-5.5cm\beta_{M_3}=&-\frac{18 g_3^2 M_3}{16\pi^2}-\frac{228}{(16\pi^2)^2} g_3^4 M_3.
\end{align*}
%%%%%%%%%%%%%%%%%%%%%%%%%%%%%%%%%%%%%%%%%%%%%%%%%%%%%%%%%%%%%%%%%%%%%%%%%%%%%%%%%
%%%%%%%%%%%%%%%%%%%%%%%%%%%%%%%%%%%%%%%%%%%%%%%%%%%%%%%%%%%%%%%%%%%%%%%%%%%%%%%%%
%%%%%%%%%%%%%%%%%%%%%%%%%%%%%%%%%%%%%%%%%%%%%%%%%%%%%%%%%%%%%%%%%%%%%%%%%%%%%%%%%
\subsubsection{Scalar masses}
%%%%%%%%%%%%%%%%%%%%%%%%%%%%%%%%%%%%%%%%%%%%%%%%%%%%%%%%%%%%%%%%%%%%%%%%%%%%%%%%%
%%%%%%%%%%%%%%%%%%%%%%%%%%%%%%%%%%%%%%%%%%%%%%%%%%%%%%%%%%%%%%%%%%%%%%%%%%%%%%%%%
%%%%%%%%%%%%%%%%%%%%%%%%%%%%%%%%%%%%%%%%%%%%%%%%%%%%%%%%%%%%%%%%%%%%%%%%%%%%%%%%%
\begin{align*}
 \beta_{m^2}=&\frac{1}{16\pi^2}\left\{m_H^2 \left(6 \lambda -\frac{9 g_1^2}{10}-\frac{9 g_2^2}{2}+\frac{1}{2}(g_{H_1}^2+g_{H_1^*}^2)+\frac{3}{2}( g_{H_2}^2+ g_{H_2^*}^2)+2 \text{Tr}[3 \ydd\yd+\yld\yl\right.\right.\\
%%%
&\left.+3 \yud\yu]\right)-\mu ^2 \left(g_{H_1}^2+g_{H_1^*}^2+3( g_{H_2}^2+g_{H_2^*}^2)\right)-\left(g_{H_1}^2+g_{H_1^*}^2\right) M_1^2-3\left( g_{H_2}^2+ g_{H_2^*}^2\right) M_2^2\\
%%%
&\left.+\mu  \left(2 g_{H_1} g_{H_1^*} M_1+6 g_{H_2} g_{H_2^*} M_2\right)\right\}
%%%%%%%%%%%%%%%%%%%%
%%%%%%%%%%%%%%%%%%%%%
+\frac{1}{(16\pi^2)^2}\left\{\mu ^2 \left(-\frac{54 g_1^4}{25}-18 g_2^4-\frac{3}{10} g_1^2 (g_{H_1}^2+g_{H_1^*}^2)\right.\right.\\
%%%
&-\frac{3}{2} g_2^2 (g_{H_1}^2+g_{H_1^*}^2)+g_{H_1}^4+g_{H_1^*}^4+\frac{21}{4} g_{H_1}^2 g_{H_1^*}^2-\frac{9}{10} g_1^2 (g_{H_2}^2+g_{H_2^*}^2)-\frac{21}{2} g_2^2 (g_{H_2}^2+g_{H_2^*}^2)\\
%%%%
&+\frac{3}{2} (g_{H_1}^2 g_{H_2}^2+g_{H_1^*}^2 g_{H_2^*}^2)+\frac{3}{4} (g_{H_1^*}^2 g_{H_2}^2+g_{H_1}^2 g_{H_2^*}^2)+\frac{9 }{2}(g_{H_2}^4+g_{H_2^*}^4)+9 g_{H_1} g_{H_1^*} g_{H_2} g_{H_2^*}\\
%%%%
&\left.+\frac{33}{4} g_{H_2}^2 g_{H_2^*}^2\right)+M_1^2\left(\frac{11 }{8}(g_{H_1}^4+g_{H_1^*}^4)+3 g_{H_1}^2 g_{H_1^*}^2+\frac{9}{8} (g_{H_1}^2 g_{H_2}^2+ g_{H_1^*}^2 g_{H_2^*}^2)\right.\\
%%%%%
&\left.
+3 g_{H_1} g_{H_1^*} g_{H_2} g_{H_2^*}\right) 
+M_1 M_2\left(\frac{3}{2} (g_{H_1}^2 g_{H_2}^2+g_{H_1^*}^2 g_{H_2^*}^2)+3 g_{H_1} g_{H_1^*} g_{H_2} g_{H_2^*}\right) +M_2^2\left(-36 g_2^4\right.\\
%%%%
&\left.\!-\!18 g_2^2 (g_{H_2}^2\!+g_{H_2^*}^2)
\!+\frac{9}{8} (g_{H_1}^2 g_{H_2}^2\!+g_{H_1^*}^2 g_{H_2^*}^2)
\!+\frac{39 }{8}(g_{H_2}^4\!+g_{H_2^*}^4)
\!+\!3 g_{H_1} g_{H_1^*} g_{H_2} g_{H_2^*}\!+\!6 g_{H_2}^2 g_{H_2^*}^2\right) \\
%%%
&+\mu  \left(\!\left(\!-6 \lambda  g_{H_1} g_{H_1^*}\!+\!\frac{3}{10} g_1^2 g_{H_1} g_{H_1^*}+\frac{3}{2} g_2^2 g_{H_1} g_{H_1^*}\!-\frac{19}{4} (g_{H_1}^3 g_{H_1^*}\!+g_{H_1} g_{H_1^*}^3)-\frac{9}{4} g_{H_1} g_{H_1^*} ( g_{H_2}^2\right.\right.\\
%%%
&\left.+g_{H_2^*}^2)
-3 g_{H_1}^2 g_{H_2} g_{H_2^*}-3 g_{H_1^*}^2 g_{H_2} g_{H_2^*}\right) M_1+M_2\left(-3 g_{H_1} g_{H_1^*} (g_{H_2}^2+g_{H_2^*}^2)
-18 \lambda  g_{H_2} g_{H_2^*}\right.\\
%%%%
&\left.\left.+\frac{9}{10} g_1^2 g_{H_2} g_{H_2^*}+\frac{57}{2} g_2^2 g_{H_2} g_{H_2^*}-\frac{9}{4} g_{H_2} g_{H_2^*}(g_{H_1}^2 +g_{H_1^*}^2)-\frac{51}{4}( g_{H_2}^3 g_{H_2^*}+ g_{H_2} g_{H_2^*}^3)\right) \right)\\
%%%%
&+m_H^2 \left(-15 \lambda ^2+\frac{36 \lambda  g_1^2}{5}+\frac{1791 g_1^4}{400}+36 \lambda  g_2^2+\frac{9}{8} g_1^2 g_2^2-\frac{25 g_2^4}{16}-3 \lambda ( g_{H_1}^2+g_{H_1^*}^2)\right.\\
%%%
&+\frac{3}{16} g_1^2 (g_{H_1}^2+g_{H_1^*}^2)+\frac{15}{16} g_2^2 (g_{H_1}^2+ g_{H_1^*}^2)\!-\frac{9}{32}( g_{H_1}^4+ g_{H_1^*}^4)-\frac{3}{8} g_{H_1}^2 g_{H_1^*}^2-9 \lambda ( g_{H_2}^2+g_{H_2^*}^2)\\
%%%
&+\frac{9}{16} g_1^2 (g_{H_2}^2+g_{H_2^*}^2)+\frac{165}{16} g_2^2 (g_{H_2}^2+g_{H_2^*}^2)-\frac{9}{16}( g_{H_1}^2 g_{H_2}^2+g_{H_1^*}^2 g_{H_2^*}^2)-\frac{45 }{32}(g_{H_2}^4+g_{H_2^*}^4)\\
%%%%
&\!-\!\frac{9}{8} g_{H_2}^2 g_{H_2^*}^2+\left(\frac{5 g_1^2}{4}+\frac{45 g_2^2}{4}+40 g_3^2-36 \lambda \right) \text{Tr}[\ydd\yd]+\left(\frac{15 g_1^2}{4}+\frac{15 g_2^2}{4}-12 \lambda \right) \text{Tr}[\yld\yl]\\
%%%%
&+\left(-36 \lambda +\frac{17 g_1^2}{4}+\frac{45 g_2^2}{4}+40 g_3^2\right) \text{Tr}[\yud\yu]-\frac{27}{2} \text{Tr}[\ydd\yd\ydd\yd]-\frac{9}{2} \text{Tr}[\yld\yl\yld\yl]\\
%%%%
&\left.\left.-9 \text{Tr}[\yud\yu\ydd\yd]-\frac{27}{2} \text{Tr}[\yud\yu\yud\yu]\right)\right\}.
\end{align*}

%%%%%%%%%%%%%%%%%%%%%%%%%%%%%%%%%%%%%%%%%%%%%%%%%%%%%%%%%%%%%%%%%%%%%%%%%%%%%%%%%
%%%%%%%%%%%%%%%%%%%%%%%%%%%%%%%%%%%%%%%%%%%%%%%%%%%%%%%%%%%%%%%%%%%%%%%%%%%%%%%%%
%%%%%%%%%%%%%%%%%%%%%%%%%%%%%%%%%%%%%%%%%%%%%%%%%%%%%%%%%%%%%%%%%%%%%%%%%%%%%%%%%
\subsubsection{Higgs anomalous dimension}
%%%%%%%%%%%%%%%%%%%%%%%%%%%%%%%%%%%%%%%%%%%%%%%%%%%%%%%%%%%%%%%%%%%%%%%%%%%%%%%%%
%%%%%%%%%%%%%%%%%%%%%%%%%%%%%%%%%%%%%%%%%%%%%%%%%%%%%%%%%%%%%%%%%%%%%%%%%%%%%%%%%
%%%%%%%%%%%%%%%%%%%%%%%%%%%%%%%%%%%%%%%%%%%%%%%%%%%%%%%%%%%%%%%%%%%%%%%%%%%%%%%%%
The anomalous dimension is gauge dependent; the standard gauge parameter is denoted by $\xi$, so that $\xi=0$ corresponds to the Landau gauge.
\begin{align*}
\gamma_H=&\frac{1}{16\pi^2}\left\{3 y_b^2+\frac{g_{H_1^*}^2}{4}+\frac{3 g_{H_2^*}^2}{4}+\frac{g_{H_1}^2}{4}+\frac{3 g_{H_2}^2}{4}+g_1^2 \left(\frac{3 \xi }{20}-\frac{9}{20}\right)+g_2^2 \left(\frac{3 \xi }{4}-\frac{9}{4}\right)+3 y_t^2+y_{\tau }^2\right\}\\
%%%%%
&+\frac{1}{(16\pi^2)^2}\left\{g_1^2 \left(\frac{5 y_b^2}{8}\!+\!\frac{3 g_{H_1^*}^2}{32}\!+\!\frac{9 g_{H_2^*}^2}{32}\!+\!\frac{3 g_{H_1}^2}{32}\!+\!\frac{9 g_{H_2}^2}{32}\!+\!\frac{27 g_2^2}{80}+\frac{17 y_t^2}{8}+\frac{15 y_{\tau }^2}{8}\right)+g_2^2 \left(\frac{45 y_b^2}{8}\right.\right.\\
%%%%
&\left.+\frac{15 g_{H_1^*}^2}{32}+\frac{165 g_{H_2^*}^2}{32}+\frac{15 g_{H_1}^2}{32}+\frac{165 g_{H_2}^2}{32}+\frac{45 y_t^2}{8}+\frac{15 y_{\tau }^2}{8}\right)+g_3^2 \left(20 y_b^2+20 y_t^2\right)+\frac{3}{2} y_b^2 y_t^2\\
%%%%
&-\frac{27 y_b^4}{4}+g_{H_1}^2 \left(-\frac{5 g_{H_1^*}^2}{16} -\frac{9 g_{H_2}^2}{32}\right)-\frac{9}{32} g_{H_1^*}^2 g_{H_2^*}^2-\frac{3}{16} g_{H_2}^2 g_{H_2^*}^2-\frac{3}{4} g_{H_1} g_{H_2} g_{H_1^*} g_{H_2^*}-\frac{9 g_{H_1^*}^4}{64}\\
%%%%
&\left.-\frac{45 g_{H_2^*}^4}{64}-\frac{9 g_{H_1}^4}{64}-\frac{45 g_{H_2}^4}{64}+g_2^4 \left(\frac{3 \xi ^2}{8}+3 \xi -\frac{151}{32}\right)+\frac{1413 g_1^4}{800}+\frac{3 \lambda ^2}{2}-\frac{27 y_t^4}{4}-\frac{9 y_{\tau }^4}{4}\right\}.
\end{align*}
\newpage
%%%%%%%%%%%%%%%%%%%%%%%%%%%%%%%%%%%%%%%%%%%%%%%%%%%%%%%%%%%%%%%%%%%%%%%%%%%%%%%%
%%%%%%%%%%%%%%%%%%%%%%%%%%%%%%%%%%%%%%%%%%%%%%%%%%%%%%%%%%%%%%%%%%%%%%%%%%%%%%%%
%%%%%%%%%%%%%%%%%%%%%%%%%%%%%%%%%%%%%%%%%%%%%%%%%%%%%%%%%%%%%%%%%%%%%%%%%%%%%%%%
%%%%%%%%%%%%%%%%%%%%%%%%%%%%%%%%%%%%%%%%%%%%%%%%%%%%%%%%%%%%%%%%%%%%%%%%%%%%%%%%
%%%%%%%%%%%%%%%%%%%%%%%%%%%%%%%%%%%%%%%%%%%%%%%%%%%%%%%%%%%%%%%%%%%%%%%%%%%%%%%%
%%%%%%%%%%%%%%%%%%%%%%%%%%%%%%%%%%%%%%%%%%%%%%%%%%%%%%%%%%%%%%%%%%%%%%%%%%%%%%%%
%%%%%%%%%%%%%%%%%%%%%%%%%%%%%%%%%%%%%%%%%%%%%%%%%%%%%%%%%%%%%%%%%%%%%%%%%%%%%%%%%
%%%%%%%%%%%%%%%%%%%%%%%%%%%%%%%%%%%%%%%%%%%%%%%%%%%%%%%%%%%%%%%%%%%%%%%%%%%%%%%%%
%%%%%%%%%%%%%%%%%%%%%%%%%%%%%%%%%%%%%%%%%%%%%%%%%%%%%%%%%%%%%%%%%%%%%%%%%%%%%%%%%
%%%%%%%%%%%%%%%%%%%%%%%%%%%%%%%%%%%%%%%%%%%%%%%%%%%%%%%%%%%%%%%%%%%%%%%%%%%%%%%%%
%%%%%%%%%%%%%%%%%%%%%%%%%%%%%%%%%%%%%%%%%%%%%%%%%%%%%%%%%%%%%%%%%%%%%%%%%%%%%%%%%
\subsection{Minimal Effective SUSY\label{subsec:Betas:MinEffSUSY}}
%%%%%%%%%%%%%%%%%%%%%%%%%%%%%%%%%%%%%%%%%%%%%%%%%%%%%%%%%%%%%%%%%%%%%%%%%%%%%%%%%
%%%%%%%%%%%%%%%%%%%%%%%%%%%%%%%%%%%%%%%%%%%%%%%%%%%%%%%%%%%%%%%%%%%%%%%%%%%%%%%%%
%%%%%%%%%%%%%%%%%%%%%%%%%%%%%%%%%%%%%%%%%%%%%%%%%%%%%%%%%%%%%%%%%%%%%%%%%%%%%%%%%
This section deals with the beta functions of the independent couplings (eq.~\eqref{eq:couplingsMES}) of the Lagrangian of eq.~\eqref{eq:LagrangianMES}. For simplicity, phases and the flavor-mixing couplings ${y_{u/d/l}}_{ij},$ $ g_{{Q/U/D}_{i,k}},$ $ {z_{q/u/d}}_i\,i,j\neq3$ have been taken to zero in most cases. Also, the more compact notation of eq.~\eqref{eq:couplingsMES:notation} is employed for the beta functions other than those of the gauge couplings. 
Some beta functions are given in comparison with those of the Split Susy scenario, which  are denoted with the superscript ``SS''. 2 loop contributions are only given for the gauge couplings, Standard Model-like Yukawas and the fermion and scalar soft mass parameters. The full formulae are available online.
\subsubsection{Gauge couplings}
%%%%%%%%%%%%%%%%%%%%%%%%%%%%%%%%%%%%%%%%%%%%%%%%%%%%%%%%%%%%%%%%%%%%%%%%%%%%%%%%%
%%%%%%%%%%%%%%%%%%%%%%%%%%%%%%%%%%%%%%%%%%%%%%%%%%%%%%%%%%%%%%%%%%%%%%%%%%%%%%%%%
%%%%%%%%%%%%%%%%%%%%%%%%%%%%%%%%%%%%%%%%%%%%%%%%%%%%%%%%%%%%%%%%%%%%%%%%%%%%%%%%%
\begin{align*}
\beta_{g_1}=&\frac{3}{10\pi^2}g_1^3-\frac{1}{(16\pi^2)^2}\Big\{-\frac{251 g_1^5}{50}-\frac{39}{10} g_1^3 g_2^2-\frac{68}{5} g_1^3 g_3^2+\frac{3}{40} g_1^3 (\bar{g}_1^2+\bar{g}_2^2)+\frac{9}{40} g_1^3 (\bar{g}_3^2+\bar{g}_4^2)\\
%%%
&+g_1^3\!\left(\Tr\!\left[\frac{1}{2} y_d^\dagger y_d+\frac{3}{2} y_l^\dagger y_l+\frac{17}{10} y_u^\dagger y_u\right]\!+\!\sum_{i=1}^3\left(\!\frac{2}{15}(g_{Q_{i,3}})^2\!+\!\frac{3}{40}(g_{Q_{i,2}})^2\!+\!\frac{1}{360}(g_{Q_{i,1}})^2\right.\right.\\
%%%
&\left.\left.+\frac{16}{15}(g_{U_{i,3}})^2+\frac{16}{45}(g_{U_{i,1}})^2+\frac{13}{10}z_{d,i}^2+z_{q,i}^2+\frac{5}{2}z_{u,i}^2\right)\right)\Big\},\\
%%%%%
\beta_{g_2}=&\!-\!\frac{1}{24\pi^2}g_2^3\!-\!\frac{1}{(16\pi^2)^2}\Big\{-\frac{13}{10} g_1^2 g_2^3-\frac{251 g_2^5}{6}-20 g_2^3 g_3^2+\frac{1}{8} g_2^3 (\bar{g}_1^2+\bar{g}_2^2)+\frac{11}{8} g_2^3 (\bar{g}_3^2+ \bar{g}_4^2)\\
%%%%
&+g_2^3\left(\Tr\left[\frac{3}{2}(\yud\yu+\ydd\yd)+\frac{1}{2}\yld\yl\right]+\sum_{i=1}^3\left(2(g_{Q_{i,3}})^2+\frac{33}{8}(g_{Q_{i,2}})^2+\frac{1}{24}(g_{Q_{i,1}})^2+\frac{3}{2}z_{d,i}^2\right.\right.\\
%%%
&\left.\left.+3z_{q,i}^2+\frac{3}{2}z_{u,i}^2\right)\right)
\Big\},\\
%%%%%%
\beta_{g_3}=&\!-\!\frac{9}{32\pi^2}g_3^3\!-\!\frac{1}{(16\pi^2)^2}\Big\{\!\!-\frac{17}{10} g_1^2 g_3^3\!-\frac{15}{2} g_2^2 g_3^3\!-\!33 g_3^5
+g_3^3\left.(2\Tr\, [y_d^\dagger y_d+y_u^\dagger y_u]+\!\sum_{i=1}^3\!\left(\!\frac{13}{3}(g_{Q_{i,3}})^2\right.\right.\\
%%%
&\left.\left.+\frac{3}{4}(g_{Q_{i,2}})^2+\frac{1}{36}(g_{Q_{i,1}})^2+\frac{13}{6}(g_{U_{i,3}})^2+\frac{2}{9}(g_{U_{i,1}})^2+z_{d,i}^2+z_{q,i}^2+z_{u,i}^2\right)\right)\Big\}.
\end{align*}
%%%%%%%%%%%%%%%%%%%%%%%%%%%%%%%%%%%%%%%%%%%%%%%%%%%%%%%%%%%%%%%%%%%%%%%%%%%%%%%%%
%%%%%%%%%%%%%%%%%%%%%%%%%%%%%%%%%%%%%%%%%%%%%%%%%%%%%%%%%%%%%%%%%%%%%%%%%%%%%%%%%
%%%%%%%%%%%%%%%%%%%%%%%%%%%%%%%%%%%%%%%%%%%%%%%%%%%%%%%%%%%%%%%%%%%%%%%%%%%%%%%%%
\subsubsection{Yukawas}
%%%%%%%%%%%%%%%%%%%%%%%%%%%%%%%%%%%%%%%%%%%%%%%%%%%%%%%%%%%%%%%%%%%%%%%%%%%%%%%%%
%%%%%%%%%%%%%%%%%%%%%%%%%%%%%%%%%%%%%%%%%%%%%%%%%%%%%%%%%%%%%%%%%%%%%%%%%%%%%%%%%
%%%%%%%%%%%%%%%%%%%%%%%%%%%%%%%%%%%%%%%%%%%%%%%%%%%%%%%%%%%%%%%%%%%%%%%%%%%%%%%%%
\begin{align*}
\hskip-0.6cm \beta_{y_t}=&\beta_{y_t}^{SS}|_{\lambda=0}+\frac{1}{16\pi^2}\left\{z_u \left(\frac{1}{6} \bar{g}_1 \bar{g}_5-\frac{3}{2} \bar{g}_3 \bar{g}_6\right)-\frac{2}{3} z_q \bar{g}_1 \bar{g}_8+y_t \left(\frac{z_q^2}{2}+z_u^2+\frac{\bar{g}_5^2}{72}+\frac{3 \bar{g}_6^2}{8}+\frac{2 \bar{g}_7^2}{3}\right.\right.\\
%%%%
&\left.\left.+\frac{2 \bar{g}_8^2}{9}+\frac{2 \bar{g}_9^2}{3}\right)\right\}+\frac{1}{(16\pi^2)^2}\left\{y_b z_d z_u \left(\frac{1}{2} \bar{g}_1 \bar{g}_2-\frac{3}{2} \bar{g}_3 \bar{g}_4\right)+z_q^2 z_u \left(-\frac{1}{4} \bar{g}_1 \bar{g}_5+\frac{9}{4} \bar{g}_3 \bar{g}_6\right)\right.
%%%%%
\end{align*}
\begin{align*}
&+z_u^3 \left(-\frac{1}{4} \bar{g}_1 \bar{g}_5+\frac{9}{4} \bar{g}_3 \bar{g}_6\right)+z_q^3 \bar{g}_1 \bar{g}_8+y_{\tau }^2 \left(z_u \left(-\frac{1}{6} \bar{g}_1 \bar{g}_5+\frac{3}{2} \bar{g}_3 \bar{g}_6\right)+\frac{2}{3} z_q \bar{g}_1 \bar{g}_8\right)\\
%%%%
\phantom{\beta_{y_t}=}&+y_b^2 \left(z_u \left(\frac{9}{2} \bar{g}_3 \bar{g}_6-\frac{1}{2} \bar{g}_1 \bar{g}_5\right)+2 z_q \bar{g}_1 \bar{g}_8\right)+y_t^2 \left(z_u \left(9 \bar{g}_3 \bar{g}_6-\bar{g}_1 \bar{g}_5\right)+4 z_q \bar{g}_1 \bar{g}_8\right)+z_q \left(z_u^2 \bar{g}_1 \bar{g}_8\right.\\
%%%%
&+\frac{5}{12} \bar{g}_1^3 \bar{g}_8+\frac{1}{2} \bar{g}_2 \bar{g}_3 \bar{g}_4 \bar{g}_8+\bar{g}_1 \left(\left(-\frac{7 g_1^2}{15}+g_2^2-\frac{16 g_3^2}{3}-\frac{2 \hat{\gamma} _3}{9}\right) \bar{g}_8+\frac{1}{2} \bar{g}_2^2 \bar{g}_8+\frac{3}{4} \bar{g}_3^2 \bar{g}_8+\frac{1}{2} \bar{g}_4^2 \bar{g}_8\right.\\
%%%%
&\left.\left.+\frac{1}{18} \bar{g}_5^2 \bar{g}_8+\frac{4 \bar{g}_8^3}{9}\right)\right)+z_u \left(-\frac{5}{48} \bar{g}_1^3 \bar{g}_5-\frac{1}{8} \bar{g}_2 \bar{g}_3 \bar{g}_4 \bar{g}_5+\frac{9}{16} \bar{g}_1^2 \bar{g}_3 \bar{g}_6+\frac{3}{8} \bar{g}_2^2 \bar{g}_3 \bar{g}_6+\frac{33}{16} \bar{g}_3^3 \bar{g}_6\right.\\
%%%%%
&+\bar{g}_3 \left(\left(\frac{3 g_1^2}{40}-\frac{69 g_2^2}{8}-12 g_3^2+\frac{\hat{\gamma} _2}{8}+\frac{3 \hat{\gamma} _7}{8}\right) \bar{g}_6+\frac{9}{8} \bar{g}_4^2 \bar{g}_6+\frac{9 \bar{g}_6^3}{8}\right)+\bar{g}_1 \left(-\frac{1}{8} \bar{g}_2^2 \bar{g}_5-\frac{3}{16} \bar{g}_3^2 \bar{g}_5\right.\\
%%%%
&\left.\left.-\frac{1}{8} \bar{g}_4^2 \bar{g}_5-\frac{\bar{g}_5^3}{72}+\frac{3}{8} \bar{g}_2 \bar{g}_4 \bar{g}_6+\bar{g}_5 \left(-\frac{g_1^2}{120}+\frac{g_2^2}{8}+\frac{4 g_3^2}{3}-\frac{\hat{\gamma} _2}{72}+\frac{\hat{\gamma} _7}{8}-\frac{\bar{g}_8^2}{9}\right)\right)\right)+y_t^3 \left(-\frac{5 z_q^2}{2}\right.\\
%%%%
&\left.-\frac{19 z_u^2}{4}-\frac{3 \hat{\gamma} _1}{2}-\frac{5 \bar{g}_5^2}{72}-\frac{15 \bar{g}_6^2}{8}-\frac{10 \bar{g}_7^2}{3}-\frac{19 \bar{g}_8^2}{18}-\frac{19 \bar{g}_9^2}{6}\right)+y_t \left(\frac{2 g_1^4}{5}+\frac{3 g_2^4}{2}+\frac{22 g_3^4}{3}-\frac{15 z_q^4}{8}\right.\\
%%%%
&-\frac{9 z_u^4}{4}+\frac{3 \hat{\gamma} _1^2}{32}+\frac{\hat{\gamma} _2^2}{48}+\frac{\hat{\gamma} _3^2}{6}+\frac{9 \hat{\gamma} _7^2}{16}-\frac{\bar{g}_5^4}{864}-\frac{1}{4} \bar{g}_2 \bar{g}_4 \bar{g}_5 \bar{g}_6-\frac{57}{64} \bar{g}_3^2 \bar{g}_6^2-\frac{81}{64} \bar{g}_4^2 \bar{g}_6^2-\frac{9 \bar{g}_6^4}{16}\\
%%%%
&+y_b z_d \left(\frac{1}{3} \bar{g}_2 \bar{g}_5+3 \bar{g}_4 \bar{g}_6\right)-\frac{3 \bar{g}_7^4}{2}+z_u^2 \left(\frac{51 g_2^2}{8}-\frac{31 g_1^2}{120}+\frac{22 g_3^2}{3}-\frac{\hat{\gamma} _2}{3}-\frac{5 \bar{g}_1^2}{8}-\frac{15 \bar{g}_3^2}{8}-\frac{11 \bar{g}_5^2}{48}\right.\\
%%%%
&\left.-\frac{99 \bar{g}_6^2}{16}-3 \bar{g}_7^2\right)+y_b^2 \left(-\frac{11 z_d^2}{4}-\frac{5 z_q^2}{4}-\frac{5 \bar{g}_5^2}{144}-\frac{15 \bar{g}_6^2}{16}-\frac{5 \bar{g}_7^2}{3}\right)+\bar{g}_6^2 \left(\frac{11 g_1^2}{320}+\frac{75 g_2^2}{64}+\frac{11 g_3^2}{4}\right.\\
%%%%
&\left.-\frac{\hat{\gamma} _2}{8}-\frac{3 \hat{\gamma} _7}{8}-\frac{3 \bar{g}_7^2}{2}\right)+z_d^2 \left(-\frac{3 z_u^2}{2}-\frac{9 \bar{g}_2^2}{16}-\frac{27 \bar{g}_4^2}{16}-\frac{\bar{g}_5^2}{48}-\frac{9 \bar{g}_6^2}{16}-\bar{g}_7^2\right)+\frac{4}{3} z_q z_u \bar{g}_5 \bar{g}_8-\frac{2 \bar{g}_8^4}{9}\\
%%%%
&+\bar{g}_1^2 \left(-\frac{19 \bar{g}_5^2}{576}-\frac{5 \bar{g}_8^2}{18}\right)+\bar{g}_2^2 \left(-\frac{11 \bar{g}_5^2}{576}-\frac{\bar{g}_8^2}{18}\right)+\bar{g}_5^2 \left(\frac{11 g_1^2}{8640}+\frac{11 g_2^2}{192}+\frac{11 g_3^2}{108}-\frac{\hat{\gamma} _2}{216}+\frac{\hat{\gamma} _7}{24}\right.\\
%%%%
&\left.-\frac{\bar{g}_6^2}{32}\!-\frac{\bar{g}_7^2}{18}\!+\!\frac{7 \bar{g}_8^2}{648}\right)\!+\!\frac{16}{27} \bar{g}_5 \bar{g}_7 \bar{g}_8 \bar{g}_9+\left(\frac{44 g_1^2}{45}+\frac{143 g_3^2}{9}+\frac{8 \hat{\gamma} _3}{9}\right) \bar{g}_9^2-\frac{17 \bar{g}_9^4}{12}+z_q^2 \left(-\frac{g_1^2}{15}-\frac{3 g_2^2}{2}\right.\\
%%%%
&\left.+\frac{11 g_3^2}{3}+\frac{7 z_u^2}{8}+\frac{2 \hat{\gamma} _3}{3}-\frac{19 \bar{g}_1^2}{32}-\frac{57 \bar{g}_3^2}{32}-\frac{11 \bar{g}_8^2}{3}-3 \bar{g}_9^2\right)+\bar{g}_8^2 \left(\frac{44 g_1^2}{135}+\frac{44 g_3^2}{27}+\frac{8 \hat{\gamma} _3}{27}-\frac{8 \bar{g}_9^2}{9}\right)\\
%%%%
&\left.\left.+\bar{g}_7^2 \left(\frac{11 g_1^2}{180}+\frac{11 g_2^2}{4}+\frac{143 g_3^2}{9}-\frac{2 \hat{\gamma} _2}{9}+2 \hat{\gamma} _7-\frac{25 \bar{g}_9^2}{36}\right)\right)\right\},
\end{align*}

\begin{align*}
 \beta_{y_b}=&\beta_{y_b}^{SS}|_{\lambda=0}+\frac{1}{16\pi^2}\left\{z_d \left(-\frac{1}{6} \bar{g}_2 \bar{g}_5-\frac{3}{2} \bar{g}_4 \bar{g}_6\right)+y_b \left(z_d^2+\frac{z_q^2}{2}+\frac{\bar{g}_5^2}{72}+\frac{3 \bar{g}_6^2}{8}+\frac{2 \bar{g}_7^2}{3}\right)\right\}\\
%%%%
&+\frac{1}{(16\pi^2)^2}\left\{y_t z_d z_u \left(\frac{1}{2} \bar{g}_1 \bar{g}_2-\frac{3}{2} \bar{g}_3 \bar{g}_4\right)+y_{\tau }^2 z_d \left(\frac{1}{6} \bar{g}_2 \bar{g}_5+\frac{3}{2} \bar{g}_4 \bar{g}_6\right)+z_d^3 \left(\frac{1}{4} \bar{g}_2 \bar{g}_5+\frac{9}{4} \bar{g}_4 \bar{g}_6\right)\right.\\
%%%%
&+y_t^2 z_d \left(\frac{1}{2} \bar{g}_2 \bar{g}_5\!+\!\frac{9}{2} \bar{g}_4 \bar{g}_6\right)\!+\!y_b^2 z_d \left(\bar{g}_2 \bar{g}_5\!+\!9 \bar{g}_4 \bar{g}_6\right)+y_b^3 \left(-\frac{19 z_d^2}{4}-\frac{5 z_q^2}{2}-\frac{3 \hat{\gamma} _1}{2}-\frac{5 \bar{g}_5^2}{72}-\frac{15 \bar{g}_6^2}{8}\right.
\end{align*}\begin{align*}
%%%%
\phantom{\beta_{y_b}=}&\left.-\frac{10 \bar{g}_7^2}{3}\right)+z_d \left(\frac{5}{48} \bar{g}_2^3 \bar{g}_5+\frac{9}{16} \bar{g}_2^2 \bar{g}_4 \bar{g}_6+\frac{9}{8} \bar{g}_3^2 \bar{g}_4 \bar{g}_6+\frac{33}{16} \bar{g}_4^3 \bar{g}_6+\bar{g}_1 \left(\frac{1}{8} \bar{g}_3 \bar{g}_4 \bar{g}_5+\frac{3}{8} \bar{g}_2 \bar{g}_3 \bar{g}_6\right)\right.\\
%%%%
&+\bar{g}_1^2 \left(\frac{1}{8} \bar{g}_2 \bar{g}_5+\frac{3}{8} \bar{g}_4 \bar{g}_6\right)+\bar{g}_4 \left(\left(\frac{21 g_1^2}{40}-\frac{69 g_2^2}{8}-12 g_3^2+\frac{\hat{\gamma} _2}{8}-\frac{3 \hat{\gamma} _7}{8}\right) \bar{g}_6+\frac{9 \bar{g}_6^3}{8}\right)\\
%%%%
&+\left.\bar{g}_2 \left(\frac{1}{8} \bar{g}_3^2 \bar{g}_5+\frac{3}{16} \bar{g}_4^2 \bar{g}_5+\frac{\bar{g}_5^3}{72}+\bar{g}_5 \left(\frac{7 g_1^2}{120}-\frac{g_2^2}{8}-\frac{4 g_3^2}{3}+\frac{\hat{\gamma} _2}{72}+\frac{\hat{\gamma} _7}{8}+\frac{\bar{g}_8^2}{9}\right)\right)\right)+y_b \left(\frac{7 g_1^4}{100}\right.\\
%%%%
&+\frac{3 g_2^4}{2}+\frac{22 g_3^4}{3}-\frac{9 z_d^4}{4}-\frac{15 z_q^4}{8}+\frac{3 \hat{\gamma} _1^2}{32}+\frac{\hat{\gamma} _2^2}{48}+\frac{\hat{\gamma} _3^2}{6}+\frac{9 \hat{\gamma} _7^2}{16}-\frac{\bar{g}_5^4}{864}+\frac{1}{4} \bar{g}_1 \bar{g}_3 \bar{g}_5 \bar{g}_6-\frac{81}{64} \bar{g}_3^2 \bar{g}_6^2\\
%%%%
&-\frac{57}{64} \bar{g}_4^2 \bar{g}_6^2-\frac{9 \bar{g}_6^4}{16}-\frac{3 \bar{g}_7^4}{2}+z_d^2 \left(-\frac{91 g_1^2}{120}+\frac{51 g_2^2}{8}+\frac{22 g_3^2}{3}-\frac{3 z_u^2}{2}-\frac{\hat{\gamma} _2}{3}-\frac{5 \bar{g}_2^2}{8}\!-\frac{15 \bar{g}_4^2}{8}-\frac{11 \bar{g}_5^2}{48}\right.\\
%%%
&\left.-\frac{99 \bar{g}_6^2}{16}-3 \bar{g}_7^2\right)+\bar{g}_6^2 \left(\frac{11 g_1^2}{320}+\frac{75 g_2^2}{64}+\frac{11 g_3^2}{4}-\frac{\hat{\gamma} _2}{8}+\frac{3 \hat{\gamma} _7}{8}-\frac{3 \bar{g}_7^2}{2}\right)+z_u^2 \left(-\frac{9 \bar{g}_1^2}{16}-\frac{27 \bar{g}_3^2}{16}\right.\\
%%%%
&\left.-\frac{\bar{g}_5^2}{48}-\frac{9 \bar{g}_6^2}{16}-\bar{g}_7^2\right)+y_t \left(z_u \left(-\frac{1}{3} \bar{g}_1 \bar{g}_5+3 \bar{g}_3 \bar{g}_6\right)+\frac{4}{3} z_q \bar{g}_1 \bar{g}_8\right)+\bar{g}_2^2 \left(-\frac{19 \bar{g}_5^2}{576}-\frac{\bar{g}_8^2}{4}\right)\\
%%%%
&+\bar{g}_1^2 \left(-\frac{11 \bar{g}_5^2}{576}-\frac{\bar{g}_8^2}{4}\right)+\bar{g}_5^2 \left(\frac{11 g_1^2}{8640}+\frac{11 g_2^2}{192}+\frac{11 g_3^2}{108}-\frac{\hat{\gamma} _2}{216}-\frac{\hat{\gamma} _7}{24}-\frac{\bar{g}_6^2}{32}-\frac{\bar{g}_7^2}{18}-\frac{\bar{g}_8^2}{216}\right)\\
%%%%
&+y_t^2 \left(-\frac{5 z_q^2}{4}-\frac{11 z_u^2}{4}-\frac{5 \bar{g}_5^2}{144}-\frac{15 \bar{g}_6^2}{16}-\frac{5 \bar{g}_7^2}{3}-\frac{11 \bar{g}_8^2}{18}-\frac{11 \bar{g}_9^2}{6}\right)+z_q^2 \left(\frac{26 g_1^2}{15}-\frac{3 g_2^2}{2}+\frac{11 g_3^2}{3}\right.\\
%%%%%
&\left.-\frac{3 z_u^2}{8}+\frac{2 \hat{\gamma} _3}{3}-\frac{35 \bar{g}_1^2}{32}-\frac{9 \bar{g}_3^2}{32}-\frac{\bar{g}_8^2}{3}-\bar{g}_9^2\right)+\bar{g}_7^2 \left(\frac{11 g_1^2}{180}+\frac{11 g_2^2}{4}+\frac{143 g_3^2}{9}-\frac{2 \hat{\gamma} _2}{9}-2 \hat{\gamma} _7\right.\\
%%%%
&\left.\left.\left.-\frac{\bar{g}_9^2}{12}\right)\right)\right\},
%%%%%
\end{align*}
\begin{align*}
 \beta_{y_\tau}=&\beta_{y_\tau}^{SS}|_{\lambda=0}+\frac{1}{(16\pi^2)^2}\left\{-\frac{3}{2} y_{\tau }^3 \hat{\gamma} _1+y_{\tau } \left(\frac{117 g_1^4}{100}+\frac{3 g_2^4}{2}+\frac{3 \hat{\gamma} _1^2}{32}+\frac{\hat{\gamma} _2^2}{48}+\frac{\hat{\gamma} _3^2}{6}+\frac{9 \hat{\gamma} _7^2}{16}+z_q^2 \left(-\frac{9 \bar{g}_1^2}{16}\right.\right.\right.\\
%%%%
&\left.-\frac{27 \bar{g}_3^2}{16}\right)\!-z_u^2 \left(\frac{9 \bar{g}_1^2}{16}\!+\frac{27 \bar{g}_3^2}{16}\right)+z_d^2 \left(-\frac{9 \bar{g}_2^2}{16}-\frac{27 \bar{g}_4^2}{16}\right)-\frac{27}{32} \bar{g}_3^2 \bar{g}_6^2-\frac{27}{32} \bar{g}_4^2 \bar{g}_6^2+y_b z_d \left(\frac{1}{2} \bar{g}_2 \bar{g}_5\right.\\
%%%%
&\left.+\frac{9}{2} \bar{g}_4 \bar{g}_6\right)\!-\!y_b^2 \left(\frac{9 z_d^2}{2}\!+\!\frac{9 z_q^2}{4}+\frac{\bar{g}_5^2}{16}+\frac{27 \bar{g}_6^2}{16}+3 \bar{g}_7^2\right)+y_t \left(z_u \left(\frac{9}{2} \bar{g}_3 \bar{g}_6-\frac{1}{2} \bar{g}_1 \bar{g}_5\right)+2 z_q \bar{g}_1 \bar{g}_8\right)\\
%%%%
&\left.\left.-\bar{g}_1^2 \left(\frac{\bar{g}_5^2}{32}\!+\!\frac{\bar{g}_8^2}{4}\right)\!-\!\bar{g}_2^2 \left(\frac{\bar{g}_5^2}{32}\!+\!\frac{\bar{g}_8^2}{4}\right)-y_t^2 \left(\frac{9 z_q^2}{4}+\frac{9 z_u^2}{2}+\frac{\bar{g}_5^2}{16}+\frac{27 \bar{g}_6^2}{16}+3 \bar{g}_7^2+\bar{g}_8^2+3 \bar{g}_9^2\right)\right)\right\},
\end{align*}
%%%%%%%%%
%%%%%%%%
\begin{align*}
 \beta_{z_q}=&\frac{1}{16\pi^2}\left\{4 z_q^3\!-\frac{2}{3} y_t \bar{g}_1 \bar{g}_8\!-z_u \left(\frac{2}{9} \bar{g}_5 \bar{g}_8+\frac{8}{3} \bar{g}_7 \bar{g}_9\right)+z_q \left(-\frac{g_1^2}{2}-\frac{9 g_2^2}{2}-4 g_3^2+\frac{y_b^2}{2}+\frac{y_t^2}{2}+\frac{3 z_u^2}{2}\right.\right.\\
%%%%
&\left.\left.+\frac{\bar{g}_1^2}{8}+\frac{3 \bar{g}_3^2}{8}+\frac{\bar{g}_5^2}{72}+\frac{3 \bar{g}_6^2}{8}+\frac{2 \bar{g}_7^2}{3}+\frac{4 \bar{g}_8^2}{9}+\frac{4 \bar{g}_9^2}{3}\right)\right\},
\end{align*}
%%%%%%%%%
%%%%%%%%
\begin{align*}
  \beta_{z_u}=&\frac{1}{16\pi^2}\left\{\frac{7 z_u^3}{2}\!+\!\frac{1}{6} y_t \bar{g}_1 \bar{g}_5\!-\frac{3}{2} y_t \bar{g}_3 \bar{g}_6+z_q \left(-\frac{2}{9} \bar{g}_5 \bar{g}_8-\frac{8}{3} \bar{g}_7 \bar{g}_9\right)+z_u \left(-\frac{5 g_1^2}{4}-\frac{9 g_2^2}{4}-4 g_3^2+y_t^2\right.\right.\\
%%%%
&\left.\left.+z_d^2+\frac{3 z_q^2}{2}+\frac{\bar{g}_1^2}{8}+\frac{3 \bar{g}_3^2}{8}+\frac{\bar{g}_5^2}{36}+\frac{3 \bar{g}_6^2}{4}+\frac{4 \bar{g}_7^2}{3}+\frac{2 \bar{g}_8^2}{9}+\frac{2 \bar{g}_9^2}{3}\right)\right\},
\end{align*}
%%%%%%%%%
%%%%%%%%
\begin{align*}
  \hskip-5pt\beta_{z_d}=&\frac{1}{16\pi^2}\left\{\frac{7 z_d^3}{2}+y_b \left(-\frac{1}{6} \bar{g}_2 \bar{g}_5-\frac{3}{2} \bar{g}_4 \bar{g}_6\right)+z_d \left(-\frac{13 g_1^2}{20}-\frac{9 g_2^2}{4}-4 g_3^2+y_b^2+z_u^2+\frac{\bar{g}_2^2}{8}+\frac{3 \bar{g}_4^2}{8}\right.\right.\\
%%%%
&\left.\left.+\frac{\bar{g}_5^2}{36}+\frac{3 \bar{g}_6^2}{4}+\frac{4 \bar{g}_7^2}{3}\right)\right\},
\end{align*}
%%%%%%%%%%%%
%%%%%%%%%%%%
\begin{align*}
 \hskip-1.9cm\beta_{\bar{g}_1}=&\beta_{\bar{g}_1}^{SS}|_{\lambda=0}+\frac{1}{16\pi^2}\left\{2 y_t z_u \bar{g}_5-8 y_t z_q \bar{g}_8+\bar{g}_1 \left(\frac{3 z_q^2}{2}+\frac{3 z_u^2}{2}+\frac{\bar{g}_5^2}{12}+\frac{2 \bar{g}_8^2}{3}\right)\right\},
\end{align*}
\begin{align*}
 \hskip-1cm\beta_{\bar{g}_2}=&\beta_{\bar{g}_1}|_{g_{H_i}\leftrightarrow g_{H_i^*}}+\frac{1}{16\pi^2}\left\{\left(\frac{3 z_d^2}{2}-\frac{3 z_q^2}{2}-\frac{3 z_u^2}{2}\right) \bar{g}_2+\left(-2 y_b z_d-2 y_t z_u\right) \bar{g}_5+8 y_t z_q \bar{g}_8\right\},
\end{align*}
\begin{align*}
\hskip-3.1cm\beta_{\bar{g}_3}=&\beta_{\bar{g}_3}^{SS}|_{\lambda=0}+\frac{1}{16\pi^2}\left\{-6 y_t z_u \bar{g}_6+\bar{g}_3 \left(\frac{3 z_q^2}{2}+\frac{3 z_u^2}{2}+\frac{3 \bar{g}_6^2}{4}\right)\right\},
\end{align*}
\begin{align*}
\hskip-1.9cm\beta_{\bar{g}_4}=&\beta_{\bar{g}_3}|_{g_{H_i}\leftrightarrow g_{H_i^*}}+\frac{1}{16\pi^2}\left\{\left(\frac{3 z_d^2}{2}-\frac{3 z_q^2}{2}-\frac{3 z_u^2}{2}\right) \bar{g}_4+\left(-6 y_b z_d+6 y_t z_u\right) \bar{g}_6\right\},
\end{align*}
\begin{align*}
 \beta_{\bar g_{5}}=&\frac{1}{16\pi^2}\left\{6 y_t z_u \bar{g}_1-6 y_b z_d \bar{g}_2+\frac{1}{4} \bar{g}_1^2 \bar{g}_5+\frac{1}{4} \bar{g}_2^2 \bar{g}_5+\frac{\bar{g}_5^3}{8}-8 z_q z_u \bar{g}_8+\bar{g}_5 \left(\frac{y_b^2}{2}-\frac{g_1^2}{20}-\frac{9 g_2^2}{4}-4 g_3^2\right.\right.\\
%%%%
&\left.\left.+\frac{y_t^2}{2}+z_d^2+\frac{z_q^2}{2}+z_u^2+\frac{9 \bar{g}_6^2}{8}+2 \bar{g}_7^2+\frac{2 \bar{g}_8^2}{3}\right)\right\},
\end{align*}
\begin{align*}
 \beta_{\bar g_{6}}=&\frac{1}{16\pi^2}\left\{-2 y_t z_u \bar{g}_3-2 y_b z_d \bar{g}_4+\frac{1}{4} \bar{g}_3^2 \bar{g}_6+\frac{1}{4} \bar{g}_4^2 \bar{g}_6+\frac{15 \bar{g}_6^3}{8}+\bar{g}_6 \left(\frac{y_t^2}{2}-\frac{g_1^2}{20}-\frac{33 g_2^2}{4}-4 g_3^2+\frac{y_b^2}{2}\right.\right.\\
%%%%
&\left.\left.+z_d^2+\frac{z_q^2}{2}+z_u^2+\frac{\bar{g}_5^2}{24}+2 \bar{g}_7^2\right)\right\},
\end{align*}
\begin{align*}
 \beta_{\bar g_{7}}=&\frac{1}{16\pi^2}\left\{\frac{5 \bar{g}_7^3}{2}-2 z_q z_u \bar{g}_9+\bar{g}_7 \left(-\frac{g_1^2}{20}-\frac{9 g_2^2}{4}-13 g_3^2+\frac{y_b^2}{2}+\frac{y_t^2}{2}+z_d^2+\frac{z_q^2}{2}+z_u^2+\frac{\bar{g}_5^2}{24}+\frac{9 \bar{g}_6^2}{8}\right.\right.\\
%%%%
&\left.\left.+\frac{\bar{g}_9^2}{4}\right)\right\},
\end{align*}
\begin{align*}
 \beta_{\bar g_{8}}=&\frac{1}{16\pi^2}\left\{-3 y_t z_q \bar{g}_1-z_q z_u \bar{g}_5+\frac{1}{4} \bar{g}_1^2 \bar{g}_8+\frac{1}{12} \bar{g}_5^2 \bar{g}_8+\frac{4 \bar{g}_8^3}{3}+\bar{g}_8 \left(-\frac{4 g_1^2}{5}-4 g_3^2+y_t^2+2 z_q^2+z_u^2\right.\right.\\
%%%%
&\left.\left.+\frac{\bar{g}_2^2}{4}+2 \bar{g}_9^2\right)\right\},
\end{align*}
\begin{align*}
\hskip-.9cm \beta_{\bar g_{9}}=&\frac{1}{16\pi^2}\left\{-4 z_q z_u \bar{g}_7+\frac{1}{2} \bar{g}_7^2 \bar{g}_9+\left(-\frac{4 g_1^2}{5}-13 g_3^2+y_t^2+2 z_q^2+z_u^2+\frac{2 \bar{g}_8^2}{3}\right) \bar{g}_9+\frac{9 \bar{g}_9^3}{4}\right\}.
\end{align*}
%%%%%%%%%%%%%%%%%%%%%%%%%%%%%%%%%%%%%%%%%%%%%%%%%%%%%%%%%%%%%%%%%%%%%%%%%%%%%%%
%%%%%%%%%%%%%%%%%%%%%%%%%%%%%%%%%%%%%%%%%%%%%%%%%%%%%%%%%%%%%%%%%%%%%%%%%%%%%%%%%
%%%%%%%%%%%%%%%%%%%%%%%%%%%%%%%%%%%%%%%%%%%%%%%%%%%%%%%%%%%%%%%%%%%%%%%%%%%%%%%%%
%%%%%%%%%%%%%%%%%%%%%%%%%%%%%%%%%%%%%%%%%%%%%%%%%%%%%%%%%%%%%%%%%%%%%%%%%%%%%%%%%
%%%%%%%%%%%%%%%%%%%%%%%%%%%%%%%%%%%%%%%%%%%%%%%%%%%%%%%%%%%%%%%%%%%%%%%%%%%%%%%%%
\subsubsection{Quartic couplings}
%%%%%%%%%%%%%%%%%%%%%%%%%%%%%%%%%%%%%%%%%%%%%%%%%%%%%%%%%%%%%%%%%%%%%%%%%%%%%%%%%
%%%%%%%%%%%%%%%%%%%%%%%%%%%%%%%%%%%%%%%%%%%%%%%%%%%%%%%%%%%%%%%%%%%%%%%%%%%%%%%%%
%%%%%%%%%%%%%%%%%%%%%%%%%%%%%%%%%%%%%%%%%%%%%%%%%%%%%%%%%%%%%%%%%%%%%%%%%%%%%%%%%
\begin{align*}
 \hskip-1.9cm\beta_{\hat{\gamma}_1}=&\frac{1}{16\pi^2}\left\{\frac{27 g_1^4}{25}+9 g_2^4-48 y_b^4-48 y_t^4-16 y_{\tau }^4+g_1^2 \left(\frac{18 g_2^2}{5}-\frac{9 \hat{\gamma} _1}{5}\right)-9 g_2^2 \hat{\gamma} _1+12 y_b^2 \hat{\gamma} _1\right.\\
%%%%
&+12 y_t^2 \hat{\gamma} _1+4 y_{\tau }^2 \hat{\gamma} _1+3 \hat{\gamma} _1^2+\frac{\hat{\gamma} _2^2}{3}+\frac{8 \hat{\gamma} _3^2}{3}+3 \hat{\gamma} _7^2-\bar{g}_1^4-\bar{g}_2^4-5 \bar{g}_3^4+\bar{g}_1^2 \left(\hat{\gamma} _1-2 \bar{g}_2^2-2 \bar{g}_3^2\right)\\
%%%%
&\left.-4 \bar{g}_1 \bar{g}_2 \bar{g}_3 \bar{g}_4+3 \hat{\gamma} _1 \bar{g}_4^2-5 \bar{g}_4^4+\bar{g}_2^2 \left(\hat{\gamma} _1-2 \bar{g}_4^2\right)+\bar{g}_3^2 \left(3 \hat{\gamma} _1-2 \bar{g}_4^2\right)\right\},
\end{align*}
%%%%%%%%
%%%%%%%%
%%%%%%%%
%%%%%%%%
%%%%%%%%
%%%%%%%%
\begin{align*}
 \beta_{\hat{\gamma}_2}=&\frac{1}{16\pi^2}\left\{\frac{9 g_1^4}{25}+27 g_2^4+\frac{\hat{\gamma} _2^2}{3}+\frac{8 \hat{\gamma} _3 \hat{\gamma} _5}{3}+9 \hat{\gamma} _7^2+z_u^2 \left(-6 \bar{g}_1^2-18 \bar{g}_3^2\right)+z_d^2 \left(-6 \bar{g}_2^2-18 \bar{g}_4^2\right)\right.\\
%%%
&-\frac{1}{3} \bar{g}_1^2 \bar{g}_5^2-\frac{1}{3} \bar{g}_2^2 \bar{g}_5^2-9 \bar{g}_3^2 \bar{g}_6^2-9 \bar{g}_4^2 \bar{g}_6^2+y_t z_u \left(-4 \bar{g}_1 \bar{g}_5+36 \bar{g}_3 \bar{g}_6\right)+y_b z_d \left(4 \bar{g}_2 \bar{g}_5+36 \bar{g}_4 \bar{g}_6\right)\\
%%%%
&+y_b^2 \left(-48 z_d^2-\frac{2 \bar{g}_5^2}{3}-18 \bar{g}_6^2-32 \bar{g}_7^2\right)+y_t^2 \left(-48 z_u^2-\frac{2 \bar{g}_5^2}{3}-18 \bar{g}_6^2-32 \bar{g}_7^2\right)+\hat{\gamma} _2 \left(-g_1^2\right.\\
%%%%
&-9 g_2^2-8 g_3^2\!+\!6 y_b^2+6 y_t^2+2 y_{\tau }^2+2 z_d^2+2 z_u^2+\frac{3 \hat{\gamma} _1}{2}+\frac{7 \hat{\gamma} _4}{18}+\frac{3 \hat{\gamma} _8}{2}+\frac{\bar{g}_1^2}{2}+\frac{\bar{g}_2^2}{2}+\frac{3 \bar{g}_3^2}{2}+\frac{3 \bar{g}_4^2}{2}\\
%%%%
&\left.\left.+\frac{\bar{g}_5^2}{18}+\frac{3 \bar{g}_6^2}{2}+\frac{8 \bar{g}_7^2}{3}\right)\right\},
\end{align*}
%%%%%%%%
%%%%%%%%
%%%%%%%%
%%%%%%%%
%%%%%%%%
%%%%%%%%
\begin{align*}
 \beta_{\hat{\gamma}_3}=&\frac{1}{16\pi^2}\left\{-\frac{36 g_1^4}{25}\!+\!12 y_b^2 z_q^2-\frac{4 \hat{\gamma} _3^2}{3}+\frac{\hat{\gamma} _2 \hat{\gamma} _5}{3}+z_q^2 \left(3 \bar{g}_1^2+9 \bar{g}_3^2\right)-8 y_t z_q \bar{g}_1 \bar{g}_8+\frac{4}{3} \bar{g}_1^2 \bar{g}_8^2+\frac{4}{3} \bar{g}_2^2 \bar{g}_8^2\right.\\
%%%%
&+\hat{\gamma} _3 \left(-\frac{5 g_1^2}{2}-\frac{9 g_2^2}{2}-8 g_3^2+6 y_b^2+6 y_t^2+2 y_{\tau }^2+4 z_q^2+\frac{3 \hat{\gamma} _1}{2}+\frac{32 \hat{\gamma} _6}{9}+\frac{\bar{g}_1^2}{2}+\frac{\bar{g}_2^2}{2}+\frac{3 \bar{g}_3^2}{2}\right.\\
%%%%
&\left.\left.+\frac{3 \bar{g}_4^2}{2}+\frac{8 \bar{g}_8^2}{9}+\frac{8 \bar{g}_9^2}{3}\right)+y_t^2 \left(12 z_q^2+\frac{16 \bar{g}_8^2}{3}+16 \bar{g}_9^2\right)\right\},
\end{align*}
%%%%%%%%
%%%%%%%%
%%%%%%%%
%%%%%%%%
%%%%%%%%
%%%%%%%%
\begin{align*}
 \beta_{\hat{\gamma}_4}=&\frac{1}{16\pi^2}\left\{\frac{3 g_1^4}{25}+81 g_2^4+\frac{6}{5} g_1^2 g_3^2+162 g_2^2 g_3^2+111 g_3^4-72 z_d^4-72 z_u^4+\hat{\gamma} _2^2+\frac{5 \hat{\gamma} _4^2}{9}+\frac{8 \hat{\gamma} _5^2}{3}\right.\\
%%%%
&+27 \hat{\gamma} _8^2+3 \hat{\gamma} _9^2-\frac{\bar{g}_5^4}{9}-27 \bar{g}_6^4-\frac{2}{3} \bar{g}_5^2 \bar{g}_7^2-54 \bar{g}_6^2 \bar{g}_7^2-46 \bar{g}_7^4+\hat{\gamma} _4 \left(-\frac{g_1^2}{5}-9 g_2^2-16 g_3^2+4 z_d^2\right.\\
%%%%
&\left.\left.+4 z_u^2+3 \hat{\gamma} _8+\frac{\bar{g}_5^2}{9}+3 \bar{g}_6^2+\frac{16 \bar{g}_7^2}{3}\right)\right\},
\end{align*}
%%%%%%%%
%%%%%%%%
%%%%%%%%
%%%%%%%%
%%%%%%%%
%%%%%%%%
\begin{align*}
 \beta_{\hat{\gamma}_5}=&\frac{1}{16\pi^2}\left\{-\frac{12 g_1^4}{25}-24 g_3^4+\hat{\gamma} _2 \hat{\gamma} _3-\frac{4 \hat{\gamma} _5^2}{9}-8 \hat{\gamma} _9^2+z_q^2 \left(36 z_u^2+\frac{\bar{g}_5^2}{3}+9 \bar{g}_6^2+16 \bar{g}_7^2\right)+\frac{4}{9} \bar{g}_5^2 \bar{g}_8^2\right.\\
%%%%
&+8 \bar{g}_7^2 \bar{g}_9^2+z_q z_u \left(-\frac{8}{3} \bar{g}_5 \bar{g}_8-32 \bar{g}_7 \bar{g}_9\right)+\hat{\gamma} _5 \left(-\frac{17 g_1^2}{10}-\frac{9 g_2^2}{2}-16 g_3^2+2 z_d^2+4 z_q^2+2 z_u^2\right.\\
%%%%
&\left.\left.+\frac{7 \hat{\gamma} _4}{18}+\frac{32 \hat{\gamma} _6}{9}+\frac{3 \hat{\gamma} _8}{2}+\frac{\bar{g}_5^2}{18}+\frac{3 \bar{g}_6^2}{2}+\frac{8 \bar{g}_7^2}{3}+\frac{8 \bar{g}_8^2}{9}+\frac{8 \bar{g}_9^2}{3}\right)+z_u^2 \left(\frac{16 \bar{g}_8^2}{3}+16 \bar{g}_9^2\right)\right\},
\end{align*}
%%%%%%%%
%%%%%%%%
%%%%%%%%
%%%%%%%%
%%%%%%%%
%%%%%%%%
\begin{align*}
 \hskip-.7cm\beta_{\hat{\gamma}_6}=&\frac{1}{16\pi^2}\left\{\frac{48 g_1^4}{25}+\frac{24}{5} g_1^2 g_3^2+\frac{39 g_3^4}{4}-18 z_q^4+\hat{\gamma} _3^2+\frac{\hat{\gamma} _5^2}{3}+\frac{56 \hat{\gamma} _6^2}{9}+\frac{3 \hat{\gamma} _9^2}{2}-\frac{16 \bar{g}_8^4}{9}-\frac{8}{3} \bar{g}_8^2 \bar{g}_9^2\right.\\
%%%%
&\left.-\frac{11 \bar{g}_9^4}{2}+\hat{\gamma} _6 \left(-\frac{16 g_1^2}{5}-16 g_3^2+8 z_q^2+\frac{16 \bar{g}_8^2}{9}+\frac{16 \bar{g}_9^2}{3}\right)\right\},
\end{align*}
%%%%%%%%
%%%%%%%%
%%%%%%%%
%%%%%%%%
%%%%%%%%
%%%%%%%%
\begin{align*}
 \beta_{\hat{\gamma}_7}=&\frac{1}{16\pi^2}\left\{\frac{6}{5} g_1^2 g_2^2+z_u^2 \left(2 \bar{g}_1^2-2 \bar{g}_3^2\right)+z_d^2 \left(-2 \bar{g}_2^2+2 \bar{g}_4^2\right)-\frac{2}{3} \bar{g}_1 \bar{g}_3 \bar{g}_5 \bar{g}_6-\frac{2}{3} \bar{g}_2 \bar{g}_4 \bar{g}_5 \bar{g}_6-2 \bar{g}_3^2 \bar{g}_6^2\right.\\
%%%%
&+2 \bar{g}_4^2 \bar{g}_6^2+y_t z_u \left(\frac{4}{3} \bar{g}_1 \bar{g}_5+4 \bar{g}_3 \bar{g}_6\right)+y_b z_d \left(\frac{4}{3} \bar{g}_2 \bar{g}_5-4 \bar{g}_4 \bar{g}_6\right)+y_b^2 \left(-\frac{2 \bar{g}_5^2}{9}+2 \bar{g}_6^2-\frac{32 \bar{g}_7^2}{3}\right)\\
%%%%
&+\hat{\gamma} _7 \left(-g_1^2-9 g_2^2-8 g_3^2+6 y_b^2+6 y_t^2+2 y_{\tau }^2+2 z_d^2+2 z_u^2+\frac{\hat{\gamma} _1}{2}+\frac{2 \hat{\gamma} _2}{3}+\frac{\hat{\gamma} _4}{18}+\frac{5 \hat{\gamma} _8}{2}+\frac{\bar{g}_1^2}{2}\right.\\
%%%%
&\left.\left.+\frac{\bar{g}_2^2}{2}+\frac{3 \bar{g}_3^2}{2}+\frac{3 \bar{g}_4^2}{2}+\frac{\bar{g}_5^2}{18}+\frac{3 \bar{g}_6^2}{2}+\frac{8 \bar{g}_7^2}{3}\right)+y_t^2 \left(\frac{2 \bar{g}_5^2}{9}-2 \bar{g}_6^2+\frac{32 \bar{g}_7^2}{3}\right)\right\},
\end{align*}
%%%%%%%%
%%%%%%%%
%%%%%%%%
%%%%%%%%
%%%%%%%%
%%%%%%%%
\begin{align*}
 \beta_{\hat{\gamma}_8}=&\frac{1}{16\pi^2}\left\{-10 g_2^2 g_3^2+5 g_3^4+g_1^2 \left(\frac{2 g_2^2}{5}+\frac{2 g_3^2}{5}\right)-8 z_d^4-8 z_u^4+\hat{\gamma} _7^2+2 \hat{\gamma} _8^2+\hat{\gamma} _9^2-2 \bar{g}_6^4\right.\\
%%%
&+\frac{10}{3} \bar{g}_6^2 \bar{g}_7^2-\frac{14 \bar{g}_7^4}{3}+\bar{g}_5^2 \left(-\frac{2 \bar{g}_6^2}{9}-\frac{2 \bar{g}_7^2}{9}\right)+\hat{\gamma} _8 \left(\frac{\bar{g}_5^2}{9}-\frac{g_1^2}{5}-9 g_2^2-16 g_3^2+4 z_d^2+4 z_u^2+\frac{\hat{\gamma} _4}{3}\right.\\
%%%%
&\left.\left.+3 \bar{g}_6^2+\frac{16 \bar{g}_7^2}{3}\right)\right\},
\end{align*}
%%%%%%%%
%%%%%%%%
%%%%%%%%
%%%%%%%%
%%%%%%%%
%%%%%%%%
\begin{align*}
 \beta_{\hat{\gamma}_9}=&\frac{1}{16\pi^2}\left\{-\frac{8}{5} g_1^2 g_3^2-5 g_3^4-\frac{5 \hat{\gamma} _9^2}{3}+z_q^2 \left(\frac{2 \bar{g}_5^2}{9}+6 \bar{g}_6^2-\frac{4 \bar{g}_7^2}{3}\right)+\frac{8}{9} \bar{g}_5 \bar{g}_7 \bar{g}_8 \bar{g}_9-\frac{4}{3} \bar{g}_7^2 \bar{g}_9^2\right.\\
%%%%
&+z_q z_u \left(-\frac{16}{9} \bar{g}_5 \bar{g}_8+\frac{8}{3} \bar{g}_7 \bar{g}_9\right)+z_u^2 \left(\frac{32 \bar{g}_8^2}{9}-\frac{4 \bar{g}_9^2}{3}\right)+\hat{\gamma} _9 \left(-\frac{17 g_1^2}{10}-\frac{9 g_2^2}{2}-16 g_3^2+2 z_d^2\right.\\
%%%%
&\left.\left.+4 z_q^2+2 z_u^2+\frac{\hat{\gamma} _4}{18}-\frac{8 \hat{\gamma} _5}{9}+\frac{8 \hat{\gamma} _6}{9}+\frac{3 \hat{\gamma} _8}{2}+\frac{\bar{g}_5^2}{18}+\frac{3 \bar{g}_6^2}{2}+\frac{8 \bar{g}_7^2}{3}+\frac{8 \bar{g}_8^2}{9}+\frac{8 \bar{g}_9^2}{3}\right)\right\}.
\end{align*}
%%%%%%%%%%%%%%%%%%%%%%%%%%%%%%%%%%%%%%%%%%%%%%%%%%%%%%%%%%%%%%%%%%%%%%%%%%%%%%%%%
%%%%%%%%%%%%%%%%%%%%%%%%%%%%%%%%%%%%%%%%%%%%%%%%%%%%%%%%%%%%%%%%%%%%%%%%%%%%%%%%%
%%%%%%%%%%%%%%%%%%%%%%%%%%%%%%%%%%%%%%%%%%%%%%%%%%%%%%%%%%%%%%%%%%%%%%%%%%%%%%%%%
\subsubsection{Fermion masses}
%%%%%%%%%%%%%%%%%%%%%%%%%%%%%%%%%%%%%%%%%%%%%%%%%%%%%%%%%%%%%%%%%%%%%%%%%%%%%%%%%
%%%%%%%%%%%%%%%%%%%%%%%%%%%%%%%%%%%%%%%%%%%%%%%%%%%%%%%%%%%%%%%%%%%%%%%%%%%%%%%%%
%%%%%%%%%%%%%%%%%%%%%%%%%%%%%%%%%%%%%%%%%%%%%%%%%%%%%%%%%%%%%%%%%%%%%%%%%%%%%%%%%
\begin{align*}
 \beta_\mu=&\beta_\mu^{SS}+\frac{1}{16\pi^2}\mu  \left(\frac{3 z_d^2}{2}+\frac{3 z_q^2}{2}+\frac{3 z_u^2}{2}\right)+\frac{1}{(16\pi^2)^2}\left\{a_u \left(\bar{g}_2 \left(\frac{1}{2} \bar{g}_5 z_q-2 \bar{g}_8 z_u\right)-\frac{9}{2} \bar{g}_4 \bar{g}_6 z_q\right.\right.\\
 %%%%%
 &\left.+6 y_b z_d z_q\right)+\mu  \left(\bar{g}_5^2 \left(-\frac{z_d^2}{16}-\frac{z_q^2}{96}-\frac{z_u^2}{16}\right)+\bar{g}_6^2 \left(-\frac{27}{16}  z_d^2-\frac{9 z_q^2}{32}-\frac{27 z_u^2}{16}\right)+\bar{g}_7^2 \left(-3 z_d^2\right.\right.\\
 %%%
 &\left.-\frac{z_q^2}{2}-3 z_u^2\right)+z_u^2 \left(-\frac{\bar{g}_8^2}{6}-\frac{\bar{g}_9^2}{2}-\frac{21 z_d^2}{2}\right)-\bar{g}_8^2 z_q^2-3 \bar{g}_9^2 z_q^2-\frac{39 z_q^4}{8}\\
 %%%%
 &+\left(-\frac{\bar{g}_5^2}{192}-\frac{\bar{g}_8^2}{24}\right) \bar{g}_1^2+\left(-\frac{\bar{g}_5^2}{192}-\frac{\bar{g}_8^2}{24}\right) \bar{g}_2^2-\frac{9}{64} \bar{g}_3^2 \bar{g}_6^2-\frac{9}{64} \bar{g}_4^2 \bar{g}_6^2+y_b^2 \left(-\frac{3}{4}  z_d^2-\frac{3 z_q^2}{8}\right)\\
 %%%%
 &+g_1^2 \left(\frac{19 z_d^2}{16}+\frac{19 z_q^2}{10}+\frac{227 z_u^2}{80}\right)+g_2^2 \left(\frac{99 z_d^2}{16}+9 z_q^2+\frac{99 z_u^2}{16}\right)+g_3^2 \left(17 z_d^2+17 z_q^2+17 z_u^2\right)\\
 %%%%
 &\left.\left.-3 z_d^4\!+\!\frac{99 g_1^4}{200}\!+\!\frac{33 g_2^4}{8}\!-y_t^2 \left(\frac{3}{8}  z_q^2+\frac{3 z_u^2}{4}\right)-3 z_u^4\right)+\frac{9}{4} M_2 \bar{g}_3 \bar{g}_4 \bar{g}_6^2+M_1 \left(\frac{\bar{g}_5^2}{12}+\frac{2 \bar{g}_8^2}{3}\right) \bar{g}_1 \bar{g}_2\right\},%%%%%%%%
\end{align*}
\begin{align*}
 \beta_{M_1}=&\beta_{M_1}^{SS}+\frac{1}{16\pi^2}M_1 \left(\frac{\bar{g}_5^2}{6}+\frac{4 \bar{g}_8^2}{3}\right)+\frac{1}{(16\pi^2)^2}\left\{a_u \left(\bar{g}_1 \left(8 \bar{g}_8 z_u-2 \bar{g}_5 z_q\right)+\frac{8}{3} \bar{g}_5 \bar{g}_8 y_t\right)\right.\\
 %%%%
 &+M_1 \left(-\frac{1}{24} \bar{g}_5^2 y_b^2+\bar{g}_5^2 \left(-\frac{7 \bar{g}_7^2}{18}-\frac{7 \bar{g}_6^2}{32}-\frac{z_d^2}{4}-\frac{z_q^2}{24}-\frac{z_u^2}{4}\right)-\frac{3}{8} \bar{g}_2^2 z_d^2+\bar{g}_1^2 \left(-\frac{3 z_q^2}{8} -\frac{3 z_u^2}{8}\right)\right.\\
 %%%
 &-4 \bar{g}_8^2 z_q^2+\frac{4\bar{g}_8^4}{27}-\frac{28}{9} \bar{g}_9^2 \bar{g}_8^2+\left(-\frac{\bar{g}_5^2}{24}-\frac{2 \bar{g}_8^2}{3}\right) y_t^2-\frac{2}{3} \bar{g}_8^2 z_u^2+\frac{\bar{g}_5^4}{864}+\frac{17}{16} g_2^2 \bar{g}_5^2\\
 %%%%
 &\left.+g_1^2 \left(\frac{17 \bar{g}_5^2}{720}+\frac{136 \bar{g}_8^2}{45}\right)+g_3^2 \left(\frac{17 \bar{g}_5^2}{9}+\frac{136 \bar{g}_8^2}{9}\right)\right)+\mu  \bar{g}_1 \bar{g}_2 \left(3 z_d^2+3 z_q^2+3 z_u^2\right)+\frac{1}{4} M_2 \bar{g}_5^2 \bar{g}_6^2\\
 %%%
 &\left.+M_3 \left(\frac{4}{9} \bar{g}_5^2 \bar{g}_7^2+\frac{32}{9} \bar{g}_8^2 \bar{g}_9^2\right)\right\},
\end{align*}
\begin{align*}
 \beta_{M_2}=&\beta_{M_2}^{SS}+\frac{3}{32\pi^2} M_2 \bar{g}_6^2+\frac{1}{(16\pi^2)^2}\left\{6 \bar{g}_3 \bar{g}_6 a_u z_q+M_2 \left(-\frac{3}{8} \bar{g}_6^2 y_b^2+\bar{g}_6^2 \left(-\frac{7 \bar{g}_7^2}{2}-\frac{ 9}{4} z_d^2-\frac{3 z_q^2}{8}\right.\right.\right.\\
 %%%%
 &\left.-\frac{9 z_u^2}{4}\right)-\frac{3}{8} \bar{g}_4^2 z_d^2-\bar{g}_3^2 \left(\frac{3 z_q^2}{8} +\frac{3 z_u^2}{8}\right)-\frac{3}{8} \bar{g}_6^2 y_t^2+\frac{273}{16} g_2^2 \bar{g}_6^2+\frac{17}{80} g_1^2 \bar{g}_6^2+17 g_3^2 \bar{g}_6^2-\frac{7}{96} \bar{g}_5^2 \bar{g}_6^2\\
 %%%%
 &\left.\left.-\frac{87 \bar{g}_6^4}{32}+11 g_2^4\right)+\mu  \bar{g}_3 \bar{g}_4 \left(3 z_d^2+3 z_q^2+3 z_u^2\right)+\frac{1}{12} M_1 \bar{g}_5^2 \bar{g}_6^2+4 M_3 \bar{g}_7^2 \bar{g}_6^2\right\},
\end{align*}
\begin{align*}
 \beta_{M_3}=&\frac{1}{16\pi^2}M_3 \left(\bar{g}_7^2+\frac{\bar{g}_9^2}{2}\right)+\frac{1}{(16\pi^2)^2}\left\{4 \bar{g}_9 \bar{g}_7 a_u y_t+M_3 \left(-\frac{1}{4} \bar{g}_7^2 y_b^2+\bar{g}_7^2 \left(-\frac{21 \bar{g}_6^2}{16}-\frac{3}{2}  z_d^2-\frac{z_q^2}{4}\right.\right.\right.\\
 %%%
 &\left.-\frac{3 z_u^2}{2}\right)-\frac{3}{2} \bar{g}_9^2 z_q^2-\frac{4\bar{g}_9^4}{3}-\frac{7 \bar{g}_8^2 \bar{g}_9^2}{18}+\left(-\frac{\bar{g}_7^2}{4}-\frac{\bar{g}_9^2}{4}\right) y_t^2-\frac{1}{4} \bar{g}_9^2 z_u^2+g_3^2 \left(\frac{113 \bar{g}_7^2}{6}\right.\\
 %%%%
 &\left.\left.+\frac{113 \bar{g}_9^2}{12}\right)+g_1^2 \left(\frac{17 \bar{g}_7^2}{120}+\frac{17 \bar{g}_9^2}{15}\right)+\frac{51}{8} g_2^2 \bar{g}_7^2-\frac{7}{144} \bar{g}_5^2 \bar{g}_7^2-\frac{8 \bar{g}_7^4}{3}+\frac{33 g_3^4}{2}\right)+\frac{3}{2} M_2 \bar{g}_6^2 \bar{g}_7^2\\
 %%%%
 &\left.+M_1 \left(\frac{1}{18} \bar{g}_5^2 \bar{g}_7^2+\frac{4}{9} \bar{g}_8^2 \bar{g}_9^2\right)\right\}.
\end{align*}
%%%%%%%%%%%%%%%%%%%%%%%%%%%%%%%%%%%%%%%%%%%%%%%%%%%%%%%%%%%%%%%%%%%%%%%%%%%%%%%%%
%%%%%%%%%%%%%%%%%%%%%%%%%%%%%%%%%%%%%%%%%%%%%%%%%%%%%%%%%%%%%%%%%%%%%%%%%%%%%%%%%
%%%%%%%%%%%%%%%%%%%%%%%%%%%%%%%%%%%%%%%%%%%%%%%%%%%%%%%%%%%%%%%%%%%%%%%%%%%%%%%%%
\subsubsection{Scalar trilinear coupling}
%%%%%%%%%%%%%%%%%%%%%%%%%%%%%%%%%%%%%%%%%%%%%%%%%%%%%%%%%%%%%%%%%%%%%%%%%%%%%%%%%
%%%%%%%%%%%%%%%%%%%%%%%%%%%%%%%%%%%%%%%%%%%%%%%%%%%%%%%%%%%%%%%%%%%%%%%%%%%%%%%%%
%%%%%%%%%%%%%%%%%%%%%%%%%%%%%%%%%%%%%%%%%%%%%%%%%%%%%%%%%%%%%%%%%%%%%%%%%%%%%%%%%
\begin{align*}
\beta_{a_u}=&\frac{1}{16\pi^2}\left\{3 M_2 z_q \bar{g}_3 \bar{g}_6+\mu  \left(4 y_b z_d z_q+z_q \left(\frac{1}{3} \bar{g}_2 \bar{g}_5-3 \bar{g}_4 \bar{g}_6\right)-\frac{4}{3} z_u \bar{g}_2 \bar{g}_8\right)+M_1 \left(-\frac{1}{3} z_q \bar{g}_1 \bar{g}_5\right.\right.\\
%%%%
&\left.+\frac{4}{3} z_u \bar{g}_1 \bar{g}_8+\frac{4}{9} y_t \bar{g}_5 \bar{g}_8\right)+\frac{16}{3} M_3 y_t \bar{g}_7 \bar{g}_9+a_u \left(-\frac{13 g_1^2}{10}-\frac{9 g_2^2}{2}-8 g_3^2+3 y_b^2+3 y_t^2+y_{\tau }^2+z_d^2\right.\\
%%%
&+2 z_q^2+z_u^2+\frac{\hat{\gamma} _2}{6}-\frac{2 \hat{\gamma} _3}{3}-\frac{2 \hat{\gamma} _5}{9}-\frac{3 \hat{\gamma} _7}{2}-\frac{8 \hat{\gamma} _9}{3}+\frac{\bar{g}_1^2}{4}+\frac{\bar{g}_2^2}{4}+\frac{3 \bar{g}_3^2}{4}+\frac{3 \bar{g}_4^2}{4}+\frac{\bar{g}_5^2}{36}+\frac{3 \bar{g}_6^2}{4}+\frac{4 \bar{g}_7^2}{3}\\
%%%%
&\left.\left.+\frac{4 \bar{g}_8^2}{9}+\frac{4 \bar{g}_9^2}{3}\right)\right\}.
\end{align*}

%%%%%%%%%%%%%%%%%%%%%%%%%%%%%%%%%%%%%%%%%%%%%%%%%%%%%%%%%%%%%%%%%%%%%%%%%%%%%%%%%
%%%%%%%%%%%%%%%%%%%%%%%%%%%%%%%%%%%%%%%%%%%%%%%%%%%%%%%%%%%%%%%%%%%%%%%%%%%%%%%%%
%%%%%%%%%%%%%%%%%%%%%%%%%%%%%%%%%%%%%%%%%%%%%%%%%%%%%%%%%%%%%%%%%%%%%%%%%%%%%%%%%
\subsubsection{Scalar masses}
%%%%%%%%%%%%%%%%%%%%%%%%%%%%%%%%%%%%%%%%%%%%%%%%%%%%%%%%%%%%%%%%%%%%%%%%%%%%%%%%%
%%%%%%%%%%%%%%%%%%%%%%%%%%%%%%%%%%%%%%%%%%%%%%%%%%%%%%%%%%%%%%%%%%%%%%%%%%%%%%%%%
%%%%%%%%%%%%%%%%%%%%%%%%%%%%%%%%%%%%%%%%%%%%%%%%%%%%%%%%%%%%%%%%%%%%%%%%%%%%%%%%% 
\begin{align*}
 \beta_{m^2_H}=&\beta_{m^2_H}|_{\lambda=0}^{SS}+\frac{1}{16\pi^2}\left\{6 a_u^2+\frac{3}{2} m_H^2 \hat{\gamma} _1+m_Q^2 \hat{\gamma} _2-2 m_U^2 \hat{\gamma} _3\right\}+\frac{1}{(16\pi^2)^2}\left\{M_2^2 \left(-18 y_t z_u \bar{g}_3 \bar{g}_6\right.\right.\\
%%%
&\left.-18 y_b z_d \bar{g}_4 \bar{g}_6+9 y_b^2 \bar{g}_6^2+9 y_t^2 \bar{g}_6^2+\bar{g}_4^2 \left(\frac{9 z_d^2}{2}+\frac{27 \bar{g}_6^2}{4}\right)+\bar{g}_3^2 \left(\frac{9 z_q^2}{2}+\frac{9 z_u^2}{2}+\frac{27 \bar{g}_6^2}{4}\right)\right)\end{align*}
\begin{align*}
\phantom{\beta_{m^2_H}=}
%%%%
&+m_Q^2 \left(\frac{3 g_1^4}{10}\!+\!\frac{45 g_2^4}{2}\!+\!\frac{2}{15} g_1^2 \hat{\gamma} _2\!+6 g_2^2 \hat{\gamma} _2+\frac{32}{3} g_3^2 \hat{\gamma} _2-2 z_d^2 \hat{\gamma} _2-2 z_u^2 \hat{\gamma} _2-\frac{\hat{\gamma} _2^2}{6}-\frac{9 \hat{\gamma} _7^2}{2}-\frac{1}{18} \hat{\gamma} _2 \bar{g}_5^2\right.\\
%%%%
&\left.-\frac{3}{2} \hat{\gamma} _2 \bar{g}_6^2+y_t \left(z_u \bar{g}_1 \bar{g}_5-9 z_u \bar{g}_3 \bar{g}_6\right)+y_b \left(-z_d \bar{g}_2 \bar{g}_5-9 z_d \bar{g}_4 \bar{g}_6\right)-\frac{8}{3} \hat{\gamma} _2 \bar{g}_7^2\right)-\frac{4}{3} a_u M_1 y_t \bar{g}_5 \bar{g}_8\\
%%%%
&+M_1^2 \left(-2 y_b z_d \bar{g}_2 \bar{g}_5\!+\frac{1}{3} y_b^2 \bar{g}_5^2\!+y_t \bar{g}_1 \left(2 z_u \bar{g}_5-8 z_q \bar{g}_8\right)+\bar{g}_2^2 \left(\frac{3 z_d^2}{2}+\frac{\bar{g}_5^2}{4}+2 \bar{g}_8^2\right)+\bar{g}_1^2 \left(\frac{3 z_q^2}{2}\right.\right.\\
%%%%
&\left.\left.+\frac{3 z_u^2}{2}+\frac{\bar{g}_5^2}{4}+2 \bar{g}_8^2\right)+y_t^2 \left(\frac{\bar{g}_5^2}{3}+\frac{16 \bar{g}_8^2}{3}\right)\right)+\mu ^2 \left(y_b^2 \left(24 z_d^2+12 z_q^2\right)+y_t^2 \left(12 z_q^2+24 z_u^2\right)\right.\\
%%%%
&+y_b \left(-2 z_d \bar{g}_2 \bar{g}_5-18 z_d \bar{g}_4 \bar{g}_6\right)+\bar{g}_4^2 \left(9 z_d^2+\frac{9 z_q^2}{2}+\frac{9 z_u^2}{2}+\frac{9 \bar{g}_6^2}{4}\right)+\bar{g}_3^2 \left(\frac{9 z_d^2}{2}+9 z_q^2+9 z_u^2\right.\\
%%%%
&\left.+\frac{9 \bar{g}_6^2}{4}\right)+\bar{g}_2^2 \left(3 z_d^2+\frac{3 z_q^2}{2}+\frac{3 z_u^2}{2}+\frac{\bar{g}_5^2}{12}+\frac{2 \bar{g}_8^2}{3}\right)+\bar{g}_1^2 \left(\frac{3 z_d^2}{2}+3 z_q^2+3 z_u^2+\frac{\bar{g}_5^2}{12}+\frac{2 \bar{g}_8^2}{3}\right)\\
%%%%%
&\left.+y_t \left(\bar{g}_1 \left(2 z_u \bar{g}_5-8 z_q \bar{g}_8\right)-18 z_u \bar{g}_3 \bar{g}_6\right)\right)+\mu  \left(-12 a_u y_b z_d z_q+M_2 \left(18 y_b z_d \bar{g}_3 \bar{g}_6+18 y_t z_u \bar{g}_4 \bar{g}_6\right.\right.\\
%%%%
&\left.+\bar{g}_3 \bar{g}_4 \left(-9 z_d^2-9 z_q^2-9 z_u^2-\frac{9 \hat{\gamma} _1}{2}-9 \bar{g}_6^2\right)\right)+M_1 \left(2 y_b z_d \bar{g}_1 \bar{g}_5+y_t \bar{g}_2 \left(-2 z_u \bar{g}_5+8 z_q \bar{g}_8\right)\right.\\
%%%%
&\left.\left.+\bar{g}_1 \bar{g}_2 \left(\!-3 z_d^2\!-3 z_q^2-3 z_u^2-\frac{3 \hat{\gamma} _1}{2}-\frac{\bar{g}_5^2}{3}-\frac{8 \bar{g}_8^2}{3}\right)\right)\right)-16 a_u M_3 y_t \bar{g}_7 \bar{g}_9+a_u^2 \left(\frac{41 g_1^2}{10}+\frac{9 g_2^2}{2}\right.\\
%%%%
&\left.+64 g_3^2-6 z_d^2-12 z_q^2-6 z_u^2-\frac{9 \hat{\gamma} _1}{2}-2 \hat{\gamma} _2+8 \hat{\gamma} _3+9 \hat{\gamma} _7-\frac{\bar{g}_5^2}{6}-\frac{9 \bar{g}_6^2}{2}-8 \bar{g}_7^2-\frac{8 \bar{g}_8^2}{3}-8 \bar{g}_9^2\right)\\
%%%%
&+m_U^2 \left(\frac{12 g_1^4}{5}-\frac{64}{15} g_1^2 \hat{\gamma} _3-\frac{64}{3} g_3^2 \hat{\gamma} _3+8 z_q^2 \hat{\gamma} _3-\frac{4 \hat{\gamma} _3^2}{3}-4 y_t z_q \bar{g}_1 \bar{g}_8+\frac{16}{9} \hat{\gamma} _3 \bar{g}_8^2+\frac{16}{3} \hat{\gamma} _3 \bar{g}_9^2\right)\\
%%%%
&+m_H^2 \left(\frac{99 g_1^4}{200}+\frac{33 g_2^4}{8}+\frac{9}{5} g_1^2 \hat{\gamma} _1+9 g_2^2 \hat{\gamma} _1-3 y_{\tau }^2 \hat{\gamma} _1-\frac{15 \hat{\gamma} _1^2}{16}-\frac{\hat{\gamma} _2^2}{24}-\frac{\hat{\gamma} _3^2}{3}-\frac{9 \hat{\gamma} _7^2}{8}+y_b \left(z_d \bar{g}_2 \bar{g}_5\right.\right.\\
%%%%
&\left.+9 z_d \bar{g}_4 \bar{g}_6\right)+\bar{g}_4^2 \left(-\frac{27 z_d^2}{8}-\frac{9 \hat{\gamma} _1}{4}-\frac{27 \bar{g}_6^2}{16}\right)+\bar{g}_3^2 \left(-\frac{27 z_q^2}{8}-\frac{27 z_u^2}{8}-\frac{9 \hat{\gamma} _1}{4}-\frac{27 \bar{g}_6^2}{16}\right)\\
%%%%
&+y_b^2 \left(-9 z_d^2-\frac{9 z_q^2}{2}-9 \hat{\gamma} _1-\frac{\bar{g}_5^2}{8}-\frac{27 \bar{g}_6^2}{8}-6 \bar{g}_7^2\right)+\bar{g}_2^2 \left(-\frac{9 z_d^2}{8}-\frac{3 \hat{\gamma} _1}{4}-\frac{\bar{g}_5^2}{16}-\frac{\bar{g}_8^2}{2}\right)\\
%%%%
&+\bar{g}_1^2 \left(-\frac{9 z_q^2}{8}-\frac{9 z_u^2}{8}-\frac{3 \hat{\gamma} _1}{4}-\frac{\bar{g}_5^2}{16}-\frac{\bar{g}_8^2}{2}\right)+y_t \left(9 z_u \bar{g}_3 \bar{g}_6+\bar{g}_1 \left(-z_u \bar{g}_5+4 z_q \bar{g}_8\right)\right)\\
%%%%
&\left.+y_t^2 \left(-\frac{9 z_q^2}{2}\!-9 z_u^2-9 \hat{\gamma} _1-\frac{\bar{g}_5^2}{8}-\frac{27 \bar{g}_6^2}{8}-6 \bar{g}_7^2-2 \bar{g}_8^2-6 \bar{g}_9^2\right)\right)+M_3^2 \left(16 y_b^2 \bar{g}_7^2+y_t^2 \left(16 \bar{g}_7^2\right.\right.\\
%%%%
&\left.\left.\left.+16 \bar{g}_9^2\right)\right)\right\},
\end{align*}
%%%%%%%%%%%%%%
%%%%%%%%%%%%%%
%%%%%%%%%%%%%%
%%%%%%%%%%%%%%
\begin{align*}
\beta_{m^2_Q}=&\frac{1}{16\pi^2}\left\{2 a_u^2+\mu ^2 \left(-4 z_d^2-4 z_u^2\right)+\frac{1}{3} m_H^2 \hat{\gamma} _2-\frac{2}{3} m_U^2 \hat{\gamma} _5-\frac{1}{9} M_1^2 \bar{g}_5^2-3 M_2^2 \bar{g}_6^2-\frac{16}{3} M_3^2 \bar{g}_7^2\right.\end{align*}
\begin{align*}
\phantom{\beta_{m^2_Q}=}&\left.m_Q^2 \left(-\frac{g_1^2}{10}-\frac{9 g_2^2}{2}-8 g_3^2+2 z_d^2+2 z_u^2+\frac{7 \hat{\gamma} _4}{18}+\frac{3 \hat{\gamma} _8}{2}+\frac{\bar{g}_5^2}{18}+\frac{3 \bar{g}_6^2}{2}+\frac{8 \bar{g}_7^2}{3}\right)\right\}\\
%%%%
&+\frac{1}{(16\pi^2)^2}\left\{m_H^2 \left(\frac{g_1^4}{10}+\frac{15 g_2^4}{2}+\frac{2}{5} g_1^2 \hat{\gamma} _2+2 g_2^2 \hat{\gamma} _2-2 y_b^2 \hat{\gamma} _2-2 y_t^2 \hat{\gamma} _2-\frac{2}{3} y_{\tau }^2 \hat{\gamma} _2-\frac{\hat{\gamma} _2^2}{18}-\frac{3 \hat{\gamma} _7^2}{2}\right.\right.\\
%%%
&-\frac{1}{6} \hat{\gamma} _2 \bar{g}_1^2-\frac{1}{6} \hat{\gamma} _2 \bar{g}_2^2-\frac{1}{2} \hat{\gamma} _2 \bar{g}_3^2-\frac{1}{2} \hat{\gamma} _2 \bar{g}_4^2+y_t \left(\frac{1}{3} z_u \bar{g}_1 \bar{g}_5-3 z_u \bar{g}_3 \bar{g}_6\right)+y_b \left(-\frac{1}{3} z_d \bar{g}_2 \bar{g}_5\right.\\
%%%%
&\left.\left.-3 z_d \bar{g}_4 \bar{g}_6\right)\right)+\mu  \left(M_1 \left(\frac{2}{3} y_b z_d \bar{g}_1 \bar{g}_5-\frac{2}{3} y_t z_u \bar{g}_2 \bar{g}_5+\bar{g}_1 \bar{g}_2 \left(-2 z_d^2-2 z_u^2-\frac{\hat{\gamma} _2}{3}-\frac{2 \bar{g}_5^2}{9}\right)\right)\right.\\
%%%%
&+a_u \left(-\frac{1}{3} z_q \bar{g}_2 \bar{g}_5+3 z_q \bar{g}_4 \bar{g}_6\right)+M_2 \left(6 y_b z_d \bar{g}_3 \bar{g}_6+6 y_t z_u \bar{g}_4 \bar{g}_6+\bar{g}_3 \bar{g}_4 \left(-6 z_d^2-6 z_u^2-\hat{\gamma} _2\right.\right.\\
%%%
&\left.\left.\left.-6 \bar{g}_6^2\right)\right)\right)+8 M_2 M_3 \bar{g}_6^2 \bar{g}_7^2+M_2^2 \left(-36 g_2^4-6 y_t z_u \bar{g}_3 \bar{g}_6-6 y_b z_d \bar{g}_4 \bar{g}_6-18 g_2^2 \bar{g}_6^2+\frac{3}{2} y_b^2 \bar{g}_6^2\right.\\
%%%%
&\left.+\frac{3}{2} y_t^2 \bar{g}_6^2+\frac{1}{8} \bar{g}_5^2 \bar{g}_6^2+\frac{75 \bar{g}_6^4}{8}+\bar{g}_4^2 \left(3 z_d^2+\frac{9 \bar{g}_6^2}{4}\right)+\bar{g}_3^2 \left(3 z_u^2+\frac{9 \bar{g}_6^2}{4}\right)+\bar{g}_6^2 \left(\frac{3 z_q^2}{2}+6 \bar{g}_7^2\right)\right)\\
%%%%
&+M_1 \left(\frac{1}{6} M_2 \bar{g}_5^2 \bar{g}_6^2\!+\frac{8}{27} M_3 \bar{g}_5^2 \bar{g}_7^2\!-\frac{4}{3} a_u z_u \bar{g}_1 \bar{g}_8\right)+M_1^2 \left(\frac{2}{3} y_t z_u \bar{g}_1 \bar{g}_5-\frac{2}{3} y_b z_d \bar{g}_2 \bar{g}_5+\frac{1}{18} y_b^2 \bar{g}_5^2\right.\\
%%%%
&+\frac{1}{18} y_t^2 \bar{g}_5^2+\frac{23 \bar{g}_5^4}{648}+\bar{g}_2^2 \left(z_d^2+\frac{\bar{g}_5^2}{12}\right)+\bar{g}_1^2 \left(z_u^2+\frac{\bar{g}_5^2}{12}\right)-\frac{8}{9} z_q z_u \bar{g}_5 \bar{g}_8+\frac{16}{9} z_u^2 \bar{g}_8^2+\bar{g}_5^2 \left(\frac{z_q^2}{18}\right.\\
%%%%
&\left.\left.+\frac{\bar{g}_6^2}{8}+\frac{2 \bar{g}_7^2}{9}+\frac{2 \bar{g}_8^2}{9}\right)\right)+a_u^2 \left(\frac{97 g_1^2}{30}+\frac{3 g_2^2}{2}+\frac{8 g_3^2}{3}-6 y_b^2-6 y_t^2-2 y_{\tau }^2-4 z_q^2-\frac{4 \hat{\gamma} _2}{3}-\frac{7 \hat{\gamma} _4}{18}\right.\\
%%%%
&\left.+\frac{16 \hat{\gamma} _5}{9}\!+\!3 \hat{\gamma} _7\!-\frac{3 \hat{\gamma} _8}{2}\!+\!\frac{16 \hat{\gamma} _9}{3}\!-\frac{\bar{g}_1^2}{2}-\frac{\bar{g}_2^2}{2}-\frac{3 \bar{g}_3^2}{2}-\frac{3 \bar{g}_4^2}{2}-\frac{8 \bar{g}_8^2}{9}-\frac{8 \bar{g}_9^2}{3}\right)+\mu ^2 \left(-\frac{6 g_1^4}{25}-18 g_2^4\right.\\
%%%%
&+4 y_b^2 z_d^2+24 z_d^4+4 y_t^2 z_u^2+12 z_q^2 z_u^2+24 z_u^4+g_2^2 \left(-6 z_d^2-6 z_u^2\right)+g_1^2 \left(-2 z_d^2-\frac{14 z_u^2}{5}\right)\\
%%%%
&+z_d^2 \left(6 z_q^2+4 z_u^2\right)+\frac{1}{9} z_q^2 \bar{g}_5^2+\bar{g}_2^2 \left(z_d^2+\frac{z_u^2}{2}+\frac{\bar{g}_5^2}{18}\right)+\bar{g}_1^2 \left(\frac{z_d^2}{2}+z_u^2+\frac{\bar{g}_5^2}{18}\right)+3 z_q^2 \bar{g}_6^2\\
%%%%
&+y_t \left(\frac{2}{3} z_u \bar{g}_1 \bar{g}_5-6 z_u \bar{g}_3 \bar{g}_6\right)+y_b \left(-\frac{2}{3} z_d \bar{g}_2 \bar{g}_5-6 z_d \bar{g}_4 \bar{g}_6\right)+\bar{g}_4^2 \left(3 z_d^2+\frac{3 z_u^2}{2}+\frac{3 \bar{g}_6^2}{2}\right)\\
%%%%
&\left.+\bar{g}_3^2 \left(\frac{3 z_d^2}{2}+3 z_u^2+\frac{3 \bar{g}_6^2}{2}\right)+\frac{16}{3} z_q^2 \bar{g}_7^2-\frac{8}{9} z_q z_u \bar{g}_5 \bar{g}_8+\frac{8}{9} z_u^2 \bar{g}_8^2-\frac{32}{3} z_q z_u \bar{g}_7 \bar{g}_9+\frac{8}{3} z_u^2 \bar{g}_9^2\right)\\
%%%%
&+m_U^2 \left(\frac{4 g_1^4}{15}+\frac{40 g_3^4}{3}-\frac{64}{45} g_1^2 \hat{\gamma} _5-\frac{64}{9} g_3^2 \hat{\gamma} _5+\frac{8}{3} z_q^2 \hat{\gamma} _5-\frac{4 \hat{\gamma} _5^2}{27}-\frac{8 \hat{\gamma} _9^2}{3}-\frac{4}{9} z_q z_u \bar{g}_5 \bar{g}_8+\frac{16}{27} \hat{\gamma} _5 \bar{g}_8^2\right.\\
%%%%
&\left.-\frac{16}{3} z_q z_u \bar{g}_7 \bar{g}_9+\frac{16}{9} \hat{\gamma} _5 \bar{g}_9^2\right)+m_Q^2 \left(\frac{1709 g_1^4}{3600}+\frac{281 g_2^4}{16}-\frac{112 g_3^4}{9}-\frac{15 z_d^4}{2}-\frac{9}{2} z_q^2 z_u^2-\frac{15 z_u^4}{2}\right.\\
%%%%
&-\frac{\hat{\gamma} _2^2}{72}-\frac{35 \hat{\gamma} _4^2}{1296}-\frac{\hat{\gamma} _5^2}{27}-\frac{3 \hat{\gamma} _7^2}{8}+z_d^2 \left(-\frac{7 \hat{\gamma} _4}{9}-3 \hat{\gamma} _8\right)+z_u^2 \left(-\frac{7 \hat{\gamma} _4}{9}-3 \hat{\gamma} _8\right)-\frac{5 \hat{\gamma} _4 \hat{\gamma} _8}{24}-\frac{75 \hat{\gamma} _8^2}{16}
\end{align*}
\begin{align*}
\phantom{\beta_{m^2_Q}=}&-\frac{2 \hat{\gamma} _9^2}{3}-\frac{7 \bar{g}_5^4}{864}-\bar{g}_2^2 \left(\frac{3 z_d^2}{8}+\frac{\bar{g}_5^2}{48}\right)-\bar{g}_1^2 \left(\frac{3 z_u^2}{8}+\frac{\bar{g}_5^2}{48}\right)-\frac{81 \bar{g}_6^4}{32}+y_t \left(3 z_u \bar{g}_3 \bar{g}_6-\frac{1}{3} z_u \bar{g}_1 \bar{g}_5\right)\\
%%%%
&+y_b \left(\frac{1}{3} z_d \bar{g}_2 \bar{g}_5+3 z_d \bar{g}_4 \bar{g}_6\right)+\bar{g}_4^2 \left(-\frac{9 z_d^2}{8}-\frac{9 \bar{g}_6^2}{16}\right)+\bar{g}_3^2 \left(-\frac{9 z_u^2}{8}-\frac{9 \bar{g}_6^2}{16}\right)-\frac{14 \bar{g}_7^4}{3}\\
%%%%
&+\bar{g}_6^2 \left(-\frac{9 z_q^2}{8}-\frac{7 \hat{\gamma} _4}{12}-\frac{9 \hat{\gamma} _8}{4}-3 \bar{g}_7^2\right)+y_b^2 \left(-3 z_d^2-\frac{\bar{g}_5^2}{24}-\frac{9 \bar{g}_6^2}{8}-2 \bar{g}_7^2\right)+y_t^2 \left(-3 z_u^2-\frac{\bar{g}_5^2}{24}\right.\\
%%%%
&\left.-\frac{9 \bar{g}_6^2}{8}-2 \bar{g}_7^2\right)+g_1^2 \left(\frac{g_2^2}{8}+\frac{2 g_3^2}{9}+\frac{13 z_d^2}{12}+\frac{25 z_u^2}{12}+\frac{7 \hat{\gamma} _4}{135}+\frac{\hat{\gamma} _8}{5}+\frac{\bar{g}_5^2}{432}+\frac{\bar{g}_6^2}{16}+\frac{\bar{g}_7^2}{9}\right)\\
%%%%
&+g_2^2 \left(10 g_3^2+\frac{15 z_d^2}{4}+\frac{15 z_u^2}{4}+\frac{7 \hat{\gamma} _4}{3}+9 \hat{\gamma} _8+\frac{5 \bar{g}_5^2}{48}+\frac{165 \bar{g}_6^2}{16}+5 \bar{g}_7^2\right)+g_3^2 \left(\frac{20 z_d^2}{3}+\frac{20 z_u^2}{3}\right.\\
%%%%
&\left.+\frac{112 \hat{\gamma} _4}{27}+16 \hat{\gamma} _8+\frac{5 \bar{g}_5^2}{27}+5 \bar{g}_6^2+\frac{260 \bar{g}_7^2}{9}\right)+\frac{4}{9} z_q z_u \bar{g}_5 \bar{g}_8-\frac{2}{3} z_u^2 \bar{g}_8^2+\bar{g}_5^2 \left(-\frac{z_q^2}{24}-\frac{7 \hat{\gamma} _4}{324}-\frac{\hat{\gamma} _8}{12}\right.\\
%%%%
&\left.\left.-\frac{\bar{g}_6^2}{16}\!-\frac{\bar{g}_7^2}{9}\!-\frac{\bar{g}_8^2}{18}\right)\!+\!\frac{16}{3} z_q z_u \bar{g}_7 \bar{g}_9\!-2 z_u^2 \bar{g}_9^2\!-\bar{g}_7^2 \left(2 z_q^2+\frac{28 \hat{\gamma} _4}{27}+4 \hat{\gamma} _8+\bar{g}_9^2\right)\right)+M_3^2 \left(-96 g_3^4\right.\\
%%%%
&-48 g_3^2 \bar{g}_7^2\!+\!\frac{8}{3} y_b^2 \bar{g}_7^2\!+\!\frac{8}{3} y_t^2 \bar{g}_7^2\!+\!\frac{2}{9} \bar{g}_5^2 \bar{g}_7^2+6 \bar{g}_6^2 \bar{g}_7^2+\frac{160 \bar{g}_7^4}{9}-\frac{32}{3} z_q z_u \bar{g}_7 \bar{g}_9+\frac{16}{3} z_u^2 \bar{g}_9^2+\bar{g}_7^2 \left(\frac{8 z_q^2}{3}\right.\\
%%%%%
&\left.\left.\left.+4 \bar{g}_9^2\right)\right)\right\},
\end{align*}
\begin{align*}
 \beta_{m^2_U}=&\frac{1}{16\pi^2}\left\{4 a_u^2-8 \mu ^2 z_q^2-\frac{4}{3} m_H^2 \hat{\gamma} _3-\frac{4}{3} m_Q^2 \hat{\gamma} _5-\frac{16}{9} M_1^2 \bar{g}_8^2-\frac{16}{3} M_3^2 \bar{g}_9^2+m_U^2 \left(-\frac{8 g_1^2}{5}-8 g_3^2\right.\right.\\
%%%
&\left.\left.+4 z_q^2+\frac{32 \hat{\gamma} _6}{9}+\frac{8 \bar{g}_8^2}{9}+\frac{8 \bar{g}_9^2}{3}\right)\right\}+\frac{1}{(16\pi^2)^2}\left\{-6 a_u M_2 z_q \bar{g}_3 \bar{g}_6+M_2^2 \left(6 z_q^2 \bar{g}_3^2+6 z_q^2 \bar{g}_6^2\right)\right.\\
%%%%
&+a_u^2 \left(-\frac{8 g_1^2}{15}+24 g_2^2+\frac{16 g_3^2}{3}-12 y_b^2-12 y_t^2-4 y_{\tau }^2-4 z_d^2-4 z_u^2+\frac{20 \hat{\gamma} _3}{3}+\frac{20 \hat{\gamma} _5}{9}-\frac{64 \hat{\gamma} _6}{9}\right.\\
%%%%
&\left.+\frac{32 \hat{\gamma} _9}{3}\!-\bar{g}_1^2\!-\bar{g}_2^2\!-3 \bar{g}_3^2\!-3 \bar{g}_4^2\!-\frac{\bar{g}_5^2}{9}\!-3 \bar{g}_6^2\!-\frac{16 \bar{g}_7^2}{3}\right)\!+\!m_H^2 \left(\frac{8 g_1^4}{5}-\frac{8}{5} g_1^2 \hat{\gamma} _3-8 g_2^2 \hat{\gamma} _3+8 y_b^2 \hat{\gamma} _3\right.\\
%%%%
&\left.+8 y_t^2 \hat{\gamma} _3\!+\!\frac{8}{3} y_{\tau }^2 \hat{\gamma} _3\!-\frac{8 \hat{\gamma} _3^2}{9}\!+\!\frac{2}{3} \hat{\gamma} _3 \bar{g}_1^2\!+\!\frac{2}{3} \hat{\gamma} _3 \bar{g}_2^2\!+\!2 \hat{\gamma} _3 \bar{g}_3^2\!+\!2 \hat{\gamma} _3 \bar{g}_4^2-\frac{8}{3} y_t z_q \bar{g}_1 \bar{g}_8\right)+\mu  \left(M_2 \left(-12 z_q^2\right.\right.\\
%%%%
&\left.\left.+4 \hat{\gamma} _3\right) \bar{g}_3 \bar{g}_4\!+\!\frac{8}{3} a_u z_u \bar{g}_2 \bar{g}_8\!+\!M_1 \left(\frac{16}{3} y_t z_q \bar{g}_2 \bar{g}_8\!+\!\bar{g}_1 \bar{g}_2 \left(\frac{4 \hat{\gamma} _3}{3}-4 z_q^2-\frac{32 \bar{g}_8^2}{9}\right)\right)\right)+m_Q^2 \left(\frac{8 g_1^4}{15}\right.\\
%%%%%
&+\frac{80 g_3^4}{3}\!-\frac{8}{45} g_1^2 \hat{\gamma} _5\!-8 g_2^2 \hat{\gamma} _5\!-\frac{128}{9} g_3^2 \hat{\gamma} _5\!+\!\frac{8}{3} z_d^2 \hat{\gamma} _5+\frac{8}{3} z_u^2 \hat{\gamma} _5-\frac{8 \hat{\gamma} _5^2}{27}-\frac{16 \hat{\gamma} _9^2}{3}+\frac{2}{27} \hat{\gamma} _5 \bar{g}_5^2+2 \hat{\gamma} _5 \bar{g}_6^2\\
%%%%
&\left.+\frac{32}{9} \hat{\gamma} _5 \bar{g}_7^2-\frac{8}{9} z_q z_u \bar{g}_5 \bar{g}_8-\frac{32}{3} z_q z_u \bar{g}_7 \bar{g}_9\right)+\mu ^2 \left(4 y_b^2 z_q^2-\frac{96 g_1^4}{25}-\frac{8}{5} g_1^2 z_q^2-24 g_2^2 z_q^2+4 y_t^2 z_q^2\right.\\
%%%
&+12 z_d^2 z_q^2+36 z_q^4+24 z_q^2 z_u^2+6 z_q^2 \bar{g}_3^2+3 z_q^2 \bar{g}_4^2+\frac{1}{9} z_q^2 \bar{g}_5^2+3 z_q^2 \bar{g}_6^2+\frac{16}{3} z_q^2 \bar{g}_7^2-\frac{16}{3} y_t z_q \bar{g}_1 \bar{g}_8\\
%%%%
&\left.-\frac{16}{9} z_q z_u \bar{g}_5 \bar{g}_8+\frac{32}{9} z_u^2 \bar{g}_8^2+\bar{g}_2^2 \left(z_q^2+\frac{8 \bar{g}_8^2}{9}\right)+\bar{g}_1^2 \left(2 z_q^2+\frac{8 \bar{g}_8^2}{9}\right)-\frac{64}{3} z_q z_u \bar{g}_7 \bar{g}_9+\frac{32}{3} z_u^2 \bar{g}_9^2\right)
\end{align*}
\begin{align*}
%%%%
\phantom{\beta_{m^2_U}=}
&+M_1 \left(\frac{2}{3} a_u z_q \bar{g}_1 \bar{g}_5\!+\!\frac{128}{27} M_3 \bar{g}_8^2 \bar{g}_9^2\right)\!+\!M_1^2 \left(\frac{16}{9} y_t^2 \bar{g}_8^2-\frac{16}{3} y_t z_q \bar{g}_1 \bar{g}_8-\frac{16}{9} z_q z_u \bar{g}_5 \bar{g}_8+\frac{4}{3} \bar{g}_2^2 \bar{g}_8^2\right.\\
%%%
&\left.+\frac{448 \bar{g}_8^4}{81}+\bar{g}_5^2 \left(\frac{2 z_q^2}{9}+\frac{4 \bar{g}_8^2}{9}\right)+\bar{g}_1^2 \left(2 z_q^2+\frac{4 \bar{g}_8^2}{3}\right)+\bar{g}_8^2 \left(\frac{16 z_u^2}{9}+\frac{32 \bar{g}_9^2}{9}\right)\right)+M_3^2 \left(-96 g_3^4\right.\\
%%%%%
&\left.-\frac{64}{3} z_q z_u \bar{g}_7 \bar{g}_9-48 g_3^2 \bar{g}_9^2+\frac{16}{3} y_t^2 \bar{g}_9^2+\frac{16}{3} z_u^2 \bar{g}_9^2+\frac{32}{9} \bar{g}_8^2 \bar{g}_9^2+\frac{124 \bar{g}_9^4}{9}+\bar{g}_7^2 \left(\frac{32 z_q^2}{3}+8 \bar{g}_9^2\right)\right)\\
%%%%
&+m_U^2 \left(\frac{2624 g_1^4}{225}-\frac{232 g_3^4}{9}+15 g_2^2 z_q^2-3 y_b^2 z_q^2-12 z_q^4-\frac{2 \hat{\gamma} _3^2}{9}-\frac{2 \hat{\gamma} _5^2}{27}-z_q^2 \left(9 z_u^2+\frac{128 \hat{\gamma} _6}{9}\right)\right.\\
%%%%%
&-\frac{320 \hat{\gamma} _6^2}{81}\!-\frac{4 \hat{\gamma} _9^2}{3}\!-\frac{9}{4} z_q^2 \bar{g}_3^2\!-\frac{9}{4} z_q^2 \bar{g}_6^2+\frac{8}{3} y_t z_q \bar{g}_1 \bar{g}_8+\frac{8}{9} z_q z_u \bar{g}_5 \bar{g}_8-\frac{1}{3} \bar{g}_2^2 \bar{g}_8^2-\frac{32 \bar{g}_8^4}{27}-\bar{g}_1^2 \left(\frac{3 z_q^2}{4}\right.\\
%%%
&\left.+\frac{\bar{g}_8^2}{3}\right)+\bar{g}_5^2 \left(-\frac{z_q^2}{12}-\frac{\bar{g}_8^2}{9}\right)+\frac{32}{3} z_q z_u \bar{g}_7 \bar{g}_9+\left(-4 z_u^2-\frac{256 \hat{\gamma} _6}{27}\right) \bar{g}_9^2-\frac{11 \bar{g}_9^4}{3}+y_t^2 \left(-3 z_q^2\right.\\
%%%
&\left.-\frac{4 \bar{g}_8^2}{3}-4 \bar{g}_9^2\right)+\bar{g}_7^2 \left(-4 z_q^2-2 \bar{g}_9^2\right)+\bar{g}_8^2 \left(-\frac{4 z_u^2}{3}-\frac{256 \hat{\gamma} _6}{81}-\frac{16 \bar{g}_9^2}{9}\right)+g_1^2 \left(\frac{32 g_3^2}{9}+\frac{5 z_q^2}{3}\right.\\
%%%%
&\left.\left.\left.+\frac{1024 \hat{\gamma} _6}{135}+\frac{16 \bar{g}_8^2}{27}+\frac{16 \bar{g}_9^2}{9}\right)+g_3^2 \left(\frac{40 z_q^2}{3}+\frac{1024 \hat{\gamma} _6}{27}+\frac{80 \bar{g}_8^2}{27}+\frac{260 \bar{g}_9^2}{9}\right)\right)\right\}.
\end{align*}
%%%%%%%%%%%%%%%%%%%%%%%%%%%%%%%%%%%%%%%%%%%%%%%%%%%%%%%%%%%%%%%%%%%%%%%%%%%%%%%%%
%%%%%%%%%%%%%%%%%%%%%%%%%%%%%%%%%%%%%%%%%%%%%%%%%%%%%%%%%%%%%%%%%%%%%%%%%%%%%%%%%
%%%%%%%%%%%%%%%%%%%%%%%%%%%%%%%%%%%%%%%%%%%%%%%%%%%%%%%%%%%%%%%%%%%%%%%%%%%%%%%%%
\subsubsection{Higgs anomalous dimension}
%%%%%%%%%%%%%%%%%%%%%%%%%%%%%%%%%%%%%%%%%%%%%%%%%%%%%%%%%%%%%%%%%%%%%%%%%%%%%%%%%
%%%%%%%%%%%%%%%%%%%%%%%%%%%%%%%%%%%%%%%%%%%%%%%%%%%%%%%%%%%%%%%%%%%%%%%%%%%%%%%%%
%%%%%%%%%%%%%%%%%%%%%%%%%%%%%%%%%%%%%%%%%%%%%%%%%%%%%%%%%%%%%%%%%%%%%%%%%%%%%%%%%
\begin{align*}
\gamma_H=&\frac{1}{16\pi^2}\left\{\frac{\bar{g}_1^2}{4}+\frac{\bar{g}_2^2}{4}+\frac{3 \bar{g}_3^2}{4}+\frac{3 \bar{g}_4^2}{4}+3 y_b^2+g_1^2 \left(\frac{3 \xi }{20}-\frac{9}{20}\right)+g_2^2 \left(\frac{3 \xi }{4}-\frac{9}{4}\right)+3 y_t^2+y_{\tau }^2\right\}\\
%%%%%
&+\frac{1}{(16\pi^2)^2}\left\{\frac{1611 g_1^4}{800}+\left(\frac{27 g_2^2}{80}+\frac{5 y_b^2}{8}+\frac{17 y_t^2}{8}+\frac{15 y_{\tau }^2}{8}+\frac{3 \bar{g}_1^2}{32}+\frac{3 \bar{g}_2^2}{32}+\frac{9 \bar{g}_3^2}{32}+\frac{9 \bar{g}_4^2}{32}\right) g_1^2\right.\\
%%%%
&+\left(\frac{3 \xi ^2}{8}+3 \xi -\frac{85}{32}\right) g_2^4+\left(\frac{45 y_b^2}{8}+\frac{45 y_t^2}{8}+\frac{15 y_{\tau }^2}{8}+\frac{15 \bar{g}_1^2}{32}+\frac{15 \bar{g}_2^2}{32}+\frac{165 \bar{g}_3^2}{32}+\frac{165 \bar{g}_4^2}{32}\right) g_2^2\\
%%%%
&+\left(-\frac{1}{16} 9 z_q^2-\frac{\bar{g}_8^2}{4}-\frac{9 z_u^2}{16}-\frac{5 \bar{g}_2^2}{16}-\frac{9 \bar{g}_3^2}{32}-\frac{\bar{g}_5^2}{32}\right) \bar{g}_1^2+\left(-\frac{1}{16} 9 z_d^2-\frac{\bar{g}_8^2}{4}-\frac{9 \bar{g}_4^2}{32}-\frac{\bar{g}_5^2}{32}\right) \bar{g}_2^2\\
%%%%
&+\left(-\frac{27}{16}  z_q^2-\frac{27 z_u^2}{16}-\frac{3 \bar{g}_4^2}{16}-\frac{27 \bar{g}_6^2}{32}\right) \bar{g}_3^2+\left(-\frac{1}{16} 27 z_d^2-\frac{27 \bar{g}_6^2}{32}\right) \bar{g}_4^2+\frac{3 \hat{\gamma }_1^2}{32}+\frac{\hat{\gamma }_2^2}{48}+\frac{\hat{\gamma }_3^2}{6}\\
%%%%
&+\frac{9 \hat{\gamma }_7^2}{16}+g_3^2 \left(20 y_b^2+20 y_t^2\right)-\frac{3}{4} \bar{g}_1 \bar{g}_2 \bar{g}_3 \bar{g}_4+y_b \left(\frac{1}{2} z_d \bar{g}_2 \bar{g}_5+\frac{9}{2} z_d \bar{g}_4 \bar{g}_6\right)+y_b^2 \left(-\frac{1}{2} 9 z_d^2-3 \bar{g}_7^2\right.\\
%%%%
&\left.-\frac{9 z_q^2}{4}-\frac{\bar{g}_5^2}{16}-\frac{27 \bar{g}_6^2}{16}\right)+y_t \left(\frac{9}{2} z_u \bar{g}_3 \bar{g}_6+\bar{g}_1 \left(2 z_q \bar{g}_8-\frac{1}{2} z_u \bar{g}_5\right)\right)+y_t^2 \left(\frac{3 y_b^2}{2}-3 \bar{g}_7^2-\bar{g}_8^2\right.\\
%%%%
&\left.\left.-3 \bar{g}_9^2\!-\!\frac{9 z_u^2}{2}\!-\!\frac{9 z_q^2}{4}-\frac{\bar{g}_5^2}{16}-\frac{27 \bar{g}_6^2}{16}\right)-\frac{27 y_b^4}{4}-\frac{27 y_t^4}{4}-\frac{9 y_{\tau }^4}{4}-\frac{9 \bar{g}_1^4}{64}-\frac{9 \bar{g}_2^4}{64}-\frac{45 \bar{g}_3^4}{64}-\frac{45 \bar{g}_4^4}{64}\right\}.
\end{align*}
\newpage
%%%%%%%%%%%%%%%%%%%%%%%%%%%%%%%%%%%%%%%%%%%%%%%%%%%%%%%%%%%%%%%%%%%%%%%%%%%%%%%%%
%%%%%%%%%%%%%%%%%%%%%%%%%%%%%%%%%%%%%%%%%%%%%%%%%%%%%%%%%%%%%%%%%%%%%%%%%%%%%%%%%
%%%%%%%%%%%%%%%%%%%%%%%%%%%%%%%%%%%%%%%%%%%%%%%%%%%%%%%%%%%%%%%%%%%%%%%%%%%%%%%%%
%%%%%%%%%%%%%%%%%%%%%%%%%%%%%%%%%%%%%%%%%%%%%%%%%%%%%%%%%%%%%%%%%%%%%%%%%%%%%%%%%
%%%%%%%%%%%%%%%%%%%%%%%%%%%%%%%%%%%%%%%%%%%%%%%%%%%%%%%%%%%%%%%%%%%%%%%%%%%%%%%%%
%%%%%%%%%%%%%%%%%%%%%%%%%%%%%%%%%%%%%%%%%%%%%%%%%%%%%%%%%%%%%%%%%%%%%%%%%%%%%%%%%
%%%%%%%%%%%%%%%%%%%%%%%%%%%%%%%%%%%%%%%%%%%%%%%%%%%%%%%%%%%%%%%%%%%%%%%%%%%%%%%%%
%%%%%%%%%%%%%%%%%%%%%%%%%%%%%%%%%%%%%%%%%%%%%%%%%%%%%%%%%%%%%%%%%%%%%%%%%%%%%%%%%
%%%%%%%%%%%%%%%%%%%%%%%%%%%%%%%%%%%%%%%%%%%%%%%%%%%%%%%%%%%%%%%%%%%%%%%%%%%%%%%%%
%%%%%%%%%%%%%%%%%%%%%%%%%%%%%%%%%%%%%%%%%%%%%%%%%%%%%%%%%%%%%%%%%%%%%%%%%%%%%%%%%
%%%%%%%%%%%%%%%%%%%%%%%%%%%%%%%%%%%%%%%%%%%%%%%%%%%%%%%%%%%%%%%%%%%%%%%%%%%%%%%%%
%%%%%%%%%%%%%%%%%%%%%%%%%%%%%%%%%%%%%%%%%%%%%%%%%%%%%%%%%%%%%%%%%%%%%%%%%%%%%%%%%
%%%%%%%%%%%%%%%%%%%%%%%%%%%%%%%%%%%%%%%%%%%%%%%%%%%%%%%%%%%%%%%%%%%%%%%%%%%%%%%%%\
\newpage
\subsection{Effective SUSY with a full generation of light scalars\label{subsec:Betas:NonMinEffSUSY}}
This section presents the beta functions of the independent couplings (eq.~\eqref{eq:couplingsNMES}) of the Lagrangian of eq.~\eqref{eq:LagrangianNMES}. Again, as in the previous section, in order to simplify the formulae the phases and the couplings ${y_{u/d/l}}_{ij}, g_{{Q/U/D}_{i,k}},$ ${z_{q/q^*/u/d/l/e}}_i$, $\,i,j\neq3$, are taken as zero. The compact notation of eq.~\eqref{eq:couplingsNMES:notation} is employed for the beta functions other than those of the gauge couplings. 
Some beta functions are given in comparison with those of the Minimal Effective Susy scenario, which  are denoted with the superscript ``MES''. The parameters $\hat\gamma$ in the minimal Effective Susy case can be expressed in terms of the parameters $\tilde\gamma$ of the nonminimal scenario with the aid of eqs.~\eqref{eq:couplingsMES:notation} and \eqref{eq:couplingsNMES:notation}. Two-loop contributions are given only for the gauge and Standard Model-like Yukawas, and for the fermion and scalar mass parameters; the full expressions are available online.
%%%%%%%%%%%%%%%%%%%%%%%%%%%%%%%%%%%%%%%%%%%%%%%%%%%%%%%%%%%%%%%%%%%%%%%%%%%%%%%%%
%%%%%%%%%%%%%%%%%%%%%%%%%%%%%%%%%%%%%%%%%%%%%%%%%%%%%%%%%%%%%%%%%%%%%%%%%%%%%%%%%
%%%%%%%%%%%%%%%%%%%%%%%%%%%%%%%%%%%%%%%%%%%%%%%%%%%%%%%%%%%%%%%%%%%%%%%%%%%%%%%%%
\subsubsection{Gauge couplings}
%%%%%%%%%%%%%%%%%%%%%%%%%%%%%%%%%%%%%%%%%%%%%%%%%%%%%%%%%%%%%%%%%%%%%%%%%%%%%%%%%
%%%%%%%%%%%%%%%%%%%%%%%%%%%%%%%%%%%%%%%%%%%%%%%%%%%%%%%%%%%%%%%%%%%%%%%%%%%%%%%%%
%%%%%%%%%%%%%%%%%%%%%%%%%%%%%%%%%%%%%%%%%%%%%%%%%%%%%%%%%%%%%%%%%%%%%%%%%%%%%%%%%
\begin{align*}
\beta_{g_1}=&\beta_{g_1}^{\rm MES}\!+\!\frac{11g_1^3}{480\pi^2}\!+\!\frac{1}{(16\pi^2)^2}\left\{\!\frac{251 g_1^5}{150}+\frac{9}{10} g_1^3 g_2^2+\frac{16}{15} g_1^3 g_3^2\!-\!\sum_{i=1}^3g_1^3\left(\!\frac{4}{15} g_{D_{i,3}}^2\!+\frac{1}{45} g_{D_{i,1}}^2\!+\frac{3}{5} g_{E_{i,1}}^2\right.\right.\\
%%%%
&\left.\left.+\frac{9}{40} g_{L_{i,2}}^2+\frac{3}{40} g_{L_{i,1}}^2+\frac{3{z_e}_i^2}{2}+\frac{3{z_l}_i^2}{5}+{z_q^*}_i^2\right)\right\},\\
%%%
\beta_{g_2}=&\beta_{g_2}^{\rm MES}\!+\frac{g_2^3}{96\pi^2}+\frac{1}{(16\pi^2)^2}\left\{\!\frac{3}{10} g_1^2 g_2^3\!+\frac{13 g_2^5}{6}\!-\!\sum_{i=1}^3g_2^3\left(\!\frac{11}{8} g_{L_{i,2}}^2\!+\frac{1}{8} g_{L_{i,1}}^2\!+\frac{{z_e}_i^2}{2}\!+\!{z_l}_i^2\!+3 {z_q^*}_i^2\right)\right\},\\
%%%%
 \beta_{g_3}=&\beta_{g_3}^{\rm MES}+\frac{g_3^3}{96\pi^2}+\frac{1}{(16\pi^2)^2}\left\{\frac{2}{15} g_1^2 g_3^3+\frac{11 g_3^5}{3}+\sum_{i=1}^3g_3^3\left(-\frac{13}{6} g_{D_{i,3}}^2 -\frac{1}{18} g_{D_{i,1}}^2 - {z_{q^*}}_i^2\right)\right\}.
\end{align*}
%%%%%%%%%%%%%%%%%%%%%%%%%%%%%%%%%%%%%%%%%%%%%%%%%%%%%%%%%%%%%%%%%%%%%%%%%%%%%%%%%
%%%%%%%%%%%%%%%%%%%%%%%%%%%%%%%%%%%%%%%%%%%%%%%%%%%%%%%%%%%%%%%%%%%%%%%%%%%%%%%%%
%%%%%%%%%%%%%%%%%%%%%%%%%%%%%%%%%%%%%%%%%%%%%%%%%%%%%%%%%%%%%%%%%%%%%%%%%%%%%%%%%
\subsubsection{Yukawas}
%%%%%%%%%%%%%%%%%%%%%%%%%%%%%%%%%%%%%%%%%%%%%%%%%%%%%%%%%%%%%%%%%%%%%%%%%%%%%%%%%
%%%%%%%%%%%%%%%%%%%%%%%%%%%%%%%%%%%%%%%%%%%%%%%%%%%%%%%%%%%%%%%%%%%%%%%%%%%%%%%%%
%%%%%%%%%%%%%%%%%%%%%%%%%%%%%%%%%%%%%%%%%%%%%%%%%%%%%%%%%%%%%%%%%%%%%%%%%%%%%%%%%
\begin{align*}
 \beta_{y_t}=&\beta_{y_t}^{\rm MES}+\frac{1}{16\pi^2}\frac{1}{2} y_t z_{q^*}^2+\frac{1}{(16\pi^2)^2}\left\{\frac{22}{45} g_1^4 y_t+\frac{1}{2} g_2^4 y_t+\frac{22}{9} g_3^4 y_t+\frac{43}{30} g_1^2 y_t z_{q^*}^2-\frac{3}{2} g_2^2 y_t z_{q^*}^2\right.\\
%%%%
&+\frac{11}{3} g_3^2 y_t z_{q^*}^2-\frac{5}{2} y_t^3 z_{q^*}^2+z_u \left(-\frac{1}{36} \bar{g}_1 \bar{g}_5 \bar{g}_{10}^2+\frac{3}{8} \bar{g}_3 \bar{g}_6 \bar{g}_{13}^2+\bar{g}_1 \bar{g}_5 \left(-\frac{\bar{g}_{12}^2}{24}-\frac{\bar{g}_{14}^2}{12}\right)\right)\\
%%%%
&+y_t \left(-\frac{15 z_{q^*}^4}{8}+\frac{\tilde{\gamma} _4^2}{24}+\frac{\tilde{\gamma} _5^2}{16}+\frac{\tilde{\gamma} _6^2}{8}+\frac{3 \tilde{\gamma} _{23}^2}{16}+z_e^2 \left(-\frac{z_{q^*}^2}{8}-\frac{3 \bar{g}_2^2}{16}-\frac{9 \bar{g}_4^2}{16}\right)+z_l^2 \left(-\frac{z_{q^*}^2}{8}-\frac{3 \bar{g}_2^2}{16}\right.\right.\\
%%%%
&\left.-\frac{9 \bar{g}_4^2}{16}\right)\!+\!\frac{2}{3} y_b z_{q^*} \bar{g}_2 \bar{g}_{10}\!-\left(\frac{\bar{g}_1^2}{16}\!+\!\frac{\bar{g}_5^2}{864}\!+\!\frac{\bar{g}_8^2}{54}\right) \bar{g}_{10}^2-\left(\frac{\bar{g}_7^2}{12}+\frac{\bar{g}_9^2}{12}\right) \bar{g}_{11}^2+y_b^2 \left(-\frac{5 z_{q^*}^2}{4}-\frac{11 \bar{g}_{10}^2}{72}\right.\\
%%%
&\left.\frac{11 \bar{g}_{11}^2}{6}\right)\!+\!z_{q^*}^2 \left(\!-\frac{3 z_d^2}{8}-\frac{\tilde{\gamma} _4}{3}-\frac{35 \bar{g}_2^2}{32}-\frac{9 \bar{g}_4^2}{32}-\frac{\bar{g}_{10}^2}{12}-\bar{g}_{11}^2\right)-\frac{9}{32} \bar{g}_4^2 \bar{g}_{13}^2-\left(\frac{9 \bar{g}_3^2}{32}+\frac{3 \bar{g}_6^2}{64}\right) \bar{g}_{13}^2
\end{align*}
\begin{align*}
%%%%
\phantom{\beta_{y_t}=}&+y_{\tau } \left(z_e \left(-\frac{1}{2} \bar{g}_2 \bar{g}_{12}+\frac{3}{2} \bar{g}_4 \bar{g}_{13}\right)+z_l \bar{g}_2 \bar{g}_{14}\right)+y_{\tau }^2 \left(-\frac{3 z_e^2}{2}-\frac{3 z_l^2}{4}-\frac{3 \bar{g}_{12}^2}{16}-\frac{9 \bar{g}_{13}^2}{16}-\frac{3 \bar{g}_{14}^2}{4}\right)\\
%%%%
&+\bar{g}_1^2 \left(-\frac{3 \bar{g}_{12}^2}{32}-\frac{3 \bar{g}_{14}^2}{16}\right)+\bar{g}_2^2 \left(-\frac{\bar{g}_{10}^2}{16}-\frac{3 \bar{g}_{12}^2}{32}-\frac{3 \bar{g}_{14}^2}{16}\right)+\bar{g}_8^2 \left(-\frac{\bar{g}_{12}^2}{36}-\frac{\bar{g}_{14}^2}{18}\right)+\bar{g}_5^2 \left(-\frac{\bar{g}_{12}^2}{576}\right.\\
%%%%
&\left.\left.\left.-\frac{\bar{g}_{14}^2}{288}\right)\right)+z_q \left(\frac{1}{9} \bar{g}_1 \bar{g}_8 \bar{g}_{10}^2+\bar{g}_1 \bar{g}_8 \left(\frac{\bar{g}_{12}^2}{6}+\frac{\bar{g}_{14}^2}{3}\right)\right)\right\},
\end{align*}
%%%%%%%%%%%
%%%%%%%%%%%
%%%%%%%%%%%
\begin{align*}
 \beta_{y_b}=&\beta_{y_b}^{\rm MES}+\frac{1}{16\pi^2}\left\{\frac{1}{2} y_b z_{q^*}^2-\frac{1}{3} z_{q^*} \bar{g}_2 \bar{g}_{10}+\frac{1}{18} y_b \bar{g}_{10}^2+\frac{2}{3} y_b \bar{g}_{11}^2\right\}+\frac{1}{(16\pi^2)^2}\left\{\frac{77}{900} g_1^4 y_b+\frac{1}{2} g_2^4 y_b\right.\\
%%%%
&+\frac{22}{9} g_3^4 y_b+2 y_b^2 z_{q^*} \bar{g}_2 \bar{g}_{10}+\frac{1}{3} y_{\tau }^2 z_{q^*} \bar{g}_2 \bar{g}_{10}+\frac{1}{2} z_d^2 z_{q^*} \bar{g}_2 \bar{g}_{10}+\frac{1}{6} z_l^2 z_{q^*} \bar{g}_2 \bar{g}_{10}+\frac{1}{2} z_{q^*}^3 \bar{g}_2 \bar{g}_{10}\\
%%%%
&+z_e^2 \left(z_d \left(\frac{1}{12} \bar{g}_2 \bar{g}_5\!+\!\frac{3}{4} \bar{g}_4 \bar{g}_6\right)\!+\frac{1}{6} z_{q^*} \bar{g}_2 \bar{g}_{10}\right)+g_2^2 \left(-\frac{3}{2} y_b z_{q^*}^2+\frac{1}{2} z_{q^*} \bar{g}_2 \bar{g}_{10}\right)+y_t^2 \left(-\frac{5}{4} y_b z_{q^*}^2\right.\\
%%%
&\left.+z_{q^*} \bar{g}_2 \bar{g}_{10}\right)+y_b^3 \left(-\frac{5 z_{q^*}^2}{2}-\frac{19 \bar{g}_{10}^2}{72}-\frac{19 \bar{g}_{11}^2}{6}\right)+g_1^2 \left(\frac{1}{15} z_{q^*} \bar{g}_2 \bar{g}_{10}+y_b \left(-\frac{11 z_{q^*}^2}{30}+\frac{11 \bar{g}_{10}^2}{540}\right.\right.\\
%%%%
&\left.\left.+\frac{11 \bar{g}_{11}^2}{45}\right)\right)+g_3^2 \left(-\frac{8}{3} z_{q^*} \bar{g}_2 \bar{g}_{10}+y_b \left(\frac{11 z_{q^*}^2}{3}+\frac{11 \bar{g}_{10}^2}{27}+\frac{143 \bar{g}_{11}^2}{9}\right)\right)+y_{\tau } \left(z_e \left(\frac{1}{3} z_{q^*} \bar{g}_{10} \bar{g}_{12}\right.\right.\\
%%%
&\left.\left.+z_d \left(\frac{1}{6} \bar{g}_5 \bar{g}_{12}-\frac{3}{2} \bar{g}_6 \bar{g}_{13}\right)\right)-\frac{1}{3} z_d z_l \bar{g}_5 \bar{g}_{14}-\frac{2}{3} z_l z_{q^*} \bar{g}_{10} \bar{g}_{14}\right)+y_b \left(-\frac{15 z_{q^*}^4}{8}+z_d^2 \left(-\frac{z_l^2}{4}\right.\right.\\
%%%%
&\left.+\frac{7 z_{q^*}^2}{8}\right)+\frac{\tilde{\gamma} _4^2}{24}+\frac{\tilde{\gamma} _5^2}{16}+\frac{\tilde{\gamma} _6^2}{8}+\frac{3 \tilde{\gamma} _{23}^2}{16}+z_l^2 \left(-\frac{z_{q^*}^2}{8}-\frac{3 \bar{g}_2^2}{16}-\frac{9 \bar{g}_4^2}{16}\right)+z_e^2 \left(-\frac{z_d^2}{4}-\frac{z_{q^*}^2}{8}-\frac{3 \bar{g}_2^2}{16}\right.\\
%%%%
&\left.-\frac{9 \bar{g}_4^2}{16}\right)-\frac{2}{3} z_d z_{q^*} \bar{g}_5 \bar{g}_{10}-\frac{\bar{g}_{10}^4}{72}-\frac{8}{27} \bar{g}_5 \bar{g}_7 \bar{g}_{10} \bar{g}_{11}-\frac{25}{36} \bar{g}_7^2 \bar{g}_{11}^2+\left(-\frac{4 \tilde{\gamma} _4}{9}-\frac{\bar{g}_9^2}{12}\right) \bar{g}_{11}^2-\frac{17 \bar{g}_{11}^4}{12}\\
%%%%
&+z_{q^*}^2 \left(-\frac{\tilde{\gamma} _4}{3}-\frac{19 \bar{g}_2^2}{32}-\frac{57 \bar{g}_4^2}{32}-\frac{11 \bar{g}_{10}^2}{12}-3 \bar{g}_{11}^2\right)-\frac{9}{32} \bar{g}_3^2 \bar{g}_{13}^2-\frac{9}{32} \bar{g}_4^2 \bar{g}_{13}^2-\frac{3}{64} \bar{g}_6^2 \bar{g}_{13}^2\\
%%%
&+y_{\tau } \left(z_e \left(-\frac{1}{2} \bar{g}_2 \bar{g}_{12}+\frac{3}{2} \bar{g}_4 \bar{g}_{13}\right)+z_l \bar{g}_2 \bar{g}_{14}\right)+y_{\tau }^2 \left(-\frac{3 z_e^2}{2}-\frac{3 z_l^2}{4}-\frac{3 \bar{g}_{12}^2}{16}-\frac{9 \bar{g}_{13}^2}{16}-\frac{3 \bar{g}_{14}^2}{4}\right)\\
%%%%
&+\bar{g}_1^2 \left(-\frac{3 \bar{g}_{12}^2}{32}-\frac{3 \bar{g}_{14}^2}{16}\right)+\bar{g}_2^2 \left(-\frac{5 \bar{g}_{10}^2}{72}-\frac{3 \bar{g}_{12}^2}{32}-\frac{3 \bar{g}_{14}^2}{16}\right)+\bar{g}_{10}^2 \left(-\frac{\tilde{\gamma} _4}{27}-\frac{\bar{g}_1^2}{72}-\frac{\bar{g}_8^2}{54}-\frac{2 \bar{g}_{11}^2}{9}\right.\\
%%%%
&\left.\left.-\frac{\bar{g}_{12}^2}{144}-\frac{\bar{g}_{14}^2}{72}\right)+\bar{g}_5^2 \left(\frac{7 \bar{g}_{10}^2}{2592}-\frac{\bar{g}_{12}^2}{576}-\frac{\bar{g}_{14}^2}{288}\right)\right)+z_d \left(z_l^2 \left(\frac{1}{12} \bar{g}_2 \bar{g}_5+\frac{3}{4} \bar{g}_4 \bar{g}_6\right)+z_{q^*}^2 \left(\frac{1}{4} \bar{g}_2 \bar{g}_5\right.\right.\\
%%%%
&\left.\left.+\frac{9}{4} \bar{g}_4 \bar{g}_6\right)+\frac{3}{8} \bar{g}_4 \bar{g}_6 \bar{g}_{13}^2+\bar{g}_2 \bar{g}_5 \left(\frac{\bar{g}_{10}^2}{36}+\frac{\bar{g}_{12}^2}{24}+\frac{\bar{g}_{14}^2}{12}\right)\right)+z_{q^*} \left(\frac{5}{24} \bar{g}_2^3 \bar{g}_{10}+\frac{1}{4} \bar{g}_1 \bar{g}_3 \bar{g}_4 \bar{g}_{10}\right.\\
%%%%
&\left.\left.+\bar{g}_2 \left(\frac{3}{8} \bar{g}_4^2 \bar{g}_{10}+\frac{1}{36} \bar{g}_5^2 \bar{g}_{10}+\frac{\bar{g}_{10}^3}{18}+\bar{g}_{10} \left(\frac{\tilde{\gamma} _4}{18}+\frac{\bar{g}_1^2}{4}+\frac{\bar{g}_3^2}{4}+\frac{2 \bar{g}_8^2}{9}+\frac{\bar{g}_{12}^2}{12}+\frac{\bar{g}_{14}^2}{6}\right)\right)\right)\right\},
\end{align*}
%%%%%%%%%%%
%%%%%%%%%%%
%%%%%%%%%%%
\begin{align*}
 \beta_{y_\tau}=&\beta_{y_\tau}^{\rm MES}+\frac{1}{16\pi^2}\left\{z_e \left(\frac{1}{2} \bar{g}_2 \bar{g}_{12}-\frac{3}{2} \bar{g}_4 \bar{g}_{13}\right)-z_l \bar{g}_2 \bar{g}_{14}+y_{\tau } \left(z_e^2+\frac{z_l^2}{2}+\frac{\bar{g}_{12}^2}{8}+\frac{3 \bar{g}_{13}^2}{8}+\frac{\bar{g}_{14}^2}{2}\right)\right\}\\
%%%%
&+\frac{1}{(16\pi^2)^2}\left\{\frac{143}{100} g_1^4 y_{\tau }+\frac{1}{2} g_2^4 y_{\tau }+z_e^3 \left(-\frac{1}{4} \bar{g}_2 \bar{g}_{12}+\frac{3}{4} \bar{g}_4 \bar{g}_{13}\right)+\frac{3}{2} z_d^2 z_l \bar{g}_2 \bar{g}_{14}+\frac{1}{2} z_e^2 z_l \bar{g}_2 \bar{g}_{14}\right.\\
%%%%
&+\frac{1}{2} z_l^3 \bar{g}_2 \bar{g}_{14}+g_2^2 \left(z_e \left(\frac{3}{8} \bar{g}_2 \bar{g}_{12}-\frac{69}{8} \bar{g}_4 \bar{g}_{13}\right)+y_{\tau } \left(\frac{51 z_e^2}{8}-\frac{3 z_l^2}{2}+\frac{33 \bar{g}_{12}^2}{64}+\frac{75 \bar{g}_{13}^2}{64}\right)\right.\\
%%%%
&\left.+\frac{3}{2} z_l \bar{g}_2 \bar{g}_{14}\right)+y_{\tau }^2 \left(z_e \left(-\bar{g}_2 \bar{g}_{12}+3 \bar{g}_4 \bar{g}_{13}\right)+2 z_l \bar{g}_2 \bar{g}_{14}\right)+y_t^2 \left(-\frac{9}{4} y_{\tau } z_{q^*}^2+z_e \left(-\frac{3}{2} \bar{g}_2 \bar{g}_{12}\right.\right.\\
%%%%
&\left.\left.+\frac{9}{2} \bar{g}_4 \bar{g}_{13}\right)+3 z_l \bar{g}_2 \bar{g}_{14}\right)+y_b^2 \left(y_{\tau } \left(-\frac{9 z_{q^*}^2}{4}-\frac{\bar{g}_{10}^2}{4}-3 \bar{g}_{11}^2\right)+z_e \left(-\frac{3}{2} \bar{g}_2 \bar{g}_{12}+\frac{9}{2} \bar{g}_4 \bar{g}_{13}\right)\right.\\
%%%%
&\left.+3 z_l \bar{g}_2 \bar{g}_{14}\right)+y_b \left(y_{\tau } z_{q^*} \bar{g}_2 \bar{g}_{10}+z_e \left(z_{q^*} \bar{g}_{10} \bar{g}_{12}+z_d \left(\frac{1}{2} \bar{g}_5 \bar{g}_{12}-\frac{9}{2} \bar{g}_6 \bar{g}_{13}\right)\right)-z_d z_l \bar{g}_5 \bar{g}_{14}\right.\\
%%%%
&\left.-2 z_l z_{q^*} \bar{g}_{10} \bar{g}_{14}\right)+y_{\tau }^3 \left(-\frac{7 z_e^2}{4}-z_l^2-\frac{\bar{g}_{12}^2}{4}-\frac{3 \bar{g}_{13}^2}{4}-\frac{7 \bar{g}_{14}^2}{8}\right)+y_{\tau } \left(-\frac{7 z_e^4}{4}-\frac{3}{8} z_d^2 z_l^2-\frac{13 z_l^4}{8}\right.\\
%%%
&+\frac{\tilde{\gamma} _4^2}{24}+\frac{\tilde{\gamma} _5^2}{16}+\frac{\tilde{\gamma} _6^2}{8}+\frac{3 \tilde{\gamma} _{23}^2}{16}+z_{q^*}^2 \left(-\frac{9 \bar{g}_2^2}{16}-\frac{27 \bar{g}_4^2}{16}\right)-\frac{\bar{g}_{12}^4}{16}-\frac{3}{4} \bar{g}_1 \bar{g}_3 \bar{g}_{12} \bar{g}_{13}-\frac{45}{64} \bar{g}_3^2 \bar{g}_{13}^2\\
%%%%
&-\frac{21}{64} \bar{g}_4^2 \bar{g}_{13}^2+\left(\frac{3 \tilde{\gamma} _5}{8}+\frac{3 \tilde{\gamma} _{23}}{8}-\frac{9 \bar{g}_6^2}{64}\right) \bar{g}_{13}^2-\frac{15 \bar{g}_{13}^4}{32}+z_e^2 \left(-\frac{3 z_d^2}{4}+\frac{13 z_l^2}{8}-\frac{3 z_{q^*}^2}{4}+\tilde{\gamma} _5-\frac{\bar{g}_2^2}{4}\right.\\
%%%
&\left.-\frac{3 \bar{g}_4^2}{4}\!-\frac{17 \bar{g}_{12}^2}{16}\!-\frac{51 \bar{g}_{13}^2}{16}\right)\!+\!2 z_e z_l \bar{g}_{12} \bar{g}_{14}\!-\tilde{\gamma} _6 \bar{g}_{14}^2\!-\frac{1}{48} \bar{g}_5^2 \bar{g}_{14}^2\!-\frac{1}{6} \bar{g}_8^2 \bar{g}_{14}^2-\frac{7 \bar{g}_{14}^4}{8}+z_l^2 \left(-\frac{3 z_{q^*}^2}{8}\right.\\
%%%
&\left.-\tilde{\gamma} _6-\frac{7 \bar{g}_2^2}{32}-\frac{21 \bar{g}_4^2}{32}-\frac{17 \bar{g}_{14}^2}{4}\right)+\bar{g}_2^2 \left(-\frac{\bar{g}_{10}^2}{16}-\frac{7 \bar{g}_{12}^2}{64}-\frac{\bar{g}_{14}^2}{4}\right)+\bar{g}_{10}^2 \left(-\frac{\bar{g}_1^2}{16}-\frac{\bar{g}_{12}^2}{96}-\frac{\bar{g}_{14}^2}{24}\right)\\
%%%
&\left.+\bar{g}_1^2 \left(\frac{\bar{g}_{12}^2}{64}\!+\!\frac{\bar{g}_{14}^2}{4}\right)\!+\!\bar{g}_{12}^2 \left(\frac{\tilde{\gamma} _5}{8}\!-\frac{3 \tilde{\gamma} _{23}}{8}\!-\frac{\bar{g}_5^2}{192}-\frac{\bar{g}_8^2}{24}-\frac{9 \bar{g}_{13}^2}{32}+\frac{13 \bar{g}_{14}^2}{32}\right)\right)+g_1^2 \left(z_e \left(\frac{21}{40} \bar{g}_2 \bar{g}_{12}\right.\right.\\
%%%
&\left.\left.-\frac{63}{40} \bar{g}_4 \bar{g}_{13}\right)\!-\frac{12}{5} z_l \bar{g}_2 \bar{g}_{14}\!+\!y_{\tau } \left(\frac{39 z_e^2}{40}\!+\!\frac{3 z_l^2}{5}\!+\!\frac{33 \bar{g}_{12}^2}{320}\!+\!\frac{99 \bar{g}_{13}^2}{320}+\frac{33 \bar{g}_{14}^2}{20}\right)\right)+z_l \left(\frac{3}{2} z_{q^*}^2 \bar{g}_2 \bar{g}_{14}\right.\\
%%%%
&+\frac{5}{8} \bar{g}_2^3 \bar{g}_{14}+\frac{3}{4} \bar{g}_1 \bar{g}_3 \bar{g}_4 \bar{g}_{14}+\bar{g}_2 \left(\frac{1}{2} \tilde{\gamma} _6 \bar{g}_{14}+\frac{3}{4} \bar{g}_1^2 \bar{g}_{14}+\frac{3}{4} \bar{g}_3^2 \bar{g}_{14}+\frac{9}{8} \bar{g}_4^2 \bar{g}_{14}+\frac{1}{12} \bar{g}_5^2 \bar{g}_{14}+\frac{2}{3} \bar{g}_8^2 \bar{g}_{14}\right.\\
%%%
&\left.\left.+\frac{1}{6} \bar{g}_{10}^2 \bar{g}_{14}+\frac{1}{4} \bar{g}_{12}^2 \bar{g}_{14}+\frac{\bar{g}_{14}^3}{2}\right)\right)+z_e \left(-\frac{5}{16} \bar{g}_2^3 \bar{g}_{12}+\frac{9}{16} \bar{g}_2^2 \bar{g}_4 \bar{g}_{13}+\frac{33}{16} \bar{g}_4^3 \bar{g}_{13}+z_l^2 \left(-\frac{1}{4} \bar{g}_2 \bar{g}_{12}\right.\right.\\
%%%%
&\left.+\frac{3}{4} \bar{g}_4 \bar{g}_{13}\right)+z_d^2 \left(-\frac{3}{4} \bar{g}_2 \bar{g}_{12}+\frac{9}{4} \bar{g}_4 \bar{g}_{13}\right)+z_{q^*}^2 \left(-\frac{3}{4} \bar{g}_2 \bar{g}_{12}+\frac{9}{4} \bar{g}_4 \bar{g}_{13}\right)+\bar{g}_4 \left(-\frac{3}{8} \bar{g}_1 \bar{g}_3 \bar{g}_{12}\right.\\
%%%%
&\left.+\frac{3}{8} \bar{g}_1^2 \bar{g}_{13}+\frac{9}{8} \bar{g}_3^2 \bar{g}_{13}+\left(-\frac{3 \tilde{\gamma} _5}{8}-\frac{3 \tilde{\gamma} _{23}}{8}+\frac{9 \bar{g}_6^2}{8}\right) \bar{g}_{13}+\frac{3 \bar{g}_{13}^3}{8}\right)+\bar{g}_2 \left(-\frac{3}{8} \bar{g}_1^2 \bar{g}_{12}-\frac{9}{16} \bar{g}_4^2 \bar{g}_{12}\right.\\
%%%
&\left.\left.\left.-\frac{1}{12} \bar{g}_{10}^2 \bar{g}_{12}-\frac{\bar{g}_{12}^3}{8}+\frac{3}{8} \bar{g}_1 \bar{g}_3 \bar{g}_{13}+\bar{g}_{12} \left(\frac{\tilde{\gamma} _5}{8}-\frac{3 \tilde{\gamma} _{23}}{8}-\frac{3 \bar{g}_3^2}{8}-\frac{\bar{g}_5^2}{24}-\frac{\bar{g}_8^2}{3}-\frac{\bar{g}_{14}^2}{4}\right)\right)\right)\right\},
\end{align*}
%%%%%%%%%%%
%%%%%%%%%%%
%%%%%%%%%%%
\begin{align*}
\hskip-6.5cm \beta_{z_q}=&\beta_{z_q}^{\rm MES}+\frac{1}{16\pi^2}\frac{1}{2} z_q z_{q^*}^2,
\end{align*}
%%%%%%%%%%%
%%%%%%%%%%%
%%%%%%%%%%%
\begin{align*}
\hskip-7.7cm  \beta_{z_u}=&\beta_{z_u}^{\rm MES},
\end{align*}
%%%%%%%%%%%
%%%%%%%%%%%
%%%%%%%%%%%
\begin{align*}
 \beta_{z_d}=&\beta_{z_d}^{\rm MES}+\frac{1}{16\pi^2}\left\{\frac{1}{2} z_d z_e^2+\frac{1}{2} z_d z_l^2+\frac{3}{2} z_d z_{q^*}^2+\frac{1}{9} z_{q^*} \bar{g}_5 \bar{g}_{10}+\frac{1}{18} z_d \bar{g}_{10}^2-\frac{8}{3} z_{q^*} \bar{g}_7 \bar{g}_{11}+\frac{2}{3} z_d \bar{g}_{11}^2\right\},
\end{align*}
%%%%%%%%%%%
%%%%%%%%%%%
%%%%%%%%%%%
\begin{align*}
\hskip-.3cm \beta_{z_q^*}=&\frac{1}{16\pi^2}\left\{4 z_{q^*}^3-\frac{1}{3} y_b \bar{g}_2 \bar{g}_{10}+z_d \left(\frac{1}{9} \bar{g}_5 \bar{g}_{10}-\frac{8}{3} \bar{g}_7 \bar{g}_{11}\right)+z_{q^*} \left(-\frac{g_1^2}{2}-\frac{9 g_2^2}{2}-4 g_3^2+\frac{y_b^2}{2}+\frac{y_t^2}{2}\right.\right.\\
%%%%%
&\left.\left.+\frac{3 z_d^2}{2}+\frac{z_e^2}{2}+\frac{z_l^2}{2}+\frac{z_q^2}{2}+\frac{\bar{g}_2^2}{8}+\frac{3 \bar{g}_4^2}{8}+\frac{\bar{g}_5^2}{72}+\frac{3 \bar{g}_6^2}{8}+\frac{2 \bar{g}_7^2}{3}+\frac{\bar{g}_{10}^2}{9}+\frac{4 \bar{g}_{11}^2}{3}\right)\right\},
\end{align*}
%%%%%%%%%%%
%%%%%%%%%%%
%%%%%%%%%%%
\begin{align*}
 \beta_{z_l}=&\frac{1}{16\pi^2}\left\{3 z_l^3-y_{\tau } \bar{g}_2 \bar{g}_{14}-z_e \bar{g}_{12} \bar{g}_{14}+z_l \left(\frac{3 \bar{g}_4^2}{8}-\frac{9 g_1^2}{10}-\frac{9 g_2^2}{2}+\frac{y_{\tau }^2}{2}+\frac{3 z_d^2}{2}+\frac{z_e^2}{2}+\frac{3 z_{q^*}^2}{2}+\frac{\bar{g}_2^2}{8}\right.\right.\\
%%%%
&\left.\left.+\frac{\bar{g}_{12}^2}{8}+\frac{3 \bar{g}_{13}^2}{8}+\bar{g}_{14}^2\right)\right\},
\end{align*}
%%%%%%%%%%%
%%%%%%%%%%%
%%%%%%%%%%%
\begin{align*}
 \beta_{z_e}=&\frac{1}{16\pi^2}\left\{\frac{5 z_e^3}{2}+y_{\tau } \left(\frac{1}{2} \bar{g}_2 \bar{g}_{12}-\frac{3}{2} \bar{g}_4 \bar{g}_{13}\right)-z_l \bar{g}_{12} \bar{g}_{14}+z_e \left(-\frac{9 g_1^2}{4}-\frac{9 g_2^2}{4}+y_{\tau }^2+\frac{3 z_d^2}{2}+\frac{z_l^2}{2}\right.\right.\\
%%%%
&\left.\left.+\frac{3 z_{q^*}^2}{2}+\frac{\bar{g}_2^2}{8}+\frac{3 \bar{g}_4^2}{8}+\frac{\bar{g}_{12}^2}{4}+\frac{3 \bar{g}_{13}^2}{4}+\frac{\bar{g}_{14}^2}{2}\right)\right\},
\end{align*}
\begin{align*}
\hskip-4.4cm \beta_{\bar{g}_1}=&\beta_{\bar{g}_1}^{\rm MES}+\frac{1}{16\pi^2}\left\{\bar{g}_1 \left(\frac{\bar{g}_{10}^2}{6}+\frac{\bar{g}_{12}^2}{4}+\frac{\bar{g}_{14}^2}{2}\right)\right\},
\end{align*}
%%%%%%%%%%%
%%%%%%%%%%%
%%%%%%%%%%%
%%%%%%%%%%%
\begin{align*}
 \hskip-.6cm\beta_{\bar{g}_2}=&\beta_{\bar{g}_1}|_{g_{H_i}\leftrightarrow g_{H_i^*}}+\frac{1}{16\pi^2}\left\{\left(\frac{3 z_d^2}{2}-\frac{3 z_q^2}{2}-\frac{3 z_u^2}{2}\right) \bar{g}_2+\left(-2 y_b z_d-2 y_t z_u\right) \bar{g}_5+8 y_t z_q \bar{g}_8\right\},
\end{align*}
%%%%%%%%%%%
%%%%%%%%%%%
%%%%%%%%%%%
\begin{align*}
  \hskip-6.1cm\beta_{\bar{g}_3}=&\beta_{\bar{g}_3}^{\rm MES}+\frac{1}{16\pi^2}\frac{1}{4} \bar{g}_3 \bar{g}_{13}^2,
\end{align*}
%%%%%%%%%%%
%%%%%%%%%%%
%%%%%%%%%%%
\begin{align*}
 \beta_{\bar{g}_4}=&\beta_{\bar{g}_3}|_{g_{H_i}\leftrightarrow g_{H_i^*}}+\frac{1}{16\pi^2}\left\{\left(\frac{3 z_d^2}{2}+\frac{z_e^2}{2}+\frac{z_l^2}{2}-\frac{3 z_q^2}{2}-\frac{3 z_u^2}{2}+\frac{3 z_{q^*}^2}{2}\right) \bar{g}_4+\left(-6 y_b z_d+6 y_t z_u\right) \bar{g}_6\right.\\
%%%%
&\left.-2 y_{\tau } z_e \bar{g}_{13}\right\},
\end{align*}
%%%%%%%%%%%
%%%%%%%%%%%
%%%%%%%%%%%
%%%%%%%%%%%
%%%%%%%%%%%
%%%%%%%%%%%
\begin{align*}
 \hskip-3cm\beta_{{\bar g_{5}}}=&\beta_{\bar{g}_5}^{\rm MES}+\frac{1}{16\pi^2}\left\{4 z_d z_{q^*} \bar{g}_{10}+\bar{g}_5 \left(\frac{z_{q^*}^2}{2}+\frac{\bar{g}_{10}^2}{6}+\frac{\bar{g}_{12}^2}{4}+\frac{\bar{g}_{14}^2}{2}\right)\right\},
\end{align*}
%%%%%%%%%%%
%%%%%%%%%%%
%%%%%%%%%%%
\begin{align*}
  \hskip-5.4cm\beta_{{\bar g_{6}}}=&\beta_{\bar{g}_6}^{\rm MES}+\frac{1}{16\pi^2}\bar{g}_6 \left(\frac{z_{q^*}^2}{2}+\frac{\bar{g}_{13}^2}{4}\right),
\end{align*}
%%%%%%%%%%%
%%%%%%%%%%%
%%%%%%%%%%%
\begin{align*}
  \hskip-3.8cm\beta_{{\bar g_{7}}}=&\beta_{\bar{g}_7}^{\rm MES}+\frac{1}{16\pi^2}\left\{-2 z_d z_{q^*} \bar{g}_{11}+\bar{g}_7 \left(\frac{z_{q^*}^2}{2}+\frac{\bar{g}_{11}^2}{4}\right)\right\},
\end{align*}
%%%%%%%%%%%
%%%%%%%%%%%
%%%%%%%%%%%
\begin{align*}
  \hskip-4.8cm\beta_{{\bar g_{8}}}=&\beta_{\bar{g}_8}^{\rm MES}+\frac{1}{16\pi^2}\bar{g}_8 \left(\frac{\bar{g}_{10}^2}{6}+\frac{\bar{g}_{12}^2}{4}+\frac{\bar{g}_{14}^2}{2}\right),
\end{align*}
%%%%%%%%%%%
%%%%%%%%%%%
%%%%%%%%%%%
\begin{align*}
 \hskip-6.2cm \beta_{{\bar g_{9}}}=&\beta_{\bar{g}_9}^{\rm MES}+\frac{1}{16\pi^2}\frac{1}{4} \bar{g}_9 \bar{g}_{11}^2,
\end{align*}
%%%%%%%%%%%
%%%%%%%%%%%
%%%%%%%%%%%
\begin{align*}
 \beta_{{\bar g_{10}}}=&\frac{1}{16\pi^2}\left\{-6 y_b z_{q^*} \bar{g}_2+2 z_d z_{q^*} \bar{g}_5+\frac{1}{4} \bar{g}_2^2 \bar{g}_{10}+\frac{1}{12} \bar{g}_5^2 \bar{g}_{10}+\frac{\bar{g}_{10}^3}{3}+\bar{g}_{10} \left(-\frac{g_1^2}{3}-4 g_3^2+y_b^2+z_d^2\right.\right.\\
%%%%
&\left.\left.+2 z_{q^*}^2+\frac{\bar{g}_1^2}{4}+\frac{2 \bar{g}_8^2}{3}+2 \bar{g}_{11}^2+\frac{\bar{g}_{12}^2}{4}+\frac{\bar{g}_{14}^2}{2}\right)\right\},
%%%%
\end{align*}
%%%%%%%%%%%
%%%%%%%%%%%
%%%%%%%%%%%
\begin{align*}
 \beta_{{\bar g_{11}}}=&\frac{1}{16\pi^2}\left\{-4 z_d z_{q^*} \bar{g}_7+\frac{1}{2} \bar{g}_7^2 \bar{g}_{11}+\left(y_b^2-\frac{g_1^2}{3}-13 g_3^2+z_d^2+2 z_{q^*}^2+\frac{\bar{g}_9^2}{4}+\frac{\bar{g}_{10}^2}{6}\right) \bar{g}_{11}+\frac{9 \bar{g}_{11}^3}{4}\right\},
\end{align*}
\begin{align*}
%%%
 \beta_{{\bar g_{12}}}=&\frac{1}{16\pi^2}\left\{2 y_{\tau } z_e \bar{g}_2+\frac{1}{4} \bar{g}_2^2 \bar{g}_{12}+\frac{5 \bar{g}_{12}^3}{8}-4 z_e z_l \bar{g}_{14}+\bar{g}_{12} \left(-\frac{3 g_1^2}{4}-\frac{9 g_2^2}{4}+\frac{y_{\tau }^2}{2}+z_e^2+\frac{z_l^2}{2}+\frac{\bar{g}_1^2}{4}\right.\right.\\
%%%%
&\left.\left.+\frac{\bar{g}_5^2}{12}+\frac{2 \bar{g}_8^2}{3}+\frac{\bar{g}_{10}^2}{6}+\frac{9 \bar{g}_{13}^2}{8}+\frac{\bar{g}_{14}^2}{2}\right)\right\},
\end{align*}
%%%%%%%%%%%
%%%%%%%%%%%
%%%%%%%%%%%
\begin{align*}
 \beta_{{\bar g_{13}}}=&\frac{1}{16\pi^2}\left\{-2 y_{\tau } z_e \bar{g}_4+\frac{1}{4} \bar{g}_4^2 \bar{g}_{13}+\left(-\frac{3 g_1^2}{4}-\frac{33 g_2^2}{4}+\frac{y_{\tau }^2}{2}+z_e^2+\frac{z_l^2}{2}+\frac{\bar{g}_3^2}{4}+\frac{3 \bar{g}_6^2}{4}+\frac{3 \bar{g}_{12}^2}{8}\right) \bar{g}_{13}\right.\\
%%%%
&\left.+\frac{11 \bar{g}_{13}^3}{8}\right\},
\end{align*}
%%%%%%%%%%%
%%%%%%%%%%%
%%%%%%%%%%%
\begin{align*}
 \hskip-.5cm \beta_{{\bar g_{14}}}=&\frac{1}{16\pi^2}\left\{-2 y_{\tau } z_l \bar{g}_2-2 z_e z_l \bar{g}_{12}+\frac{1}{4} \bar{g}_2^2 \bar{g}_{14}+\left(-3 g_1^2+y_{\tau }^2+z_e^2+2 z_l^2+\frac{\bar{g}_1^2}{4}+\frac{\bar{g}_5^2}{12}+\frac{2 \bar{g}_8^2}{3}\right.\right.\\
%%%
&\left.\left.+\frac{\bar{g}_{10}^2}{6}\right) \bar{g}_{14}+\frac{1}{4} \bar{g}_{12}^2 \bar{g}_{14}+2 \bar{g}_{14}^3\right\}.
\end{align*}
%%%%%%%%%%%%%%%%%%%%%%%%%%%%%%%%%%%%%%%%%%%%%%%%%%%%%%%%%%%%%%%%%%%%%%%%%%%%%%%%%
%%%%%%%%%%%%%%%%%%%%%%%%%%%%%%%%%%%%%%%%%%%%%%%%%%%%%%%%%%%%%%%%%%%%%%%%%%%%%%%%%
%%%%%%%%%%%%%%%%%%%%%%%%%%%%%%%%%%%%%%%%%%%%%%%%%%%%%%%%%%%%%%%%%%%%%%%%%%%%%%%%%
\subsubsection{Quartic couplings}
%%%%%%%%%%%%%%%%%%%%%%%%%%%%%%%%%%%%%%%%%%%%%%%%%%%%%%%%%%%%%%%%%%%%%%%%%%%%%%%%%
%%%%%%%%%%%%%%%%%%%%%%%%%%%%%%%%%%%%%%%%%%%%%%%%%%%%%%%%%%%%%%%%%%%%%%%%%%%%%%%%%
%%%%%%%%%%%%%%%%%%%%%%%%%%%%%%%%%%%%%%%%%%%%%%%%%%%%%%%%%%%%%%%%%%%%%%%%%%%%%%%%%
\begin{align*}
 \beta_{\tilde{\gamma}_1}=&\beta_{\tilde{\gamma}_{1}}^{\rm MES}+\frac{1}{16\pi^2}\left\{\frac{2 \tilde{\gamma} _4^2}{3}+\tilde{\gamma} _5^2+2 \tilde{\gamma} _6^2+\tilde{\gamma} _{23}^2\right\},
\end{align*}
%%%%%%%%%%%%%
%%%%%%%%%%%%%
%%%%%%%%%%%%%
%%%%%%%%%%%%%
%%%%%%%%%%%%%
\begin{align*}
 \beta_{\tilde{\gamma}_2}=&\beta_{\tilde{\gamma}_{2}}^{\rm MES}+\frac{1}{16\pi^2}\left\{\frac{2 \tilde{\gamma} _4 \tilde{\gamma} _9}{3}+\tilde{\gamma} _5 \tilde{\gamma} _{10}+2 \tilde{\gamma} _6 \tilde{\gamma} _{11}\right\},
\end{align*}
%%%%%%%%%%%%%
%%%%%%%%%%%%%
%%%%%%%%%%%%%
%%%%%%%%%%%%%
\begin{align*}
 \beta_{\tilde{\gamma}_3}=&\beta_{\tilde{\gamma}_{3}}^{\rm MES}+\frac{1}{16\pi^2}\left\{\frac{2 \tilde{\gamma} _4 \tilde{\gamma} _{13}}{3}+\tilde{\gamma} _5 \tilde{\gamma} _{14}+2 \tilde{\gamma} _6 \tilde{\gamma} _{15}\right\},
\end{align*}
%%%%%%%%%%%%%
%%%%%%%%%%%%%
%%%%%%%%%%%%%
%%%%%%%%%%%%%
\begin{align*}
 \beta_{\tilde{\gamma}_7}=&\beta_{\tilde{\gamma}_{7}}^{\rm MES}+\frac{1}{16\pi^2}\left\{\frac{2 \tilde{\gamma} _9^2}{3}+\tilde{\gamma} _{10}^2+2 \tilde{\gamma} _{11}^2+3 \tilde{\gamma} _{27}^2\right\},
\end{align*}
%%%%%%%%%%%%%
%%%%%%%%%%%%%
%%%%%%%%%%%%%
%%%%%%%%%%%%% 
\begin{align*}
 \beta_{\tilde{\gamma}_8}&=\beta_{\tilde{\gamma}_{8}}^{\rm MES}+\frac{1}{16\pi^2}\left\{\frac{2 \tilde{\gamma} _9 \tilde{\gamma} _{13}}{3}+\tilde{\gamma} _{10} \tilde{\gamma} _{14}+2 \tilde{\gamma} _{11} \tilde{\gamma} _{15}\right\},
\end{align*}
%%%%%%%%%%%%%
%%%%%%%%%%%%%
%%%%%%%%%%%%%
%%%%%%%%%%%%%
\begin{align*}
 \beta_{\tilde{\gamma}_{12}}=&\beta_{\tilde{\gamma}_{12}}^{\rm MES}+\frac{1}{16\pi^2}\left\{\frac{2 \tilde{\gamma} _{13}^2}{3}+\tilde{\gamma} _{14}^2+2 \tilde{\gamma} _{15}^2+\frac{3 \tilde{\gamma} _{28}^2}{4}\right\},
\end{align*}
%%%%%%%%%%%%%
%%%%%%%%%%%%%
%%%%%%%%%%%%%
%%%%%%%%%%%%%
 \begin{align*}
 \beta_{\tilde{\gamma}_{22}}=&\beta_{\tilde{\gamma}_{22}}^{\rm MES}+\frac{1}{16\pi^2}\tilde{\gamma} _{23} \tilde{\gamma} _{25},
\end{align*}
%%%%%%%%%%%%%
%%%%%%%%%%%%%
%%%%%%%%%%%%%
%%%%%%%%%%%%%
\begin{align*}
 \beta_{\tilde{\gamma}_{24}}=&\beta_{\tilde{\gamma}_{24}}^{\rm MES}+\frac{1}{16\pi^2}\left\{\tilde{\gamma} _{25}^2+\tilde{\gamma} _{27}^2\right\},
\end{align*}
%%%%%%%%%%%%%
%%%%%%%%%%%%%
%%%%%%%%%%%%%
%%%%%%%%%%%%%
\begin{align*}
 \beta_{\tilde{\gamma}_{26}}=&\beta_{\tilde{\gamma}_{26}}^{\rm MES}+\frac{1}{16\pi^2}\tilde{\gamma}_{27}\tilde{\gamma}_{28},
\end{align*}
%%%%%%%%%%%%%
%%%%%%%%%%%%%
%%%%%%%%%%%%%
%%%%%%%%%%%%%
\begin{align*}
 \beta_{\tilde{\gamma}_4}=&\frac{1}{16\pi^2}\left\{2 g_1^4-\frac{13}{6} g_1^2 \tilde{\gamma} _4-\frac{9}{2} g_2^2 \tilde{\gamma} _4-8 g_3^2 \tilde{\gamma} _4+6 y_b^2 \tilde{\gamma} _4+6 y_t^2 \tilde{\gamma} _4+2 y_{\tau }^2 \tilde{\gamma} _4+\frac{2 \tilde{\gamma} _4^2}{3}+z_{q^*}^2 \left(-24 y_b^2\right.\right.\\
%%%%
&\left.-24 y_t^2\!+\!4 \tilde{\gamma} _4\right)\!+\!\frac{\tilde{\gamma} _2 \tilde{\gamma} _9}{3}+\frac{8 \tilde{\gamma} _3 \tilde{\gamma} _{13}}{3}+\tilde{\gamma} _4 \left(\frac{3 \tilde{\gamma} _1}{2}+\frac{8 \tilde{\gamma} _{16}}{9}\right)+\tilde{\gamma} _5 \tilde{\gamma} _{17}+2 \tilde{\gamma} _6 \tilde{\gamma} _{18}+\frac{3}{2} \tilde{\gamma} _4 \bar{g}_3^2+\left(-18 z_{q^*}^2\right.\\
%%%%
&\left.+\frac{3 \tilde{\gamma} _4}{2}\right) \bar{g}_4^2+8 y_b z_{q^*} \bar{g}_2 \bar{g}_{10}+\left(-\frac{8 y_b^2}{3}+\frac{2 \tilde{\gamma} _4}{9}\right) \bar{g}_{10}^2+\bar{g}_1^2 \left(\frac{\tilde{\gamma} _4}{2}-\frac{2 \bar{g}_{10}^2}{3}\right)+\bar{g}_2^2 \left(-6 z_{q^*}^2+\frac{\tilde{\gamma} _4}{2}\right.\\
%%%%
&\left.\left.-\frac{2 \bar{g}_{10}^2}{3}\right)+\left(-32 y_b^2+\frac{8 \tilde{\gamma} _4}{3}\right) \bar{g}_{11}^2\right\},
\end{align*}
%%%%%%%%%%%%%
%%%%%%%%%%%%%
%%%%%%%%%%%%%
%%%%%%%%%%%%%
\begin{align*}
 \beta_{\tilde{\gamma}_5}=&\frac{1}{16\pi^2}\left\{-3 g_1^4-9 g_2^4-3 g_1^2 \tilde{\gamma} _5-9 g_2^2 \tilde{\gamma} _5+6 y_b^2 \tilde{\gamma} _5+6 y_t^2 \tilde{\gamma} _5+2 y_{\tau }^2 \tilde{\gamma} _5-\tilde{\gamma} _5^2+z_e^2 \left(16 y_{\tau }^2+2 \tilde{\gamma} _5\right)\right.\\
%%%%
&+\frac{\tilde{\gamma} _2 \tilde{\gamma} _{10}}{3}+\frac{8 \tilde{\gamma} _3 \tilde{\gamma} _{14}}{3}+\frac{2 \tilde{\gamma} _4 \tilde{\gamma} _{17}}{3}+\tilde{\gamma} _5 \left(\frac{3 \tilde{\gamma} _1}{2}+\frac{3 \tilde{\gamma} _{19}}{2}\right)+2 \tilde{\gamma} _6 \tilde{\gamma} _{20}-3 \tilde{\gamma} _{23}^2+4 y_{\tau } z_e \bar{g}_2 \bar{g}_{12}+\left(2 y_{\tau }^2\right.\\
%%%%
&\left.+\frac{\tilde{\gamma} _5}{2}\right) \bar{g}_{12}^2+\bar{g}_1^2 \left(\frac{\tilde{\gamma} _5}{2}+\bar{g}_{12}^2\right)+\bar{g}_2^2 \left(2 z_e^2+\frac{\tilde{\gamma} _5}{2}+\bar{g}_{12}^2\right)-12 y_{\tau } z_e \bar{g}_4 \bar{g}_{13}+\left(6 y_{\tau }^2+\frac{3 \tilde{\gamma} _5}{2}\right) \bar{g}_{13}^2\\
%%%%
&\left.+\bar{g}_3^2 \left(\frac{3 \tilde{\gamma} _5}{2}+3 \bar{g}_{13}^2\right)+\bar{g}_4^2 \left(6 z_e^2+\frac{3 \tilde{\gamma} _5}{2}+3 \bar{g}_{13}^2\right)\right\},
\end{align*}
%%%%%%%%%%%%%
%%%%%%%%%%%%%
%%%%%%%%%%%%%
%%%%%%%%%%%%%
\begin{align*}
 \beta_{\tilde{\gamma}_6}=&\frac{1}{16\pi^2}\left\{6 g_1^4-\frac{15}{2} g_1^2 \tilde{\gamma} _6-\frac{9}{2} g_2^2 \tilde{\gamma} _6+6 y_b^2 \tilde{\gamma} _6+6 y_t^2 \tilde{\gamma} _6+2 y_{\tau }^2 \tilde{\gamma} _6+2 \tilde{\gamma} _6^2+z_l^2 \left(4 \tilde{\gamma} _6-8 y_{\tau }^2\right)+\frac{\tilde{\gamma} _2 \tilde{\gamma} _{11}}{3}\right.\\
%%%%
&+\frac{8 \tilde{\gamma} _3 \tilde{\gamma} _{15}}{3}\!+\!\frac{2 \tilde{\gamma} _4 \tilde{\gamma} _{18}}{3}\!+\tilde{\gamma} _5 \tilde{\gamma} _{20}+\tilde{\gamma} _6 \left(\frac{3 \tilde{\gamma} _1}{2}+4 \tilde{\gamma} _{21}\right)+\frac{3}{2} \tilde{\gamma} _6 \bar{g}_3^2+\left(\frac{3 \tilde{\gamma} _6}{2}-6 z_l^2\right) \bar{g}_4^2+8 y_{\tau } z_l \bar{g}_2 \bar{g}_{14}\\
%%%%
&\left.+\left(-8 y_{\tau }^2+2 \tilde{\gamma} _6\right) \bar{g}_{14}^2+\bar{g}_1^2 \left(\frac{\tilde{\gamma} _6}{2}-2 \bar{g}_{14}^2\right)+\bar{g}_2^2 \left(-2 z_l^2+\frac{\tilde{\gamma} _6}{2}-2 \bar{g}_{14}^2\right)\right\},
\end{align*}
%%%%%%%%%%%%%
%%%%%%%%%%%%%
%%%%%%%%%%%%%
%%%%%%%%%%%%%
\begin{align*}
 \beta_{\tilde{\gamma}_9}=&\frac{1}{16\pi^2}\left\{\frac{2 g_1^4}{3}+48 g_3^4+\tilde{\gamma} _2 \tilde{\gamma} _4-\frac{5}{6} g_1^2 \tilde{\gamma} _9-\frac{9}{2} g_2^2 \tilde{\gamma} _9-16 g_3^2 \tilde{\gamma} _9+2 z_u^2 \tilde{\gamma} _9+4 z_{q^*}^2 \tilde{\gamma} _9+\frac{2 \tilde{\gamma} _9^2}{9}\right.\\
%%%%
&+z_d^2 \left(-72 z_{q^*}^2+2 \tilde{\gamma} _9\right)+\frac{8 \tilde{\gamma} _8 \tilde{\gamma} _{13}}{3}+\tilde{\gamma} _{10} \tilde{\gamma} _{17}+2 \tilde{\gamma} _{11} \tilde{\gamma} _{18}+\tilde{\gamma} _9 \left(\frac{7 \tilde{\gamma} _7}{18}+\frac{8 \tilde{\gamma} _{16}}{9}+\frac{3 \tilde{\gamma} _{24}}{2}\right)+16 \tilde{\gamma} _{27}^2\\
%%%%
&+\left(-18 z_{q^*}^2+\frac{3 \tilde{\gamma} _9}{2}\right) \bar{g}_6^2-\frac{8}{3} z_d z_{q^*} \bar{g}_5 \bar{g}_{10}+\left(-\frac{8 z_d^2}{3}+\frac{2 \tilde{\gamma} _9}{9}\right) \bar{g}_{10}^2+\bar{g}_5^2 \left(-\frac{2 z_{q^*}^2}{3}+\frac{\tilde{\gamma} _9}{18}-\frac{2 \bar{g}_{10}^2}{9}\right)\\
%%%%
&\left.+64 z_d z_{q^*} \bar{g}_7 \bar{g}_{11}+\left(-32 z_d^2+\frac{8 \tilde{\gamma} _9}{3}\right) \bar{g}_{11}^2+\bar{g}_7^2 \left(-32 z_{q^*}^2+\frac{8 \tilde{\gamma} _9}{3}-16 \bar{g}_{11}^2\right)\right\},
\end{align*}
%%%%%%%%%%%%%
%%%%%%%%%%%%%
%%%%%%%%%%%%%
%%%%%%%%%%%%%
\begin{align*}
 \beta_{\tilde{\gamma}_{10}}=&\frac{1}{16\pi^2}\left\{-g_1^4-27 g_2^4+\tilde{\gamma} _2 \tilde{\gamma} _5-\frac{5}{3} g_1^2 \tilde{\gamma} _{10}-9 g_2^2 \tilde{\gamma} _{10}-8 g_3^2 \tilde{\gamma} _{10}+2 z_e^2 \tilde{\gamma} _{10}+2 z_u^2 \tilde{\gamma} _{10}-\frac{\tilde{\gamma} _{10}^2}{3}\right.\\
%%%%
&+z_d^2 \left(24 z_e^2+2 \tilde{\gamma} _{10}\right)+\frac{8 \tilde{\gamma} _8 \tilde{\gamma} _{14}}{3}+\frac{2 \tilde{\gamma} _9 \tilde{\gamma} _{17}}{3}+2 \tilde{\gamma} _{11} \tilde{\gamma} _{20}+\tilde{\gamma} _{10} \left(\frac{7 \tilde{\gamma} _7}{18}+\frac{3 \tilde{\gamma} _{19}}{2}+\frac{3 \tilde{\gamma} _{24}}{2}\right)-9 \tilde{\gamma} _{25}^2\\
%%%%
&\left.+\frac{8}{3} \tilde{\gamma} _{10} \bar{g}_7^2+\frac{1}{2} \tilde{\gamma} _{10} \bar{g}_{12}^2+\bar{g}_5^2 \left(\frac{\tilde{\gamma} _{10}}{18}+\frac{\bar{g}_{12}^2}{3}\right)+\frac{3}{2} \tilde{\gamma} _{10} \bar{g}_{13}^2+\bar{g}_6^2 \left(\frac{3 \tilde{\gamma} _{10}}{2}+9 \bar{g}_{13}^2\right)\right\},
\end{align*}
%%%%%%%%%%%%%
%%%%%%%%%%%%%
%%%%%%%%%%%%%
%%%%%%%%%%%%%
\begin{align*}
 \beta_{\tilde{\gamma}_{11}}&=\frac{1}{16\pi^2}\left\{2 g_1^4\!+\!\tilde{\gamma} _2 \tilde{\gamma} _6-\frac{37}{6} g_1^2 \tilde{\gamma} _{11}-\frac{9}{2} g_2^2 \tilde{\gamma} _{11}-8 g_3^2 \tilde{\gamma} _{11}+4 z_l^2 \tilde{\gamma} _{11}+2 z_u^2 \tilde{\gamma} _{11}+\frac{2 \tilde{\gamma} _{11}^2}{3}+z_d^2 \left(-24 z_l^2\right.\right.\\
%%%%
&\left.+2 \tilde{\gamma} _{11}\right)+\frac{8 \tilde{\gamma} _8 \tilde{\gamma} _{15}}{3}+\frac{2 \tilde{\gamma} _9 \tilde{\gamma} _{18}}{3}+\tilde{\gamma} _{10} \tilde{\gamma} _{20}+\tilde{\gamma} _{11} \left(\frac{7 \tilde{\gamma} _7}{18}+4 \tilde{\gamma} _{21}+\frac{3 \tilde{\gamma} _{24}}{2}\right)+\frac{3}{2} \tilde{\gamma} _{11} \bar{g}_6^2+\frac{8}{3} \tilde{\gamma} _{11} \bar{g}_7^2\\
%%%%
&\left.+2 \tilde{\gamma} _{11} \bar{g}_{14}^2+\bar{g}_5^2 \left(\frac{\tilde{\gamma} _{11}}{18}-\frac{2 \bar{g}_{14}^2}{3}\right)\right\},
\end{align*}
%%%%%%%%%%%%%
%%%%%%%%%%%%%
%%%%%%%%%%%%%
%%%%%%%%%%%%%
\begin{align*}
 \beta_{\tilde{\gamma}_{13}}=&\frac{1}{16\pi^2}\left\{-\frac{8 g_1^4}{3}-12 g_3^4+\tilde{\gamma} _3 \tilde{\gamma} _4+\frac{\tilde{\gamma} _8 \tilde{\gamma} _9}{3}-\frac{10}{3} g_1^2 \tilde{\gamma} _{13}-16 g_3^2 \tilde{\gamma} _{13}+4 z_{q^*}^2 \tilde{\gamma} _{13}-\frac{8 \tilde{\gamma} _{13}^2}{9}+z_q^2 \left(12 z_{q^*}^2\right.\right.\\
%%%%
&\left.+4 \tilde{\gamma} _{13}\right)\!+\!\tilde{\gamma} _{13} \left(\frac{32 \tilde{\gamma} _{12}}{9}\!+\!\frac{8 \tilde{\gamma} _{16}}{9}\right)+\tilde{\gamma} _{14} \tilde{\gamma} _{17}+2 \tilde{\gamma} _{15} \tilde{\gamma} _{18}-4 \tilde{\gamma} _{28}^2+\frac{2}{9} \tilde{\gamma} _{13} \bar{g}_{10}^2+\bar{g}_8^2 \left(\frac{8 \tilde{\gamma} _{13}}{9}+\frac{8 \bar{g}_{10}^2}{9}\right)\\
%%%%
&\left.+\frac{8}{3} \tilde{\gamma} _{13} \bar{g}_{11}^2+\bar{g}_9^2 \left(\frac{8 \tilde{\gamma} _{13}}{3}+4 \bar{g}_{11}^2\right)\right\},
\end{align*}
%%%%%%%%%%%%%
%%%%%%%%%%%%%
%%%%%%%%%%%%%
%%%%%%%%%%%%%
\begin{align*}
 \beta_{\tilde{\gamma}_{14}}=&\frac{1}{16\pi^2}\left\{4 g_1^4+\tilde{\gamma} _3 \tilde{\gamma} _5+\frac{\tilde{\gamma} _8 \tilde{\gamma} _{10}}{3}-\frac{25}{6} g_1^2 \tilde{\gamma} _{14}-\frac{9}{2} g_2^2 \tilde{\gamma} _{14}-8 g_3^2 \tilde{\gamma} _{14}+2 z_e^2 \tilde{\gamma} _{14}+4 z_q^2 \tilde{\gamma} _{14}+\frac{4 \tilde{\gamma} _{14}^2}{3}\right.\\
%%%
&+\frac{2 \tilde{\gamma} _{13} \tilde{\gamma} _{17}}{3}+\tilde{\gamma} _{14} \left(\frac{32 \tilde{\gamma} _{12}}{9}+\frac{3 \tilde{\gamma} _{19}}{2}\right)+2 \tilde{\gamma} _{15} \tilde{\gamma} _{20}+\frac{8}{3} \tilde{\gamma} _{14} \bar{g}_9^2+\frac{1}{2} \tilde{\gamma} _{14} \bar{g}_{12}^2+\bar{g}_8^2 \left(\frac{8 \tilde{\gamma} _{14}}{9}-\frac{4 \bar{g}_{12}^2}{3}\right)\\
%%%%
&\left.+\frac{3}{2} \tilde{\gamma} _{14} \bar{g}_{13}^2\right\},
\end{align*}
%%%%%%%%%%%%%
%%%%%%%%%%%%%
%%%%%%%%%%%%%
%%%%%%%%%%%%%
\begin{align*}
 \beta_{\tilde{\gamma}_{15}}=&\frac{1}{16\pi^2}\left\{-8 g_1^4+\tilde{\gamma} _3 \tilde{\gamma} _6+\frac{\tilde{\gamma} _8 \tilde{\gamma} _{11}}{3}-\frac{26}{3} g_1^2 \tilde{\gamma} _{15}-8 g_3^2 \tilde{\gamma} _{15}+4 z_l^2 \tilde{\gamma} _{15}+4 z_q^2 \tilde{\gamma} _{15}-\frac{8 \tilde{\gamma} _{15}^2}{3}+\frac{2 \tilde{\gamma} _{13} \tilde{\gamma} _{18}}{3}\right.\\
%%%%
&\left.+\tilde{\gamma} _{14} \tilde{\gamma} _{20}+\tilde{\gamma} _{15} \left(\frac{32 \tilde{\gamma} _{12}}{9}+4 \tilde{\gamma} _{21}\right)+\frac{8}{3} \tilde{\gamma} _{15} \bar{g}_9^2+2 \tilde{\gamma} _{15} \bar{g}_{14}^2+\bar{g}_8^2 \left(\frac{8 \tilde{\gamma} _{15}}{9}+\frac{8 \bar{g}_{14}^2}{3}\right)\right\},
\end{align*}
%%%%%%%%%%%%%
%%%%%%%%%%%%%
%%%%%%%%%%%%%
%%%%%%%%%%%%%
\begin{align*}
 \beta_{\tilde{\gamma}_{16}}=&\frac{1}{16\pi^2}\left\{\frac{4 g_1^4}{3}+39 g_3^4-72 z_{q^*}^4+\tilde{\gamma} _4^2+\frac{\tilde{\gamma} _9^2}{3}+\frac{8 \tilde{\gamma} _{13}^2}{3}+g_1^2 \left(8 g_3^2-\frac{4 \tilde{\gamma} _{16}}{3}\right)-16 g_3^2 \tilde{\gamma} _{16}+8 z_{q^*}^2 \tilde{\gamma} _{16}\right.\\
%%%%
&\left.+\frac{14 \tilde{\gamma} _{16}^2}{9}\!+\!\tilde{\gamma} _{17}^2+2 \tilde{\gamma} _{18}^2+6 \tilde{\gamma} _{27}^2+3 \tilde{\gamma} _{28}^2-\frac{4 \bar{g}_{10}^4}{9}+\frac{16}{3} \tilde{\gamma} _{16} \bar{g}_{11}^2-22 \bar{g}_{11}^4+\bar{g}_{10}^2 \left(\frac{4 \tilde{\gamma} _{16}}{9}-\frac{8 \bar{g}_{11}^2}{3}\right)\right\},
\end{align*}
%%%%%%%%%%%%%
%%%%%%%%%%%%%
%%%%%%%%%%%%%
%%%%%%%%%%%%%
\begin{align*}
 \beta_{\tilde{\gamma}_{17}}=&\frac{1}{16\pi^2}\left\{-2 g_1^4+\tilde{\gamma} _4 \tilde{\gamma} _5+\frac{\tilde{\gamma} _9 \tilde{\gamma} _{10}}{3}+\frac{8 \tilde{\gamma} _{13} \tilde{\gamma} _{14}}{3}-\frac{13}{6} g_1^2 \tilde{\gamma} _{17}-\frac{9}{2} g_2^2 \tilde{\gamma} _{17}-8 g_3^2 \tilde{\gamma} _{17}+4 z_{q^*}^2 \tilde{\gamma} _{17}-\frac{2 \tilde{\gamma} _{17}^2}{3}\right.\\
%%%%
&+z_e^2 \left(24 z_{q^*}^2+2 \tilde{\gamma} _{17}\right)+\tilde{\gamma} _{17} \left(\frac{8 \tilde{\gamma} _{16}}{9}+\frac{3 \tilde{\gamma} _{19}}{2}\right)+2 \tilde{\gamma} _{18} \tilde{\gamma} _{20}+\frac{8}{3} \tilde{\gamma} _{17} \bar{g}_{11}^2+\frac{1}{2} \tilde{\gamma} _{17} \bar{g}_{12}^2+\bar{g}_{10}^2 \left(\frac{2 \tilde{\gamma} _{17}}{9}\right.\\
%%%
&\left.\left.+\frac{2 \bar{g}_{12}^2}{3}\right)+\frac{3}{2} \tilde{\gamma} _{17} \bar{g}_{13}^2\right\},
\end{align*}
%%%%%%%%%%%%%
%%%%%%%%%%%%%
%%%%%%%%%%%%%
%%%%%%%%%%%%%
\begin{align*}
 \beta_{\tilde{\gamma}_{18}}=&\frac{1}{16\pi^2}\left\{4 g_1^4\!+\!\tilde{\gamma} _4 \tilde{\gamma} _6\!+\!\frac{\tilde{\gamma} _9 \tilde{\gamma} _{11}}{3}+\frac{8 \tilde{\gamma} _{13} \tilde{\gamma} _{15}}{3}-\frac{20}{3} g_1^2 \tilde{\gamma} _{18}-8 g_3^2 \tilde{\gamma} _{18}+4 z_{q^*}^2 \tilde{\gamma} _{18}+\frac{4 \tilde{\gamma} _{18}^2}{3}+z_l^2 \left(-24 z_{q^*}^2\right.\right.\\
%%%%
&\left.\left.+4 \tilde{\gamma} _{18}\right)+\tilde{\gamma} _{17} \tilde{\gamma} _{20}+\tilde{\gamma} _{18} \left(\frac{8 \tilde{\gamma} _{16}}{9}+4 \tilde{\gamma} _{21}\right)+\frac{8}{3} \tilde{\gamma} _{18} \bar{g}_{11}^2+2 \tilde{\gamma} _{18} \bar{g}_{14}^2+\bar{g}_{10}^2 \left(\frac{2 \tilde{\gamma} _{18}}{9}-\frac{4 \bar{g}_{14}^2}{3}\right)\right\},
\end{align*}
%%%%%%%%%%%%%
%%%%%%%%%%%%%
%%%%%%%%%%%%%
%%%%%%%%%%%%%
\begin{align*}
 \beta_{\tilde{\gamma}_{19}}=&\frac{1}{16\pi^2}\left\{3 g_1^4\!+\!9 g_2^4-16 z_e^4+\tilde{\gamma} _5^2+\frac{\tilde{\gamma} _{10}^2}{3}+\frac{8 \tilde{\gamma} _{14}^2}{3}+\frac{2 \tilde{\gamma} _{17}^2}{3}+g_1^2 \left(6 g_2^2-3 \tilde{\gamma} _{19}\right)-9 g_2^2 \tilde{\gamma} _{19}+4 z_e^2 \tilde{\gamma} _{19}\right.\\
%%%%
&\left.+3 \tilde{\gamma} _{19}^2+2 \tilde{\gamma} _{20}^2+\tilde{\gamma} _{23}^2+3 \tilde{\gamma} _{25}^2-\bar{g}_{12}^4+3 \tilde{\gamma} _{19} \bar{g}_{13}^2-5 \bar{g}_{13}^4+\bar{g}_{12}^2 \left(\tilde{\gamma} _{19}-2 \bar{g}_{13}^2\right)\right\},
\end{align*}
%%%%%%%%%%%%%
%%%%%%%%%%%%%
%%%%%%%%%%%%%
%%%%%%%%%%%%%
\begin{align*}
 \beta_{\tilde{\gamma}_{20}}=&\frac{1}{16\pi^2}\left\{-6 g_1^4\!+\tilde{\gamma} _5 \tilde{\gamma} _6+\frac{\tilde{\gamma} _{10} \tilde{\gamma} _{11}}{3}+\frac{8 \tilde{\gamma} _{14} \tilde{\gamma} _{15}}{3}+\frac{2 \tilde{\gamma} _{17} \tilde{\gamma} _{18}}{3}-\frac{15}{2} g_1^2 \tilde{\gamma} _{20}-\frac{9}{2} g_2^2 \tilde{\gamma} _{20}+4 z_l^2 \tilde{\gamma} _{20}-2 \tilde{\gamma} _{20}^2\right.\\
%%%%
&+z_e^2 \left(8 z_l^2+2 \tilde{\gamma} _{20}\right)+\tilde{\gamma} _{20} \left(\frac{3 \tilde{\gamma} _{19}}{2}+4 \tilde{\gamma} _{21}\right)+\left(6 z_l^2+\frac{3 \tilde{\gamma} _{20}}{2}\right) \bar{g}_{13}^2-8 z_e z_l \bar{g}_{12} \bar{g}_{14}+\left(8 z_e^2\right.\\
%%%%
&\left.\left.+2 \tilde{\gamma} _{20}\right) \bar{g}_{14}^2+\bar{g}_{12}^2 \left(2 z_l^2+\frac{\tilde{\gamma} _{20}}{2}+2 \bar{g}_{14}^2\right)\right\},
\end{align*}
%%%%%%%%%%%%%
%%%%%%%%%%%%%
%%%%%%%%%%%%%
%%%%%%%%%%%%%
\begin{align*}
 \beta_{\tilde{\gamma}_{21}}=&\frac{1}{16\pi^2}\left\{12 g_1^4\!-8 z_l^4\!+\!\tilde{\gamma} _6^2\!+\!\frac{\tilde{\gamma} _{11}^2}{3}+\frac{8 \tilde{\gamma} _{15}^2}{3}+\frac{2 \tilde{\gamma} _{18}^2}{3}+\tilde{\gamma} _{20}^2-12 g_1^2 \tilde{\gamma} _{21}+8 z_l^2 \tilde{\gamma} _{21}+10 \tilde{\gamma} _{21}^2+4 \tilde{\gamma} _{21} \bar{g}_{14}^2\right.\\
%%%%
&\left.-4 \bar{g}_{14}^4\right\},
\end{align*}
%%%%%%%%%%%%%
%%%%%%%%%%%%%
%%%%%%%%%%%%%
%%%%%%%%%%%%%
\begin{align*}
 \beta_{\tilde{\gamma}_{23}}=&\frac{1}{16\pi^2}\left\{g_1^2 \left(-6 g_2^2-3 \tilde{\gamma} _{23}\right)-9 g_2^2 \tilde{\gamma} _{23}+6 y_b^2 \tilde{\gamma} _{23}+6 y_t^2 \tilde{\gamma} _{23}+2 y_{\tau }^2 \tilde{\gamma} _{23}+2 z_e^2 \tilde{\gamma} _{23}+\left(\frac{\tilde{\gamma} _1}{2}-2 \tilde{\gamma} _5\right.\right.\\
%%%%
&\left.+\frac{\tilde{\gamma} _{19}}{2}\right) \tilde{\gamma} _{23}+3 \tilde{\gamma} _{22} \tilde{\gamma} _{25}+\frac{1}{2} \tilde{\gamma} _{23} \bar{g}_1^2+\left(-2 z_e^2+\frac{\tilde{\gamma} _{23}}{2}\right) \bar{g}_2^2+\left(-2 y_{\tau }^2+\frac{\tilde{\gamma} _{23}}{2}\right) \bar{g}_{12}^2-4 y_{\tau } z_e \bar{g}_4 \bar{g}_{13}\\
%%%%
&+2 \bar{g}_1 \bar{g}_3 \bar{g}_{12} \bar{g}_{13}+\left(2 y_{\tau }^2+\frac{3 \tilde{\gamma} _{23}}{2}\right) \bar{g}_{13}^2+\bar{g}_2 \left(-4 y_{\tau } z_e \bar{g}_{12}+2 \bar{g}_4 \bar{g}_{12} \bar{g}_{13}\right)+\bar{g}_3^2 \left(\frac{3 \tilde{\gamma} _{23}}{2}-2 \bar{g}_{13}^2\right)\\
%%%%
&\left.+\bar{g}_4^2 \left(2 z_e^2+\frac{3 \tilde{\gamma} _{23}}{2}+2 \bar{g}_{13}^2\right)\right\},
\end{align*}
%%%%%%%%%%%%%
%%%%%%%%%%%%%
%%%%%%%%%%%%%
%%%%%%%%%%%%%
\begin{align*}
 \beta_{\tilde{\gamma}_{25}}=&\frac{1}{16\pi^2}\left\{\tilde{\gamma} _{22} \tilde{\gamma} _{23}-g_1^2 \left(2 g_2^2+\frac{5 \tilde{\gamma} _{25}}{3}\right)-9 g_2^2 \tilde{\gamma} _{25}-8 g_3^2 \tilde{\gamma} _{25}+2 z_e^2 \tilde{\gamma} _{25}+2 z_u^2 \tilde{\gamma} _{25}+\left(\frac{\tilde{\gamma} _7}{18}-\frac{2 \tilde{\gamma} _{10}}{3}\right.\right.\\
%%%%
&\left.+\frac{\tilde{\gamma} _{19}}{2}+\frac{5 \tilde{\gamma} _{24}}{2}\right) \tilde{\gamma} _{25}+z_d^2 \left(-8 z_e^2+2 \tilde{\gamma} _{25}\right)+\frac{1}{18} \tilde{\gamma} _{25} \bar{g}_5^2+\frac{8}{3} \tilde{\gamma} _{25} \bar{g}_7^2+\frac{1}{2} \tilde{\gamma} _{25} \bar{g}_{12}^2+\frac{2}{3} \bar{g}_5 \bar{g}_6 \bar{g}_{12} \bar{g}_{13}\\
%%%
&\left.+\frac{3}{2} \tilde{\gamma} _{25} \bar{g}_{13}^2+\bar{g}_6^2 \left(\frac{3 \tilde{\gamma} _{25}}{2}-2 \bar{g}_{13}^2\right)\right\},
\end{align*}
%%%%%%%%%%%%%
%%%%%%%%%%%%%
%%%%%%%%%%%%%
%%%%%%%%%%%%%
\begin{align*}
 \beta_{\tilde{\gamma}_{27}}=&\frac{1}{16\pi^2}\left\{-5 g_3^4+g_1^2 \left(\frac{4 g_3^2}{3}-\frac{5 \tilde{\gamma} _{27}}{6}\right)-\frac{9}{2} g_2^2 \tilde{\gamma} _{27}-16 g_3^2 \tilde{\gamma} _{27}+2 z_d^2 \tilde{\gamma} _{27}+2 z_u^2 \tilde{\gamma} _{27}+4 z_{q^*}^2 \tilde{\gamma} _{27}\right.\\
%%%%
&+\left(\frac{\tilde{\gamma} _7}{18}+\frac{4 \tilde{\gamma} _9}{9}+\frac{2 \tilde{\gamma} _{16}}{9}+\frac{3 \tilde{\gamma} _{24}}{2}\right) \tilde{\gamma} _{27}-\frac{5 \tilde{\gamma} _{27}^2}{3}+\tilde{\gamma} _{26} \tilde{\gamma} _{28}+\left(\frac{2 z_{q^*}^2}{9}+\frac{\tilde{\gamma} _{27}}{18}\right) \bar{g}_5^2+\left(6 z_{q^*}^2\right.\\
%%%%
&\left.+\frac{3 \tilde{\gamma} _{27}}{2}\right) \bar{g}_6^2+\left(\frac{8 z_d^2}{9}+\frac{2 \tilde{\gamma} _{27}}{9}\right) \bar{g}_{10}^2+\frac{8}{3} z_d z_{q^*} \bar{g}_7 \bar{g}_{11}+\left(-\frac{4 z_d^2}{3}+\frac{8 \tilde{\gamma} _{27}}{3}\right) \bar{g}_{11}^2+\bar{g}_5 \left(\frac{8}{9} z_d z_{q^*} \bar{g}_{10}\right.
\end{align*}
\begin{align*}
 \hskip-4cm\phantom{\beta_{\tilde{\gamma}_{27}}=}&\left.\left.-\frac{4}{9} \bar{g}_7 \bar{g}_{10} \bar{g}_{11}\right)+\bar{g}_7^2 \left(-\frac{4 z_{q^*}^2}{3}+\frac{8 \tilde{\gamma} _{27}}{3}-\frac{4 \bar{g}_{11}^2}{3}\right)\right\},
\end{align*}
%%%%%%%%%%%%%
%%%%%%%%%%%%%
%%%%%%%%%%%%%
%%%%%%%%%%%%%
\begin{align*}
 \beta_{\tilde{\gamma}_{28}}=&\frac{1}{16\pi^2}\left\{5 g_3^4+2 \tilde{\gamma} _{26} \tilde{\gamma} _{27}+g_1^2 \left(-\frac{16 g_3^2}{3}-\frac{10 \tilde{\gamma} _{28}}{3}\right)-16 g_3^2 \tilde{\gamma} _{28}+4 z_{q^*}^2 \tilde{\gamma} _{28}+\left(\frac{8 \tilde{\gamma} _{12}}{9}-\frac{16 \tilde{\gamma} _{13}}{9}\right.\right.\\
%%%%
&\left.+\frac{2 \tilde{\gamma} _{16}}{9}\right) \tilde{\gamma} _{28}+\frac{5 \tilde{\gamma} _{28}^2}{3}+z_q^2 \left(4 \tilde{\gamma} _{28}-16 z_{q^*}^2\right)+\frac{8}{9} \tilde{\gamma} _{28} \bar{g}_8^2+\frac{2}{9} \tilde{\gamma} _{28} \bar{g}_{10}^2+\frac{16}{9} \bar{g}_8 \bar{g}_9 \bar{g}_{10} \bar{g}_{11}+\frac{8}{3} \tilde{\gamma} _{28} \bar{g}_{11}^2\\
%%%
&\left.+\bar{g}_9^2 \left(\frac{8 \tilde{\gamma} _{28}}{3}-\frac{14 \bar{g}_{11}^2}{3}\right)\right\},
\end{align*}
%%%%%%%%%%%%%
%%%%%%%%%%%%%
%%%%%%%%%%%%%
%%%%%%%%%%%%%
\begin{align*}
 \beta_{\tilde{\gamma}_{29}}=&\frac{1}{16\pi^2}\left\{-\frac{3}{2} z_l z_{q^*} \bar{g}_6 \bar{g}_{13}+\bar{g}_{10} \left(\frac{1}{3} z_d z_l \bar{g}_{12}-\frac{2}{3} z_d z_e \bar{g}_{14}\right)+\bar{g}_5 \left(\frac{1}{6} z_l z_{q^*} \bar{g}_{12}-\frac{1}{3} z_e z_{q^*} \bar{g}_{14}\right)\right.\\
%%%%
&+\tilde{\gamma} _{29} \left(z_d^2\!-\frac{25 g_1^2}{6}\!-\frac{9 g_2^2}{2}\!-8 g_3^2+z_e^2+2 z_l^2+z_u^2+2 z_{q^*}^2+\frac{\tilde{\gamma} _9}{9}-\frac{\tilde{\gamma} _{10}}{6}+\frac{\tilde{\gamma} _{11}}{3}-\frac{\tilde{\gamma} _{17}}{3}+\frac{2 \tilde{\gamma} _{18}}{3}\right.\\
%%%%
&\left.\left.-\tilde{\gamma} _{20}+\frac{3 \tilde{\gamma} _{25}}{2}-\frac{8 \tilde{\gamma} _{27}}{3}+\frac{\bar{g}_5^2}{36}+\frac{3 \bar{g}_6^2}{4}+\frac{4 \bar{g}_7^2}{3}+\frac{\bar{g}_{10}^2}{9}+\frac{4 \bar{g}_{11}^2}{3}+\frac{\bar{g}_{12}^2}{4}+\frac{3 \bar{g}_{13}^2}{4}+\bar{g}_{14}^2\right)\right\},
\end{align*}
%%%%%%%%%%%%%
%%%%%%%%%%%%%
%%%%%%%%%%%%%
%%%%%%%%%%%%%
\begin{align*}
 \beta_{\tilde{\gamma}_{30}}=&\frac{1}{16\pi^2}\left\{\tilde{\gamma} _{30} \left(-\frac{31 g_1^2}{3 \sqrt{2}}-\frac{9 g_2^2}{\sqrt{2}}-8 \sqrt{2} g_3^2+\sqrt{2} z_d^2+\sqrt{2} z_e^2+2 \sqrt{2} z_l^2+2 \sqrt{2} z_q^2+\sqrt{2} z_u^2\right.\right.\\
%%%%%
&-\frac{2 \sqrt{2} \tilde{\gamma} _8}{9}-\frac{\tilde{\gamma} _{10}}{3 \sqrt{2}}+\frac{\sqrt{2} \tilde{\gamma} _{11}}{3}+\frac{2 \sqrt{2} \tilde{\gamma} _{14}}{3}-\frac{4 \sqrt{2} \tilde{\gamma} _{15}}{3}-\sqrt{2} \tilde{\gamma} _{20}-\frac{3 \tilde{\gamma} _{25}}{\sqrt{2}}-\frac{8 \sqrt{2} \tilde{\gamma} _{26}}{3}+\frac{\bar{g}_5^2}{18 \sqrt{2}}\\
%%%
&\left.\left.+\frac{3 \bar{g}_6^2}{2 \sqrt{2}}+\frac{4}{3} \sqrt{2} \bar{g}_7^2+\frac{4}{9} \sqrt{2} \bar{g}_8^2+\frac{4}{3} \sqrt{2} \bar{g}_9^2+\frac{\bar{g}_{12}^2}{2 \sqrt{2}}+\frac{3 \bar{g}_{13}^2}{2 \sqrt{2}}+\sqrt{2} \bar{g}_{14}^2\right)\right\}.
\end{align*}
%%%%%%%%%%%%%%%%%%%%%%%%%%%%%%%%%%%%%%%%%%%%%%%%%%%%%%%%%%%%%%%%%%%%%%%%%%%%%%%%%
%%%%%%%%%%%%%%%%%%%%%%%%%%%%%%%%%%%%%%%%%%%%%%%%%%%%%%%%%%%%%%%%%%%%%%%%%%%%%%%%%
%%%%%%%%%%%%%%%%%%%%%%%%%%%%%%%%%%%%%%%%%%%%%%%%%%%%%%%%%%%%%%%%%%%%%%%%%%%%%%%%%
\subsubsection{Fermion masses}
%%%%%%%%%%%%%%%%%%%%%%%%%%%%%%%%%%%%%%%%%%%%%%%%%%%%%%%%%%%%%%%%%%%%%%%%%%%%%%%%%
%%%%%%%%%%%%%%%%%%%%%%%%%%%%%%%%%%%%%%%%%%%%%%%%%%%%%%%%%%%%%%%%%%%%%%%%%%%%%%%%%
%%%%%%%%%%%%%%%%%%%%%%%%%%%%%%%%%%%%%%%%%%%%%%%%%%%%%%%%%%%%%%%%%%%%%%%%%%%%%%%%%
\begin{align*}
 \beta_\mu=&\beta_\mu^{\rm MES}+\frac{1}{16\pi^2}\mu  \left(\frac{z_e^2}{2}+\frac{z_l^2}{2}+\frac{3 z_{q^*}^2}{2}\right)+\frac{1}{(16\pi^2)^2}\left\{c_d \left(6 y_t z_u z_{q^*}-\frac{9}{2} z_{q^*} \bar{g}_3 \bar{g}_6+\bar{g}_1 \left(-\frac{1}{2} z_{q^*} \bar{g}_5\right.\right.\right.\\
 %%%%
 &\left.\left.-z_d \bar{g}_{10}\right)\right)+\frac{3}{4} M_2 \bar{g}_3 \bar{g}_4 \bar{g}_{13}^2+M_1 \bar{g}_1 \bar{g}_2 \left(\frac{\bar{g}_{10}^2}{6}+\frac{\bar{g}_{12}^2}{4}+\frac{\bar{g}_{14}^2}{2}\right)+c_l \left(-\frac{3}{2} z_l \bar{g}_3 \bar{g}_{13}+\bar{g}_1 \left(\frac{1}{2} z_l \bar{g}_{12}\right.\right.\\
 %%%%
 &\left.\left.-z_e \bar{g}_{14}\right)\right)\!+\!\mu  \left(\frac{121 g_1^4}{200}\!+\frac{11 g_2^4}{8}-z_e^4-\frac{13 z_l^4}{8}-y_{\tau }^2 \left(\frac{z_e^2}{4}+\frac{z_l^2}{8}\right)+17 g_3^2 z_{q^*}^2-\frac{3}{8} y_b^2 z_{q^*}^2-\frac{3}{8} y_t^2 z_{q^*}^2\right.\\
 %%%%
 &-\frac{27}{4} z_q^2 z_{q^*}^2-\frac{39 z_{q^*}^4}{8}+g_1^2 \left(\frac{33 z_e^2}{16}+\frac{3 z_l^2}{2}+z_{q^*}^2\right)+g_2^2 \left(\frac{33 z_e^2}{16}+3 z_l^2+9 z_{q^*}^2\right)-\frac{1}{96} z_{q^*}^2 \bar{g}_5^2\\
 %%%%
 &-\frac{9}{32} z_{q^*}^2 \bar{g}_6^2-\frac{1}{2} z_{q^*}^2 \bar{g}_7^2+\left(-\frac{z_d^2}{24}-\frac{z_{q^*}^2}{4}\right) \bar{g}_{10}^2+\left(-\frac{z_d^2}{2}-3 z_{q^*}^2\right) \bar{g}_{11}^2+\left(-\frac{3 z_e^2}{16}-\frac{z_l^2}{32}\right) \bar{g}_{12}^2\end{align*}
 \begin{align*}\phantom{ \beta_\mu=}&
 +\left(-\frac{9 z_e^2}{16}-\frac{3 z_l^2}{32}\right) \bar{g}_{13}^2-\frac{3}{64} \bar{g}_3^2 \bar{g}_{13}^2-\frac{3}{64} \bar{g}_4^2 \bar{g}_{13}^2+\left(-\frac{z_e^2}{8}-\frac{3 z_l^2}{4}\right) \bar{g}_{14}^2+\bar{g}_1^2 \left(-\frac{\bar{g}_{10}^2}{96}-\frac{\bar{g}_{12}^2}{64}\right.\\
 %%%%
 &\left.\left.\left.-\frac{\bar{g}_{14}^2}{32}\right)+\bar{g}_2^2 \left(-\frac{\bar{g}_{10}^2}{96}-\frac{\bar{g}_{12}^2}{64}-\frac{\bar{g}_{14}^2}{32}\right)\right)
 \right\},
\end{align*}
%%%%%%%%%%%%%%%%%%%%%%%%%%%%%%%%%%%%%%%%%%%%%%%%%%%%%%%%%%%%%%%%%%%%%%%%%%%%%%%%%
%%%%%%%%%%%%%%%%%%%%%%%%%%%%%%%%%%%%%%%%%%%%%%%%%%%%%%%%%%%%%%%%%%%%%%%%%%%%%%%%%
%%%%%%%%%%%%%%%%%%%%%%%%%%%%%%%%%%%%%%%%%%%%%%%%%%%%%%%%%%%%%%%%%%%%%%%%%%%%%%%%%
%%%%%%%%%%%%%%%%%%%%%%%%%%%%%%%%%%%%%%%%%%%%%%%%%%%%%%%%%%%%%%%%%%%%%%%%%%%%%%%%%
%%%%%%%%%%%%%%%%%%%%%%%%%%%%%%%%%%%%%%%%%%%%%%%%%%%%%%%%%%%%%%%%%%%%%%%%%%%%%%%%%
\begin{align*}
 \beta_{M_1}=&\beta_{M_1}^{\rm MES}+\frac{1}{16\pi^2}M_1 \left(\frac{\bar{g}_{10}^2}{3}+\frac{\bar{g}_{12}^2}{2}+\bar{g}_{14}^2\right)+\frac{1}{(16\pi^2)^2}\left\{\mu  \left(z_e^2+z_l^2+3 z_{q^*}^2\right) \bar{g}_1 \bar{g}_2\right.\\
 %%%
 &+c_d \left(-\frac{4}{3} y_b \bar{g}_5 \bar{g}_{10}+\bar{g}_2 \left(2 z_{q^*} \bar{g}_5+4 z_d \bar{g}_{10}\right)\right)+\frac{8}{9} M_3 \bar{g}_{10}^2 \bar{g}_{11}^2+\frac{3}{4} M_2 \bar{g}_{12}^2 \bar{g}_{13}^2+c_l \left(4 y_{\tau } \bar{g}_{12} \bar{g}_{14}\right.\\
 %%%
 &\left.+\bar{g}_2 \left(4 z_e \bar{g}_{14}-2 z_l \bar{g}_{12}\right)\right)+M_1 \left(\left(-\frac{z_e^2}{8}-\frac{z_l^2}{8}-\frac{3 z_{q^*}^2}{8}\right) \bar{g}_2^2-\frac{1}{24} z_{q^*}^2 \bar{g}_5^2+\frac{34}{9} g_3^2 \bar{g}_{10}^2-\frac{1}{6} y_b^2 \bar{g}_{10}^2\right.\\
 %%%
 &+\frac{\bar{g}_{10}^4}{108}+\bar{g}_{10}^2 \left(-\frac{z_d^2}{6}-z_{q^*}^2-\frac{7 \bar{g}_{11}^2}{9}\right)+\frac{51}{16} g_2^2 \bar{g}_{12}^2+\frac{\bar{g}_{12}^4}{32}+\bar{g}_{12}^2 \left(-\frac{3 z_e^2}{4}-\frac{z_l^2}{8}-\frac{21 \bar{g}_{13}^2}{32}\right)\\
 %%%
 &\left.\left.+\left(-\frac{z_e^2}{2}-3 z_l^2\right) \bar{g}_{14}^2+\frac{\bar{g}_{14}^4}{4}+y_{\tau }^2 \left(-\frac{\bar{g}_{12}^2}{8}-\frac{\bar{g}_{14}^2}{2}\right)+g_1^2 \left(\frac{17 \bar{g}_{10}^2}{90}+\frac{51 \bar{g}_{12}^2}{80}+\frac{51 \bar{g}_{14}^2}{10}\right)\right)\right\},
\end{align*}
%%%%%%%%%%%%%%%%%%%%%%%%%%%%%%%%%%%%%%%%%%%%%%%%%%%%%%%%%%%%%%%%%%%%%%%%%%%%%%%%%
%%%%%%%%%%%%%%%%%%%%%%%%%%%%%%%%%%%%%%%%%%%%%%%%%%%%%%%%%%%%%%%%%%%%%%%%%%%%%%%%%
%%%%%%%%%%%%%%%%%%%%%%%%%%%%%%%%%%%%%%%%%%%%%%%%%%%%%%%%%%%%%%%%%%%%%%%%%%%%%%%%%
%%%%%%%%%%%%%%%%%%%%%%%%%%%%%%%%%%%%%%%%%%%%%%%%%%%%%%%%%%%%%%%%%%%%%%%%%%%%%%%%%
%%%%%%%%%%%%%%%%%%%%%%%%%%%%%%%%%%%%%%%%%%%%%%%%%%%%%%%%%%%%%%%%%%%%%%%%%%%%%%%%%
\begin{align*}
 \beta_{M_2}=&\beta_{M_2}^{\rm MES}+\frac{1}{32\pi^2} M_2 \bar{g}_{13}^2+\frac{1}{(16\pi^2)^2}\left\{\mu  \left(z_e^2+z_l^2+3 z_{q^*}^2\right) \bar{g}_3 \bar{g}_4+6 c_d z_{q^*} \bar{g}_4 \bar{g}_6+2 c_l z_l \bar{g}_4 \bar{g}_{13}\right.\\
 %%%%
 &+\frac{1}{4} M_1 \bar{g}_{12}^2 \bar{g}_{13}^2+M_2 \left(\frac{11 g_2^4}{3}+\left(-\frac{z_e^2}{8}-\frac{z_l^2}{8}-\frac{3 z_{q^*}^2}{8}\right) \bar{g}_4^2-\frac{3}{8} z_{q^*}^2 \bar{g}_6^2+\frac{51}{80} g_1^2 \bar{g}_{13}^2+\frac{91}{16} g_2^2 \bar{g}_{13}^2\right.\\
 %%%
 &\left.\left.-\frac{1}{8} y_{\tau }^2 \bar{g}_{13}^2+\left(-\frac{3 z_e^2}{4}-\frac{z_l^2}{8}\right) \bar{g}_{13}^2-\frac{7}{32} \bar{g}_{12}^2 \bar{g}_{13}^2-\frac{29 \bar{g}_{13}^4}{32}\right)\right\},
\end{align*}
%%%%%%%%%%%%%%%%%%%%%%%%%%%%%%%%%%%%%%%%%%%%%%%%%%%%%%%%%%%%%%%%%%%%%%%%%%%%%%%%%
%%%%%%%%%%%%%%%%%%%%%%%%%%%%%%%%%%%%%%%%%%%%%%%%%%%%%%%%%%%%%%%%%%%%%%%%%%%%%%%%%
%%%%%%%%%%%%%%%%%%%%%%%%%%%%%%%%%%%%%%%%%%%%%%%%%%%%%%%%%%%%%%%%%%%%%%%%%%%%%%%%%
%%%%%%%%%%%%%%%%%%%%%%%%%%%%%%%%%%%%%%%%%%%%%%%%%%%%%%%%%%%%%%%%%%%%%%%%%%%%%%%%%
%%%%%%%%%%%%%%%%%%%%%%%%%%%%%%%%%%%%%%%%%%%%%%%%%%%%%%%%%%%%%%%%%%%%%%%%%%%%%%%%%
\begin{align*}
 \beta_{M_3}=&\beta_{M_3}^{\rm MES}+\frac{1}{32\pi^2}M_3\bar{g}_{11}^2+\frac{1}{(16\pi^2)^2}\left\{4 c_d y_b \bar{g}_7 \bar{g}_{11}+\frac{1}{9} M_1 \bar{g}_{10}^2 \bar{g}_{11}^2+M_3 \left(\frac{11 g_3^4}{2}-\frac{1}{4} z_{q^*}^2 \bar{g}_7^2\right.\right.\\
 %%%%
 &\left.\left.+\frac{17}{60} g_1^2 \bar{g}_{11}^2+\frac{113}{12} g_3^2 \bar{g}_{11}^2-\frac{1}{4} y_b^2 \bar{g}_{11}^2+\left(-\frac{z_d^2}{4}-\frac{3 z_{q^*}^2}{2}\right) \bar{g}_{11}^2-\frac{7}{72} \bar{g}_{10}^2 \bar{g}_{11}^2-\frac{4 \bar{g}_{11}^4}{3}\right)\right\}.
\end{align*}
%%%%%%%%%%%%%%%%%%%%%%%%%%%%%%%%%%%%%%%%%%%%%%%%%%%%%%%%%%%%%%%%%%%%%%%%%%%%%%%%%
%%%%%%%%%%%%%%%%%%%%%%%%%%%%%%%%%%%%%%%%%%%%%%%%%%%%%%%%%%%%%%%%%%%%%%%%%%%%%%%%%
%%%%%%%%%%%%%%%%%%%%%%%%%%%%%%%%%%%%%%%%%%%%%%%%%%%%%%%%%%%%%%%%%%%%%%%%%%%%%%%%%
\subsubsection{Scalar trilinear couplings}
%%%%%%%%%%%%%%%%%%%%%%%%%%%%%%%%%%%%%%%%%%%%%%%%%%%%%%%%%%%%%%%%%%%%%%%%%%%%%%%%%
%%%%%%%%%%%%%%%%%%%%%%%%%%%%%%%%%%%%%%%%%%%%%%%%%%%%%%%%%%%%%%%%%%%%%%%%%%%%%%%%%
%%%%%%%%%%%%%%%%%%%%%%%%%%%%%%%%%%%%%%%%%%%%%%%%%%%%%%%%%%%%%%%%%%%%%%%%%%%%%%%%%
\begin{align*}
 \beta_{a_u}=&\beta_{a_u}^{\rm MES}+\frac{1}{16\pi^2}4 c_l \tilde{\gamma} _{30},
\end{align*}
%%%%%%%%%%%%%%%
%%%%%%%%%%%%%%%
%%%%%%%%%%%%%%%
%%%%%%%%%%%%%%%
%%%%%%%%%%%%%%%
%%%%%%%%%%%%%%%
\begin{align*}
 \beta_{c_d}=&\frac{1}{16\pi^2}\left\{4 c_l \tilde{\gamma} _{29}+3 M_2 z_{q^*} \bar{g}_4 \bar{g}_6+\mu  \left(4 y_t z_u z_{q^*}-3 z_{q^*} \bar{g}_3 \bar{g}_6+\bar{g}_1 \left(-\frac{1}{3} z_{q^*} \bar{g}_5-\frac{2}{3} z_d \bar{g}_{10}\right)\right)\right.\\
%%%%
&+M_1 \left(-\frac{2}{9} y_b \bar{g}_5 \bar{g}_{10}+\bar{g}_2 \left(\frac{1}{3} z_{q^*} \bar{g}_5+\frac{2}{3} z_d \bar{g}_{10}\right)\right)+\frac{16}{3} M_3 y_b \bar{g}_7 \bar{g}_{11}+c_d \left(-\frac{7 g_1^2}{6}-\frac{9 g_2^2}{2}-8 g_3^2\right.\\
%%%
&+3 y_b^2+3 y_t^2+y_{\tau }^2+z_d^2+z_u^2+2 z_{q^*}^2+\frac{\tilde{\gamma} _2}{6}+\frac{\tilde{\gamma} _4}{3}+\frac{\tilde{\gamma} _9}{9}+\frac{3 \tilde{\gamma} _{22}}{2}-\frac{8 \tilde{\gamma} _{27}}{3}+\frac{\bar{g}_1^2}{4}+\frac{\bar{g}_2^2}{4}+\frac{3 \bar{g}_3^2}{4}\\
%%%
&\left.\left.+\frac{3 \bar{g}_4^2}{4}+\frac{\bar{g}_5^2}{36}+\frac{3 \bar{g}_6^2}{4}+\frac{4 \bar{g}_7^2}{3}+\frac{\bar{g}_{10}^2}{9}+\frac{4 \bar{g}_{11}^2}{3}\right)\right\},
\end{align*}
%%%%%%%%%%%%%%%
%%%%%%%%%%%%%%%
%%%%%%%%%%%%%%%
%%%%%%%%%%%%%%%
%%%%%%%%%%%%%%%
%%%%%%%%%%%%%%%
\begin{align*}
 \beta_{c_l}=&\frac{1}{16\pi^2}\left\{12 c_d \tilde{\gamma} _{29}+12 a_u \tilde{\gamma} _{30}+3 M_2 z_l \bar{g}_4 \bar{g}_{13}+c_l \left(3 y_b^2-\frac{9 g_1^2}{2}-\frac{9 g_2^2}{2}+3 y_t^2+y_{\tau }^2+z_e^2+2 z_l^2\right.\right.\\
%%%%
&\left.-\frac{\tilde{\gamma} _5}{2}+\tilde{\gamma} _6-\tilde{\gamma} _{20}+\frac{3 \tilde{\gamma} _{23}}{2}+\frac{\bar{g}_1^2}{4}+\frac{\bar{g}_2^2}{4}+\frac{3 \bar{g}_3^2}{4}+\frac{3 \bar{g}_4^2}{4}+\frac{\bar{g}_{12}^2}{4}+\frac{3 \bar{g}_{13}^2}{4}+\bar{g}_{14}^2\right)+\mu  \left(-3 z_l \bar{g}_3 \bar{g}_{13}\right.\\
%%%%
&\left.\left.+\bar{g}_1 \left(z_l \bar{g}_{12}-2 z_e \bar{g}_{14}\right)\right)+M_1 \left(2 y_{\tau } \bar{g}_{12} \bar{g}_{14}+\bar{g}_2 \left(-z_l \bar{g}_{12}+2 z_e \bar{g}_{14}\right)\right)\right\}.
\end{align*}
%%%%%%%%%%%%%%%%%%%%%%%%%%%%%%%%%%%%%%%%%%%%%%%%%%%%%%%%%%%%%%%%%%%%%%%%%%%%%%%%%
%%%%%%%%%%%%%%%%%%%%%%%%%%%%%%%%%%%%%%%%%%%%%%%%%%%%%%%%%%%%%%%%%%%%%%%%%%%%%%%%%
%%%%%%%%%%%%%%%%%%%%%%%%%%%%%%%%%%%%%%%%%%%%%%%%%%%%%%%%%%%%%%%%%%%%%%%%%%%%%%%%%
\subsubsection{Scalar masses}
%%%%%%%%%%%%%%%%%%%%%%%%%%%%%%%%%%%%%%%%%%%%%%%%%%%%%%%%%%%%%%%%%%%%%%%%%%%%%%%%%
%%%%%%%%%%%%%%%%%%%%%%%%%%%%%%%%%%%%%%%%%%%%%%%%%%%%%%%%%%%%%%%%%%%%%%%%%%%%%%%%%
%%%%%%%%%%%%%%%%%%%%%%%%%%%%%%%%%%%%%%%%%%%%%%%%%%%%%%%%%%%%%%%%%%%%%%%%%%%%%%%%%
\begin{align*}
 \beta_{m^2_H}=&\beta_{m^2_H}^{\rm MES}+\frac{1}{16\pi^2}\left\{6 c_d^2+2 c_l^2+m_D^2 \tilde{\gamma} _4-m_L^2 \tilde{\gamma} _5+m_e^2 \tilde{\gamma} _6\right\}+\frac{1}{(16\pi^2)^2}\left\{16 M_3^2 y_b^2 \bar{g}_{11}^2\right.\\
%%%
&+c_d \left(\frac{2}{3} M_1 y_b \bar{g}_5 \bar{g}_{10}-16 M_3 y_b \bar{g}_7 \bar{g}_{11}\right)+c_d^2 \left(-\frac{7 g_1^2}{10}+\frac{9 g_2^2}{2}+64 g_3^2-6 z_d^2-6 z_u^2-12 z_{q^*}^2\right.\\
%%%%
&\left.-\frac{9 \tilde{\gamma} _1}{2}-2 \tilde{\gamma} _2-4 \tilde{\gamma} _4-9 \tilde{\gamma} _{22}-\frac{\bar{g}_5^2}{6}-\frac{9 \bar{g}_6^2}{2}-8 \bar{g}_7^2-\frac{2 \bar{g}_{10}^2}{3}-8 \bar{g}_{11}^2\right)+m_D^2 \left(\frac{3 g_1^4}{5}+\frac{8}{15} g_1^2 \tilde{\gamma} _4\right.\\
%%%%
&\left.+\frac{32}{3} g_3^2 \tilde{\gamma} _4-4 z_{q^*}^2 \tilde{\gamma} _4-\frac{\tilde{\gamma} _4^2}{3}-2 y_b z_{q^*} \bar{g}_2 \bar{g}_{10}-\frac{2}{9} \tilde{\gamma} _4 \bar{g}_{10}^2-\frac{8}{3} \tilde{\gamma} _4 \bar{g}_{11}^2\right)+M_2^2 \left(\frac{3}{2} z_e^2 \bar{g}_4^2+\frac{3}{2} z_l^2 \bar{g}_4^2\right.\\
%%%
&\left.+\frac{9}{2} z_{q^*}^2 \bar{g}_4^2-6 y_{\tau } z_e \bar{g}_4 \bar{g}_{13}+3 y_{\tau }^2 \bar{g}_{13}^2+\frac{9}{4} \bar{g}_3^2 \bar{g}_{13}^2+\frac{9}{4} \bar{g}_4^2 \bar{g}_{13}^2\right)+m_L^2 \left(\frac{9 g_1^4}{10}+\frac{15 g_2^4}{2}-\frac{6}{5} g_1^2 \tilde{\gamma} _5\right.\\
%%%%
&\left.-6 g_2^2 \tilde{\gamma} _5+2 z_e^2 \tilde{\gamma} _5-\frac{\tilde{\gamma} _5^2}{2}-\frac{3 \tilde{\gamma} _{23}^2}{2}+\frac{1}{2} \tilde{\gamma} _5 \bar{g}_{12}^2+\frac{3}{2} \tilde{\gamma} _5 \bar{g}_{13}^2+y_{\tau } z_e \left(\bar{g}_2 \bar{g}_{12}-3 \bar{g}_4 \bar{g}_{13}\right)\right)\\
%%%%
&-2 c_l M_1 y_{\tau } \bar{g}_{12} \bar{g}_{14}\!+\!c_l^2 \left(\frac{51 g_1^2}{10}\!+\!\frac{3 g_2^2}{2}\!-2 z_e^2\!-4 z_l^2-\frac{3 \tilde{\gamma} _1}{2}+2 \tilde{\gamma} _5-4 \tilde{\gamma} _6-3 \tilde{\gamma} _{23}-\frac{\bar{g}_{12}^2}{2}-\frac{3 \bar{g}_{13}^2}{2}\right.
\end{align*}
\begin{align*}
%%%%
\phantom{\beta_{m^2_H}=}&\left.-2 \bar{g}_{14}^2\right)+m_e^2 \left(\frac{9 g_1^4}{5}+\frac{24}{5} g_1^2 \tilde{\gamma} _6-4 z_l^2 \tilde{\gamma} _6-\tilde{\gamma} _6^2-2 y_{\tau } z_l \bar{g}_2 \bar{g}_{14}-2 \tilde{\gamma} _6 \bar{g}_{14}^2\right)+m_H^2 \left(\frac{121 g_1^4}{200}\right.\\
%%%
&+\frac{11 g_2^4}{8}-\frac{9}{2} y_t^2 z_{q^*}^2-\frac{\tilde{\gamma} _4^2}{12}-\frac{\tilde{\gamma} _5^2}{8}-\frac{\tilde{\gamma} _6^2}{4}-\frac{3 \tilde{\gamma} _{23}^2}{8}+z_{q^*}^2 \left(-\frac{9 \bar{g}_2^2}{8}-\frac{27 \bar{g}_4^2}{8}\right)+z_e^2 \left(-\frac{3 \bar{g}_2^2}{8}-\frac{9 \bar{g}_4^2}{8}\right)\\
%%%%
&+z_l^2 \left(-\frac{3 \bar{g}_2^2}{8}-\frac{9 \bar{g}_4^2}{8}\right)+2 y_b z_{q^*} \bar{g}_2 \bar{g}_{10}+y_b^2 \left(-\frac{9 z_{q^*}^2}{2}-\frac{\bar{g}_{10}^2}{2}-6 \bar{g}_{11}^2\right)-\frac{9}{16} \bar{g}_3^2 \bar{g}_{13}^2-\frac{9}{16} \bar{g}_4^2 \bar{g}_{13}^2\\
%%%%
&+y_{\tau } \left(z_e \left(-\bar{g}_2 \bar{g}_{12}+3 \bar{g}_4 \bar{g}_{13}\right)+2 z_l \bar{g}_2 \bar{g}_{14}\right)+y_{\tau }^2 \left(-3 z_e^2-\frac{3 z_l^2}{2}-\frac{3 \bar{g}_{12}^2}{8}-\frac{9 \bar{g}_{13}^2}{8}-\frac{3 \bar{g}_{14}^2}{2}\right)\\
%%%%
&\left.+\bar{g}_1^2 \left(-\frac{\bar{g}_{10}^2}{8}-\frac{3 \bar{g}_{12}^2}{16}-\frac{3 \bar{g}_{14}^2}{8}\right)+\bar{g}_2^2 \left(-\frac{\bar{g}_{10}^2}{8}-\frac{3 \bar{g}_{12}^2}{16}-\frac{3 \bar{g}_{14}^2}{8}\right)\right)+\mu ^2 \left(y_{\tau }^2 \left(8 z_e^2+4 z_l^2\right)\right.\\
%%%%
&+12 y_b^2 z_{q^*}^2+12 y_t^2 z_{q^*}^2+z_e^2 \left(\frac{\bar{g}_1^2}{2}+\bar{g}_2^2+\frac{3 \bar{g}_3^2}{2}+3 \bar{g}_4^2\right)+z_l^2 \left(\frac{\bar{g}_1^2}{2}+\bar{g}_2^2+\frac{3 \bar{g}_3^2}{2}+3 \bar{g}_4^2\right)\\
%%%
&+z_{q^*}^2 \left(\frac{3 \bar{g}_1^2}{2}+3 \bar{g}_2^2+\frac{9 \bar{g}_3^2}{2}+9 \bar{g}_4^2\right)-4 y_b z_{q^*} \bar{g}_2 \bar{g}_{10}+\frac{3}{4} \bar{g}_3^2 \bar{g}_{13}^2+\frac{3}{4} \bar{g}_4^2 \bar{g}_{13}^2+y_{\tau } \left(z_e \left(2 \bar{g}_2 \bar{g}_{12}\right.\right.\\
%%%%
&\left.\left.\left.-6 \bar{g}_4 \bar{g}_{13}\right)\!-4 z_l \bar{g}_2 \bar{g}_{14}\right)\!+\!\bar{g}_1^2 \left(\frac{\bar{g}_{10}^2}{6}\!+\!\frac{\bar{g}_{12}^2}{4}\!+\!\frac{\bar{g}_{14}^2}{2}\right)+\bar{g}_2^2 \left(\frac{\bar{g}_{10}^2}{6}+\frac{\bar{g}_{12}^2}{4}+\frac{\bar{g}_{14}^2}{2}\right)\right)+M_1^2 \left(\frac{1}{2} z_e^2 \bar{g}_2^2\right.\\
%%%%
&+\frac{1}{2} z_l^2 \bar{g}_2^2+\frac{3}{2} z_{q^*}^2 \bar{g}_2^2-4 y_b z_{q^*} \bar{g}_2 \bar{g}_{10}+\frac{4}{3} y_b^2 \bar{g}_{10}^2+y_{\tau } \left(2 z_e \bar{g}_2 \bar{g}_{12}-4 z_l \bar{g}_2 \bar{g}_{14}\right)+\bar{g}_1^2 \left(\frac{\bar{g}_{10}^2}{2}+\frac{3 \bar{g}_{12}^2}{4}\right.\\
%%%%
&\left.\left.+\frac{3 \bar{g}_{14}^2}{2}\right)+\bar{g}_2^2 \left(\frac{\bar{g}_{10}^2}{2}+\frac{3 \bar{g}_{12}^2}{4}+\frac{3 \bar{g}_{14}^2}{2}\right)+y_{\tau }^2 \left(\bar{g}_{12}^2+4 \bar{g}_{14}^2\right)\right)+\mu  \left(-12 c_d y_t z_u z_{q^*}\right.\\
%%%%
&+M_2 \left(-3 z_e^2 \bar{g}_3 \bar{g}_4-3 z_l^2 \bar{g}_3 \bar{g}_4-9 z_{q^*}^2 \bar{g}_3 \bar{g}_4+6 y_{\tau } z_e \bar{g}_3 \bar{g}_{13}-3 \bar{g}_3 \bar{g}_4 \bar{g}_{13}^2\right)+M_1 \left(-z_e^2 \bar{g}_1 \bar{g}_2\right.\\
%%%%
&-z_l^2 \bar{g}_1 \bar{g}_2-3 z_{q^*}^2 \bar{g}_1 \bar{g}_2+4 y_b z_{q^*} \bar{g}_1 \bar{g}_{10}+y_{\tau } \left(-2 z_e \bar{g}_1 \bar{g}_{12}+4 z_l \bar{g}_1 \bar{g}_{14}\right)+\bar{g}_1 \bar{g}_2 \left(-\frac{2 \bar{g}_{10}^2}{3}-\bar{g}_{12}^2\right.\\
%%%%
&\left.\left.\left.\left.-2 \bar{g}_{14}^2\right)\right)\right)\right\},
\end{align*}
%%%%%%%%%%%%%%%
%%%%%%%%%%%%%%%
%%%%%%%%%%%%%%%
%%%%%%%%%%%%%%%
%%%%%%%%%%%%%%%
%%%%%%%%%%%%%%%
\begin{align*}
 \beta_{m^2_Q}=&\beta_{m^2_Q}^{\rm MES}+\frac{1}{16\pi^2}\left\{2 c_d^2+\frac{1}{3} m_D^2 \tilde{\gamma} _9-\frac{1}{3} m_L^2 \tilde{\gamma} _{10}+\frac{1}{3} m_e^2 \tilde{\gamma} _{11}\right\}+\frac{1}{(16\pi^2)^2}\left\{c_l^2 \left(-\frac{\tilde{\gamma} _2}{3}+\frac{\tilde{\gamma} _{10}}{3}\right.\right.\\
%%%%
&\left.-\frac{2 \tilde{\gamma} _{11}}{3}\right)\!-4 a_u c_l \tilde{\gamma} _{30}\!-2 m_U^2 \tilde{\gamma} _{30}^2\!+\mu  c_d z_{q^*} \left(\frac{1}{3} \bar{g}_1 \bar{g}_5+3 \bar{g}_3 \bar{g}_6\right)-c_d \left(4 c_l \tilde{\gamma} _{29}+\frac{2}{3} M_1 z_d \bar{g}_2 \bar{g}_{10}\right)\\
%%%%
&+c_d^2 \left(\frac{49 g_1^2}{30}+\frac{3 g_2^2}{2}+\frac{8 g_3^2}{3}-6 y_b^2-6 y_t^2-2 y_{\tau }^2-4 z_{q^*}^2-\frac{4 \tilde{\gamma} _2}{3}-\frac{7 \tilde{\gamma} _7}{18}-\frac{8 \tilde{\gamma} _9}{9}-3 \tilde{\gamma} _{22}-\frac{3 \tilde{\gamma} _{24}}{2}\right.\\
%%%%
&\left.+\frac{16 \tilde{\gamma} _{27}}{3}-\frac{\bar{g}_1^2}{2}-\frac{\bar{g}_2^2}{2}-\frac{3 \bar{g}_3^2}{2}-\frac{3 \bar{g}_4^2}{2}-\frac{2 \bar{g}_{10}^2}{9}-\frac{8 \bar{g}_{11}^2}{3}\right)+M_3^2 \left(\frac{8}{3} z_{q^*}^2 \bar{g}_7^2-\frac{32}{3} z_d z_{q^*} \bar{g}_7 \bar{g}_{11}\right.\\
%%%%
&\left.+\frac{16}{3} z_d^2 \bar{g}_{11}^2+4 \bar{g}_7^2 \bar{g}_{11}^2\right)+m_D^2 \left(\frac{g_1^4}{15}+\frac{40 g_3^4}{3}+\frac{8}{45} g_1^2 \tilde{\gamma} _9+\frac{32}{9} g_3^2 \tilde{\gamma} _9-\frac{4}{3} z_{q^*}^2 \tilde{\gamma} _9-\frac{\tilde{\gamma} _9^2}{27}-\frac{8 \tilde{\gamma} _{27}^2}{3}\right.\\
%%%%
&\left.-2 \tilde{\gamma} _{29}^2\!-\frac{2}{27} \tilde{\gamma} _9 \bar{g}_{10}^2\!-\frac{8}{9} \tilde{\gamma} _9 \bar{g}_{11}^2\!+\!z_d z_{q^*} \left(\frac{2}{9} \bar{g}_5 \bar{g}_{10}-\frac{16}{3} \bar{g}_7 \bar{g}_{11}\right)\right)+\mu ^2 \left(2 z_e^2 z_u^2+2 z_l^2 z_u^2+6 z_u^2 z_{q^*}^2\right.
\end{align*}
\begin{align*}
%%%
\phantom{\beta_{m^2_Q}=}&+z_{q^*}^2 \left(\frac{\bar{g}_5^2}{9}\!+\!3 \bar{g}_6^2\!+\!\frac{16 \bar{g}_7^2}{3}\right)\!+\!z_d z_{q^*} \left(\frac{4}{9} \bar{g}_5 \bar{g}_{10}-\frac{32}{3} \bar{g}_7 \bar{g}_{11}\right)+z_d^2 \left(4 z_e^2+4 z_l^2+12 z_{q^*}^2+\frac{2 \bar{g}_{10}^2}{9}\right.\\
%%%%
&\left.\left.+\frac{8 \bar{g}_{11}^2}{3}\right)\right)+m_L^2 \left(\frac{g_1^4}{10}+\frac{15 g_2^4}{2}-\frac{2}{5} g_1^2 \tilde{\gamma} _{10}-2 g_2^2 \tilde{\gamma} _{10}+\frac{2}{3} z_e^2 \tilde{\gamma} _{10}-\frac{\tilde{\gamma} _{10}^2}{18}-\frac{3 \tilde{\gamma} _{25}^2}{2}-2 \tilde{\gamma} _{29}^2-2 \tilde{\gamma} _{30}^2\right.\\
%%%%
&\left.+\frac{1}{6} \tilde{\gamma} _{10} \bar{g}_{12}^2+\frac{1}{2} \tilde{\gamma} _{10} \bar{g}_{13}^2\right)+M_2^2 \left(\frac{3}{2} z_{q^*}^2 \bar{g}_6^2+\frac{9}{4} \bar{g}_6^2 \bar{g}_{13}^2\right)+m_e^2 \left(\frac{g_1^4}{5}+\frac{8}{5} g_1^2 \tilde{\gamma} _{11}-\frac{4}{3} z_l^2 \tilde{\gamma} _{11}-\frac{\tilde{\gamma} _{11}^2}{9}\right.\\
%%%%
&\left.-2 \tilde{\gamma} _{29}^2-2 \tilde{\gamma} _{30}^2-\frac{2}{3} \tilde{\gamma} _{11} \bar{g}_{14}^2\right)+m_Q^2 \left(\frac{121 g_1^4}{1800}+\frac{11 g_2^4}{8}+\frac{22 g_3^4}{9}-\frac{\tilde{\gamma} _9^2}{108}-\frac{\tilde{\gamma} _{10}^2}{72}-\frac{\tilde{\gamma} _{11}^2}{36}-\frac{3 \tilde{\gamma} _{25}^2}{8}\right.\\
%%%
&-\frac{2 \tilde{\gamma} _{27}^2}{3}+\tilde{\gamma} _{29}^2+\tilde{\gamma} _{30}^2+z_{q^*}^2 \left(-\frac{\bar{g}_5^2}{24}-\frac{9 \bar{g}_6^2}{8}-2 \bar{g}_7^2\right)-\bar{g}_7^2 \bar{g}_{11}^2+z_d z_{q^*} \left(-\frac{2}{9} \bar{g}_5 \bar{g}_{10}+\frac{16}{3} \bar{g}_7 \bar{g}_{11}\right)\\
%%%%
&\left.+z_d^2 \left(-\frac{3 z_e^2}{2}-\frac{3 z_l^2}{2}-\frac{9 z_{q^*}^2}{2}-\frac{\bar{g}_{10}^2}{6}-2 \bar{g}_{11}^2\right)-\frac{9}{16} \bar{g}_6^2 \bar{g}_{13}^2+\bar{g}_5^2 \left(-\frac{\bar{g}_{10}^2}{72}-\frac{\bar{g}_{12}^2}{48}-\frac{\bar{g}_{14}^2}{24}\right)\right)\\
%%%
&\left.+M_1^2 \left(\frac{1}{18} z_{q^*}^2 \bar{g}_5^2+\frac{4}{9} z_d z_{q^*} \bar{g}_5 \bar{g}_{10}+\frac{4}{9} z_d^2 \bar{g}_{10}^2+\bar{g}_5^2 \left(\frac{\bar{g}_{10}^2}{18}+\frac{\bar{g}_{12}^2}{12}+\frac{\bar{g}_{14}^2}{6}\right)\right)\right\},
\end{align*}
%%%%%%%%%%%%%%%
%%%%%%%%%%%%%%%
%%%%%%%%%%%%%%%
%%%%%%%%%%%%%%%
%%%%%%%%%%%%%%%
%%%%%%%%%%%%%%%
\begin{align*}
 \beta_{m^2_U}=&\beta_{m^2_U}^{\rm MES}+\frac{1}{16\pi^2}\left\{-\frac{4}{3} m_D^2 \tilde{\gamma} _{13}+\frac{4}{3} m_L^2 \tilde{\gamma} _{14}-\frac{4}{3} m_e^2 \tilde{\gamma} _{15}\right\}+\frac{1}{(16\pi^2)^2}\left\{\mu ^2 \left(4 z_e^2 z_q^2+z_q^2 \left(4 z_l^2\right.\right.\right.\\
%%%%
&\left.\left.+16 z_{q^*}^2\right)\right)+c_d^2 \left(4 \tilde{\gamma} _3+\frac{4 \tilde{\gamma} _8}{3}+\frac{8 \tilde{\gamma} _{13}}{3}\right)+c_l^2 \left(\frac{4 \tilde{\gamma} _3}{3}-\frac{4 \tilde{\gamma} _{14}}{3}+\frac{8 \tilde{\gamma} _{15}}{3}\right)-8 a_u c_l \tilde{\gamma} _{30}-4 m_Q^2 \tilde{\gamma} _{30}^2\\
%%%%
&+4 M_3^2 \bar{g}_9^2 \bar{g}_{11}^2+m_D^2 \left(\frac{16 g_1^4}{15}+\frac{40 g_3^4}{3}-\frac{32}{45} g_1^2 \tilde{\gamma} _{13}-\frac{128}{9} g_3^2 \tilde{\gamma} _{13}+\frac{16}{3} z_{q^*}^2 \tilde{\gamma} _{13}-\frac{16 \tilde{\gamma} _{13}^2}{27}-\frac{8 \tilde{\gamma} _{28}^2}{3}\right.\\
%%%%
&\left.+\frac{8}{27} \tilde{\gamma} _{13} \bar{g}_{10}^2+\frac{32}{9} \tilde{\gamma} _{13} \bar{g}_{11}^2\right)+m_L^2 \left(\frac{8 g_1^4}{5}+\frac{8}{5} g_1^2 \tilde{\gamma} _{14}+8 g_2^2 \tilde{\gamma} _{14}-\frac{8}{3} z_e^2 \tilde{\gamma} _{14}-\frac{8 \tilde{\gamma} _{14}^2}{9}-4 \tilde{\gamma} _{30}^2\right.\\
%%%%
&\left.-\frac{2}{3} \tilde{\gamma} _{14} \bar{g}_{12}^2-2 \tilde{\gamma} _{14} \bar{g}_{13}^2\right)+m_e^2 \left(\frac{16 g_1^4}{5}-\frac{32}{5} g_1^2 \tilde{\gamma} _{15}+\frac{16}{3} z_l^2 \tilde{\gamma} _{15}-\frac{16 \tilde{\gamma} _{15}^2}{9}-4 \tilde{\gamma} _{30}^2+\frac{8}{3} \tilde{\gamma} _{15} \bar{g}_{14}^2\right)\\
%%%
\phantom{\beta_{m^2_U}=}&+m_U^2 \left(\frac{242 g_1^4}{225}+\frac{22 g_3^4}{9}-3 z_q^2 z_{q^*}^2-\frac{4 \tilde{\gamma} _{13}^2}{27}-\frac{2 \tilde{\gamma} _{14}^2}{9}-\frac{4 \tilde{\gamma} _{15}^2}{9}-\frac{2 \tilde{\gamma} _{28}^2}{3}+2 \tilde{\gamma} _{30}^2-\frac{2}{9} \bar{g}_8^2 \bar{g}_{10}^2-\bar{g}_9^2 \bar{g}_{11}^2\right.\\
%%%%
&\left.\left.+\bar{g}_8^2 \left(-\frac{\bar{g}_{12}^2}{3}-\frac{2 \bar{g}_{14}^2}{3}\right)\right)+M_1^2 \left(\frac{8}{9} \bar{g}_8^2 \bar{g}_{10}^2+\bar{g}_8^2 \left(\frac{4 \bar{g}_{12}^2}{3}+\frac{8 \bar{g}_{14}^2}{3}\right)\right)\right\},
\end{align*}
%%%%%%%%%%%%%%%
%%%%%%%%%%%%%%%
%%%%%%%%%%%%%%%
%%%%%%%%%%%%%%%
%%%%%%%%%%%%%%%
%%%%%%%%%%%%%%%
\begin{align*}
 \beta_{m^2_D}=&\frac{1}{16\pi^2}\left\{4 c_d^2-8 \mu ^2 z_{q^*}^2+\frac{2}{3} m_H^2 \tilde{\gamma} _4+\frac{2}{3} m_Q^2 \tilde{\gamma} _9-\frac{4}{3} m_U^2 \tilde{\gamma} _{13}-\frac{2}{3} m_L^2 \tilde{\gamma} _{17}+\frac{2}{3} m_e^2 \tilde{\gamma} _{18}-\frac{4}{9} M_1^2 \bar{g}_{10}^2\right.\\
%%%
&\left.-\frac{16}{3} M_3^2 \bar{g}_{11}^2+m_D^2 \left(4 z_{q^*}^2-\frac{2 g_1^2}{5}-8 g_3^2+\frac{8 \tilde{\gamma} _{16}}{9}+\frac{2 \bar{g}_{10}^2}{9}+\frac{8 \bar{g}_{11}^2}{3}\right)\right\}+\frac{1}{(16\pi^2)^2}\left\{a_u^2 \left(-2 \tilde{\gamma} _4\right.\right.\\
%%%%
&\left.-\frac{2 \tilde{\gamma} _9}{3}\!+\!\frac{8 \tilde{\gamma} _{13}}{3}\right)\!+c_l^2 \left(\frac{2 \tilde{\gamma} _{17}}{3}-\frac{2 \tilde{\gamma} _4}{3}-\frac{4 \tilde{\gamma} _{18}}{3}\right)-c_d \left(8 c_l \tilde{\gamma} _{29}+\frac{2}{3} M_1 z_{q^*} \bar{g}_2 \bar{g}_5+6 M_2 z_{q^*} \bar{g}_4 \bar{g}_6\right)
\end{align*}
\begin{align*}
%%%%
\phantom{\beta_{m^2_H}=}&+M_2^2 \left(6 z_{q^*}^2 \bar{g}_4^2+6 z_{q^*}^2 \bar{g}_6^2\right)+c_d^2 \left(\frac{28 g_1^2}{15}+24 g_2^2+\frac{16 g_3^2}{3}-12 y_b^2-12 y_t^2-4 y_{\tau }^2-4 z_d^2-4 z_u^2\right.\\
%%%%
&\left.-\frac{10 \tilde{\gamma} _4}{3}-\frac{10 \tilde{\gamma} _9}{9}-\frac{16 \tilde{\gamma} _{16}}{9}+\frac{32 \tilde{\gamma} _{27}}{3}-\bar{g}_1^2-\bar{g}_2^2-3 \bar{g}_3^2-3 \bar{g}_4^2-\frac{\bar{g}_5^2}{9}-3 \bar{g}_6^2-\frac{16 \bar{g}_7^2}{3}\right)\\
%%%%
&+m_U^2 \left(\frac{16 g_1^4}{15}\!+\!\frac{40 g_3^4}{3}\!-\frac{16 \tilde{\gamma} _{13}^2}{27}\!-\frac{8 \tilde{\gamma} _{28}^2}{3}\!+\!\tilde{\gamma} _{13} \left(\frac{16 z_q^2}{3}-\frac{128 g_1^2}{45}-\frac{128 g_3^2}{9}+\frac{32 \bar{g}_8^2}{27}+\frac{32 \bar{g}_9^2}{9}\right)\right)\\
%%%%
&+m_H^2 \left(\frac{2 g_1^4}{5}-\frac{2 \tilde{\gamma} _4^2}{9}+\tilde{\gamma} _4 \left(\frac{4 g_1^2}{5}+4 g_2^2-4 y_b^2-4 y_t^2-\frac{4 y_{\tau }^2}{3}-\frac{\bar{g}_1^2}{3}-\frac{\bar{g}_2^2}{3}-\bar{g}_3^2-\bar{g}_4^2\right)\right.\\
%%%
&\left.-\frac{4}{3} y_b z_{q^*} \bar{g}_2 \bar{g}_{10}\right)+\mu  \left(M_2 \left(-12 z_{q^*}^2 \bar{g}_3 \bar{g}_4-2 \tilde{\gamma} _4 \bar{g}_3 \bar{g}_4\right)+\frac{4}{3} c_d z_d \bar{g}_1 \bar{g}_{10}+M_1 \left(-\frac{2}{3} \tilde{\gamma} _4 \bar{g}_1 \bar{g}_2\right.\right.\\
%%%%
&\left.\left.+\bar{g}_1 \left(\frac{8}{3} y_b z_{q^*} \bar{g}_{10}\!-\bar{g}_2 \left(4 z_{q^*}^2\!+\!\frac{8 \bar{g}_{10}^2}{9}\right)\right)\right)\right)\!+\!\frac{32}{27} M_1 M_3 \bar{g}_{10}^2 \bar{g}_{11}^2+m_Q^2 \left(\frac{2 g_1^4}{15}+\frac{80 g_3^4}{3}-\frac{2 \tilde{\gamma} _9^2}{27}\right.\\
%%%%
&-\frac{16 \tilde{\gamma} _{27}^2}{3}\!-4 \tilde{\gamma} _{29}^2\!+\!\tilde{\gamma} _9 \left(\frac{4 g_1^2}{45}\!+\!4 g_2^2\!+\!\frac{64 g_3^2}{9}\!-\frac{4 z_d^2}{3}\!-\frac{4 z_u^2}{3}-\frac{\bar{g}_5^2}{27}-\bar{g}_6^2-\frac{16 \bar{g}_7^2}{9}\right)+\frac{4}{9} z_d z_{q^*} \bar{g}_5 \bar{g}_{10}\\
%%%%
&\left.-\frac{32}{3} z_d z_{q^*} \bar{g}_7 \bar{g}_{11}\right)\!+\!\mu ^2 \left(4 y_b^2 z_{q^*}^2\!-\frac{24 g_1^4}{25}\!-\frac{16}{5} g_1^2 z_{q^*}^2\!-24 g_2^2 z_{q^*}^2\!+4 y_t^2 z_{q^*}^2+24 z_d^2 z_{q^*}^2+8 z_e^2 z_{q^*}^2\right.\\
%%%%
&+8 z_l^2 z_{q^*}^2+16 z_q^2 z_{q^*}^2+12 z_u^2 z_{q^*}^2+36 z_{q^*}^4+3 z_{q^*}^2 \bar{g}_3^2+6 z_{q^*}^2 \bar{g}_4^2+\frac{1}{9} z_{q^*}^2 \bar{g}_5^2+3 z_{q^*}^2 \bar{g}_6^2+\frac{16}{3} z_{q^*}^2 \bar{g}_7^2\\
%%%%
&-\frac{8}{3} y_b z_{q^*} \bar{g}_2 \bar{g}_{10}+\frac{8}{9} z_d z_{q^*} \bar{g}_5 \bar{g}_{10}+\frac{8}{9} z_d^2 \bar{g}_{10}^2+\bar{g}_1^2 \left(z_{q^*}^2+\frac{2 \bar{g}_{10}^2}{9}\right)+\bar{g}_2^2 \left(2 z_{q^*}^2+\frac{2 \bar{g}_{10}^2}{9}\right)\\
%%%
&\left.-\frac{64}{3} z_d z_{q^*} \bar{g}_7 \bar{g}_{11}\!+\!\frac{32}{3} z_d^2 \bar{g}_{11}^2\right)\!+\!M_3^2 \left(\!-96 g_3^4\!-\!\frac{64}{3} z_d z_{q^*} \bar{g}_7 \bar{g}_{11}\!-48 g_3^2 \bar{g}_{11}^2\!+\!\left(\frac{16 y_b^2}{3}\!+\!\frac{16 z_d^2}{3}\right) \bar{g}_{11}^2\right.\\
%%%%
&\left.+4 \bar{g}_9^2 \bar{g}_{11}^2+\frac{8}{9} \bar{g}_{10}^2 \bar{g}_{11}^2+\frac{124 \bar{g}_{11}^4}{9}+\bar{g}_7^2 \left(\frac{32 z_{q^*}^2}{3}+8 \bar{g}_{11}^2\right)\right)+m_L^2 \left(\frac{2 g_1^4}{5}-\frac{2 \tilde{\gamma} _{17}^2}{9}-4 \tilde{\gamma} _{29}^2\right.\\
%%%%
&\left.+\tilde{\gamma} _{17} \left(\frac{4 z_e^2}{3}\!-\frac{4 g_1^2}{5}\!-4 g_2^2+\frac{\bar{g}_{12}^2}{3}+\bar{g}_{13}^2\right)\right)+m_e^2 \left(\frac{4 g_1^4}{5}-\frac{4 \tilde{\gamma} _{18}^2}{9}-4 \tilde{\gamma} _{29}^2+\tilde{\gamma} _{18} \left(\frac{16 g_1^2}{5}-\frac{8 z_l^2}{3}\right.\right.\\
%%%%
\phantom{\beta_{m^2_D}=}&\left.\left.-\frac{4 \bar{g}_{14}^2}{3}\right)\right)+m_D^2 \left(\frac{1043 g_1^4}{450}-\frac{70 g_3^4}{3}+15 g_2^2 z_{q^*}^2-3 y_b^2 z_{q^*}^2-3 y_t^2 z_{q^*}^2-9 z_d^2 z_{q^*}^2-3 z_e^2 z_{q^*}^2\right.\\
%%%
&-3 z_l^2 z_{q^*}^2-3 z_q^2 z_{q^*}^2-12 z_{q^*}^4-\frac{\tilde{\gamma} _4^2}{18}-\frac{\tilde{\gamma} _9^2}{54}-\frac{4 \tilde{\gamma} _{13}^2}{27}-\frac{20 \tilde{\gamma} _{16}^2}{81}-\frac{\tilde{\gamma} _{17}^2}{18}-\frac{\tilde{\gamma} _{18}^2}{9}-\frac{4 \tilde{\gamma} _{27}^2}{3}-\frac{2 \tilde{\gamma} _{28}^2}{3}\\
%%%%
&+2 \tilde{\gamma} _{29}^2-\frac{9}{4} z_{q^*}^2 \bar{g}_4^2-\frac{9}{4} z_{q^*}^2 \bar{g}_6^2+\frac{4}{3} y_b z_{q^*} \bar{g}_2 \bar{g}_{10}-\frac{4}{9} z_d z_{q^*} \bar{g}_5 \bar{g}_{10}-\frac{1}{12} \bar{g}_1^2 \bar{g}_{10}^2-\frac{2}{9} \bar{g}_8^2 \bar{g}_{10}^2-\frac{2 \bar{g}_{10}^4}{27}\\
%%%%
&+\bar{g}_2^2 \left(-\frac{3 z_{q^*}^2}{4}-\frac{\bar{g}_{10}^2}{12}\right)+\bar{g}_5^2 \left(-\frac{z_{q^*}^2}{12}-\frac{\bar{g}_{10}^2}{36}\right)+\frac{32}{3} z_d z_{q^*} \bar{g}_7 \bar{g}_{11}+\left(-4 y_b^2-4 z_d^2\right) \bar{g}_{11}^2-\bar{g}_9^2 \bar{g}_{11}^2\\
%%%%
&-\frac{11 \bar{g}_{11}^4}{3}+\tilde{\gamma} _{16} \left(\frac{64 g_1^2}{135}+\frac{256 g_3^2}{27}-\frac{32 z_{q^*}^2}{9}-\frac{16 \bar{g}_{10}^2}{81}-\frac{64 \bar{g}_{11}^2}{27}\right)+\bar{g}_7^2 \left(-4 z_{q^*}^2-2 \bar{g}_{11}^2\right)
%%%%
\end{align*}
\begin{align*}
&+g_1^2 \left(\frac{8 g_3^2}{9}+\frac{5 z_{q^*}^2}{3}+\frac{\bar{g}_{10}^2}{27}+\frac{4 \bar{g}_{11}^2}{9}\right)+g_3^2 \left(\frac{40 z_{q^*}^2}{3}+\frac{20 \bar{g}_{10}^2}{27}+\frac{260 \bar{g}_{11}^2}{9}\right)+\bar{g}_{10}^2 \left(-\frac{y_b^2}{3}-\frac{z_d^2}{3}\right.\\
%%%%
&\left.\left.-\frac{4 \bar{g}_{11}^2}{9}-\frac{\bar{g}_{12}^2}{12}-\frac{\bar{g}_{14}^2}{6}\right)\right)+M_1^2 \left(\frac{8}{9} z_d z_{q^*} \bar{g}_5 \bar{g}_{10}-\frac{8}{3} y_b z_{q^*} \bar{g}_2 \bar{g}_{10}+\frac{1}{3} \bar{g}_1^2 \bar{g}_{10}^2+\frac{8}{9} \bar{g}_8^2 \bar{g}_{10}^2+\frac{28 \bar{g}_{10}^4}{81}\right.\\
%%%%
&\left.\left.+\bar{g}_5^2 \left(\frac{2 z_{q^*}^2}{9}+\frac{\bar{g}_{10}^2}{9}\right)+\bar{g}_2^2 \left(2 z_{q^*}^2+\frac{\bar{g}_{10}^2}{3}\right)+\bar{g}_{10}^2 \left(\frac{4 y_b^2}{9}+\frac{4 z_d^2}{9}+\frac{8 \bar{g}_{11}^2}{9}+\frac{\bar{g}_{12}^2}{3}+\frac{2 \bar{g}_{14}^2}{3}\right)\right)\right\},
\end{align*}
%%%%%%%%%%%%%%%
%%%%%%%%%%%%%%%
%%%%%%%%%%%%%%%
%%%%%%%%%%%%%%%
%%%%%%%%%%%%%%%
%%%%%%%%%%%%%%%
\begin{align*}
 \beta_{m^2_L}=&\frac{1}{16\pi^2}\left\{2 c_l^2-4 \mu ^2 z_e^2-m_H^2 \tilde{\gamma} _5-m_Q^2 \tilde{\gamma} _{10}+2 m_U^2 \tilde{\gamma} _{14}-m_D^2 \tilde{\gamma} _{17}-m_e^2 \tilde{\gamma} _{20}-M_1^2 \bar{g}_{12}^2-3 M_2^2 \bar{g}_{13}^2\right.\\
%%%%
&\left.+m_L^2 \left(-\frac{9 g_1^2}{10}-\frac{9 g_2^2}{2}+2 z_e^2+\frac{3 \tilde{\gamma} _{19}}{2}+\frac{\bar{g}_{12}^2}{2}+\frac{3 \bar{g}_{13}^2}{2}\right)\right\}+\frac{1}{(16\pi^2)^2}\left\{a_u^2 \left(3 \tilde{\gamma} _5+\tilde{\gamma} _{10}-4 \tilde{\gamma} _{14}\right)\right.\\
%%%
&+c_d^2 \left(3 \tilde{\gamma} _5\!+\!\tilde{\gamma} _{10}\!+\!2 \tilde{\gamma} _{17}\right)\!-12 c_d c_l \tilde{\gamma} _{29}\!-12 a_u c_l \tilde{\gamma} _{30}\!+m_Q^2 \left(\frac{3 g_1^4}{10}+\frac{45 g_2^4}{2}-\frac{2}{15} g_1^2 \tilde{\gamma} _{10}-6 g_2^2 \tilde{\gamma} _{10}\right.\\
%%%%
&\left.-\frac{32}{3} g_3^2 \tilde{\gamma} _{10}\!+\!2 z_d^2 \tilde{\gamma} _{10}\!+\!2 z_u^2 \tilde{\gamma} _{10}\!-\frac{\tilde{\gamma} _{10}^2}{6}\!-\frac{9 \tilde{\gamma} _{25}^2}{2}-6 \tilde{\gamma} _{29}^2-6 \tilde{\gamma} _{30}^2+\frac{1}{18} \tilde{\gamma} _{10} \bar{g}_5^2+\frac{3}{2} \tilde{\gamma} _{10} \bar{g}_6^2+\frac{8}{3} \tilde{\gamma} _{10} \bar{g}_7^2\right)\\
%%%
&+m_U^2 \left(\frac{12 g_1^4}{5}+\frac{64}{15} g_1^2 \tilde{\gamma} _{14}+\frac{64}{3} g_3^2 \tilde{\gamma} _{14}-8 z_q^2 \tilde{\gamma} _{14}-\frac{4 \tilde{\gamma} _{14}^2}{3}-6 \tilde{\gamma} _{30}^2-\frac{16}{9} \tilde{\gamma} _{14} \bar{g}_8^2-\frac{16}{3} \tilde{\gamma} _{14} \bar{g}_9^2\right)\\
%%%%
&+m_D^2 \left(\frac{3 g_1^4}{5}-\frac{8}{15} g_1^2 \tilde{\gamma} _{17}-\frac{32}{3} g_3^2 \tilde{\gamma} _{17}+4 z_{q^*}^2 \tilde{\gamma} _{17}-\frac{\tilde{\gamma} _{17}^2}{3}-6 \tilde{\gamma} _{29}^2+\frac{2}{9} \tilde{\gamma} _{17} \bar{g}_{10}^2+\frac{8}{3} \tilde{\gamma} _{17} \bar{g}_{11}^2\right)\\
%%%%
&+\frac{3}{2} M_1 M_2 \bar{g}_{12}^2 \bar{g}_{13}^2+M_2^2 \left(-36 g_2^4+3 z_e^2 \bar{g}_4^2-6 y_{\tau } z_e \bar{g}_4 \bar{g}_{13}-18 g_2^2 \bar{g}_{13}^2+\frac{3}{2} y_{\tau }^2 \bar{g}_{13}^2+\frac{3}{2} z_l^2 \bar{g}_{13}^2\right.\\
%%%
&\left.+\frac{9}{4} \bar{g}_3^2 \bar{g}_{13}^2\!+\!\frac{9}{4} \bar{g}_4^2 \bar{g}_{13}^2\!+\!\frac{27}{4} \bar{g}_6^2 \bar{g}_{13}^2\!+\!\frac{9}{8} \bar{g}_{12}^2 \bar{g}_{13}^2+\frac{39 \bar{g}_{13}^4}{8}\right)+m_H^2 \left(\frac{9 g_1^4}{10}+\frac{15 g_2^4}{2}-\frac{6}{5} g_1^2 \tilde{\gamma} _5-6 g_2^2 \tilde{\gamma} _5\right.\\
%%%%
&+6 y_b^2 \tilde{\gamma} _5\!+\!6 y_t^2 \tilde{\gamma} _5\!+\!2 y_{\tau }^2 \tilde{\gamma} _5-\frac{\tilde{\gamma} _5^2}{2}-\frac{3 \tilde{\gamma} _{23}^2}{2}+\frac{1}{2} \tilde{\gamma} _5 \bar{g}_1^2+\frac{1}{2} \tilde{\gamma} _5 \bar{g}_2^2+\frac{3}{2} \tilde{\gamma} _5 \bar{g}_3^2+\frac{3}{2} \tilde{\gamma} _5 \bar{g}_4^2+y_{\tau } z_e \left(\bar{g}_2 \bar{g}_{12}\right.\\
%%%%
&\left.\left.-3 \bar{g}_4 \bar{g}_{13}\right)\right)+\mu  \left(M_1 \left(-2 z_e^2 \bar{g}_1 \bar{g}_2-2 y_{\tau } z_e \bar{g}_1 \bar{g}_{12}+\bar{g}_1 \bar{g}_2 \left(\tilde{\gamma} _5-2 \bar{g}_{12}^2\right)\right)+c_l z_l \left(3 \bar{g}_3 \bar{g}_{13}-\bar{g}_1 \bar{g}_{12}\right)\right.\\
%%%%
&\left.+M_2 \left(-6 z_e^2 \bar{g}_3 \bar{g}_4+6 y_{\tau } z_e \bar{g}_3 \bar{g}_{13}+\bar{g}_3 \bar{g}_4 \left(3 \tilde{\gamma} _5-6 \bar{g}_{13}^2\right)\right)\right)-2 c_l M_1 z_e \bar{g}_2 \bar{g}_{14}+c_l^2 \left(\frac{51 g_1^2}{10}+\frac{3 g_2^2}{2}\right.\\
%%%%
&\left.-6 y_b^2-6 y_t^2-2 y_{\tau }^2-4 z_l^2+2 \tilde{\gamma} _5-\frac{3 \tilde{\gamma} _{19}}{2}+4 \tilde{\gamma} _{20}-3 \tilde{\gamma} _{23}-\frac{\bar{g}_1^2}{2}-\frac{\bar{g}_2^2}{2}-\frac{3 \bar{g}_3^2}{2}-\frac{3 \bar{g}_4^2}{2}-2 \bar{g}_{14}^2\right)\\
%%%%
&+m_e^2 \left(\frac{9 g_1^4}{5}-\frac{24}{5} g_1^2 \tilde{\gamma} _{20}+4 z_l^2 \tilde{\gamma} _{20}-\tilde{\gamma} _{20}^2-6 \tilde{\gamma} _{29}^2-6 \tilde{\gamma} _{30}^2-2 z_e z_l \bar{g}_{12} \bar{g}_{14}+2 \tilde{\gamma} _{20} \bar{g}_{14}^2\right)\\
%%%%
&+m_L^2 \left(\frac{2231 g_1^4}{400}+\frac{63 g_2^4}{16}-\frac{9 z_e^4}{2}-\frac{\tilde{\gamma} _5^2}{8}-\frac{\tilde{\gamma} _{10}^2}{24}-\frac{\tilde{\gamma} _{14}^2}{3}-\frac{\tilde{\gamma} _{17}^2}{12}-\frac{15 \tilde{\gamma} _{19}^2}{16}-\frac{\tilde{\gamma} _{20}^2}{4}-\frac{3 \tilde{\gamma} _{23}^2}{8}-\frac{9 \tilde{\gamma} _{25}^2}{8}\right.\\
%%%%
&+3 \tilde{\gamma} _{29}^2+3 \tilde{\gamma} _{30}^2-\frac{3}{16} \bar{g}_1^2 \bar{g}_{12}^2-\frac{3}{16} \bar{g}_2^2 \bar{g}_{12}^2-\frac{1}{16} \bar{g}_5^2 \bar{g}_{12}^2-\frac{1}{2} \bar{g}_8^2 \bar{g}_{12}^2-\frac{1}{8} \bar{g}_{10}^2 \bar{g}_{12}^2-\frac{9 \bar{g}_{12}^4}{32}-\frac{9}{4} \tilde{\gamma} _{19} \bar{g}_{13}^2\\
%%%%
\end{align*}
\begin{align*}
\phantom{\beta_{m^2_L}=}&-\frac{9}{16} \bar{g}_3^2 \bar{g}_{13}^2-\frac{9}{16} \bar{g}_4^2 \bar{g}_{13}^2-\frac{27}{16} \bar{g}_6^2 \bar{g}_{13}^2-\frac{45 \bar{g}_{13}^4}{32}+y_{\tau } z_e \left(3 \bar{g}_4 \bar{g}_{13}-\bar{g}_2 \bar{g}_{12}\right)-z_l^2 \left(\frac{3 \bar{g}_{12}^2}{8}+\frac{9 \bar{g}_{13}^2}{8}\right)\\
%%%%
&+y_{\tau }^2 \left(-3 z_e^2-\frac{3 \bar{g}_{12}^2}{8}-\frac{9 \bar{g}_{13}^2}{8}\right)+g_1^2 \left(\frac{9 g_2^2}{8}+\frac{15 z_e^2}{4}+\frac{9 \tilde{\gamma} _{19}}{5}+\frac{3 \bar{g}_{12}^2}{16}+\frac{9 \bar{g}_{13}^2}{16}\right)+g_2^2 \left(\frac{15 z_e^2}{4}\right.\\
%%%%
&\left.+9 \tilde{\gamma} _{19}+\frac{15 \bar{g}_{12}^2}{16}+\frac{165 \bar{g}_{13}^2}{16}\right)+2 z_e z_l \bar{g}_{12} \bar{g}_{14}-z_e^2 \left(\frac{9 z_d^2}{2}+\frac{3 z_l^2}{2}+\frac{9 z_{q^*}^2}{2}+3 \tilde{\gamma} _{19}+\frac{3 \bar{g}_2^2}{8}+\frac{9 \bar{g}_4^2}{8}\right.\\
%%%
&\left.\left.+\frac{3 \bar{g}_{14}^2}{2}\right)-\bar{g}_{12}^2 \left(\frac{3 \tilde{\gamma} _{19}}{4}+\frac{9 \bar{g}_{13}^2}{16}+\frac{3 \bar{g}_{14}^2}{8}\right)\right)+\mu ^2 \left(4 y_{\tau }^2 z_e^2-\frac{54 g_1^4}{25}-18 g_2^4-\frac{18}{5} g_1^2 z_e^2-6 g_2^2 z_e^2\right.\\
%%%%
&+16 z_e^4\!+\!\frac{1}{2} \bar{g}_1^2 \bar{g}_{12}^2\!+\!\frac{1}{2} \bar{g}_2^2 \bar{g}_{12}^2+\frac{3}{2} \bar{g}_3^2 \bar{g}_{13}^2+\frac{3}{2} \bar{g}_4^2 \bar{g}_{13}^2+y_{\tau } z_e \left(2 \bar{g}_2 \bar{g}_{12}-6 \bar{g}_4 \bar{g}_{13}\right)+z_l^2 \left(\bar{g}_{12}^2+3 \bar{g}_{13}^2\right)\\
%%%%
&\left.-4 z_e z_l \bar{g}_{12} \bar{g}_{14}+z_e^2 \left(12 z_d^2+4 z_l^2+6 z_q^2+6 z_u^2+12 z_{q^*}^2+\frac{\bar{g}_1^2}{2}+\bar{g}_2^2+\frac{3 \bar{g}_3^2}{2}+3 \bar{g}_4^2+2 \bar{g}_{14}^2\right)\right)\\
%%%%
&+M_1^2 \left(2 y_{\tau } z_e \bar{g}_2 \bar{g}_{12}+\frac{1}{2} y_{\tau }^2 \bar{g}_{12}^2+\frac{1}{2} z_l^2 \bar{g}_{12}^2+\frac{3}{4} \bar{g}_1^2 \bar{g}_{12}^2+\frac{3}{4} \bar{g}_2^2 \bar{g}_{12}^2+\frac{1}{4} \bar{g}_5^2 \bar{g}_{12}^2+2 \bar{g}_8^2 \bar{g}_{12}^2+\frac{1}{2} \bar{g}_{10}^2 \bar{g}_{12}^2\right.\\
%%%%
&\left.\left.+\frac{11 \bar{g}_{12}^4}{8}-4 z_e z_l \bar{g}_{12} \bar{g}_{14}+\bar{g}_{12}^2 \left(\frac{9 \bar{g}_{13}^2}{8}+\frac{3 \bar{g}_{14}^2}{2}\right)+z_e^2 \left(\bar{g}_2^2+4 \bar{g}_{14}^2\right)\right)\right\},
\end{align*}
%%%%%%%%%%%%%%%
%%%%%%%%%%%%%%%
%%%%%%%%%%%%%%%
%%%%%%%%%%%%%%%
%%%%%%%%%%%%%%%
%%%%%%%%%%%%%%%
\begin{align*}
 \beta_{m^2_E}=&\frac{1}{16\pi^2}\left\{4 c_l^2-8 \mu ^2 z_l^2+2 m_H^2 \tilde{\gamma} _6+2 m_Q^2 \tilde{\gamma} _{11}-4 m_U^2 \tilde{\gamma} _{15}+2 m_D^2 \tilde{\gamma} _{18}-2 m_L^2 \tilde{\gamma} _{20}-4 M_1^2 \bar{g}_{14}^2\right.\\
%%%%
&\left.+m_e^2 \left(-\frac{18 g_1^2}{5}+4 z_l^2+4 \tilde{\gamma} _{21}+2 \bar{g}_{14}^2\right)\right\}+\frac{1}{(16\pi^2)^2}\left\{a_u^2 \left(-6 \tilde{\gamma} _6-2 \tilde{\gamma} _{11}+8 \tilde{\gamma} _{15}\right)+c_d^2 \left(-6 \tilde{\gamma} _6\right.\right.\\
%%%
&\left.-2 \tilde{\gamma} _{11}-4 \tilde{\gamma} _{18}\right)-24 c_d c_l \tilde{\gamma} _{29}-24 a_u c_l \tilde{\gamma} _{30}+m_Q^2 \left(\frac{6 g_1^4}{5}+\frac{4}{15} g_1^2 \tilde{\gamma} _{11}+12 g_2^2 \tilde{\gamma} _{11}+\frac{64}{3} g_3^2 \tilde{\gamma} _{11}\right.\\
%%%%
&\left.-4 z_d^2 \tilde{\gamma} _{11}\!-4 z_u^2 \tilde{\gamma} _{11}\!-\frac{2 \tilde{\gamma} _{11}^2}{3}\!-12 \tilde{\gamma} _{29}^2\!-12 \tilde{\gamma} _{30}^2\!-\frac{1}{9} \tilde{\gamma} _{11} \bar{g}_5^2-3 \tilde{\gamma} _{11} \bar{g}_6^2-\frac{16}{3} \tilde{\gamma} _{11} \bar{g}_7^2\right)+m_U^2 \left(\frac{48 g_1^4}{5}\right.\\
%%%%
&\left.-\frac{128}{15} g_1^2 \tilde{\gamma} _{15}\!-\frac{128}{3} g_3^2 \tilde{\gamma} _{15}\!+\!16 z_q^2 \tilde{\gamma} _{15}\!-\frac{16 \tilde{\gamma} _{15}^2}{3}\!-12 \tilde{\gamma} _{30}^2\!+\!\frac{32}{9} \tilde{\gamma} _{15} \bar{g}_8^2+\frac{32}{3} \tilde{\gamma} _{15} \bar{g}_9^2\right)+m_D^2 \left(\frac{12 g_1^4}{5}\right.\\
%%%
&\left.+\frac{16}{15} g_1^2 \tilde{\gamma} _{18}\!+\!\frac{64}{3} g_3^2 \tilde{\gamma} _{18}\!-8 z_{q^*}^2 \tilde{\gamma} _{18}\!-\frac{4 \tilde{\gamma} _{18}^2}{3}-12 \tilde{\gamma} _{29}^2-\frac{4}{9} \tilde{\gamma} _{18} \bar{g}_{10}^2-\frac{16}{3} \tilde{\gamma} _{18} \bar{g}_{11}^2\right)+c_l \left(2 M_1 z_l \bar{g}_2 \bar{g}_{12}\right.\\
%%%
&\left.-6 M_2 z_l \bar{g}_4 \bar{g}_{13}\right)+c_l^2 \left(24 g_2^2-\frac{12 g_1^2}{5}-12 y_b^2-12 y_t^2-4 y_{\tau }^2-4 z_e^2-6 \tilde{\gamma} _6\!+6 \tilde{\gamma} _{20}-8 \tilde{\gamma} _{21}-\bar{g}_1^2\right.\\
%%%%
&\left.-\bar{g}_2^2-3 \bar{g}_3^2-3 \bar{g}_4^2-\bar{g}_{12}^2-3 \bar{g}_{13}^2\right)+M_2^2 z_l^2 \left(6 \bar{g}_4^2+6 \bar{g}_{13}^2\right)+m_H^2 \left(\frac{18 g_1^4}{5}+\frac{12}{5} g_1^2 \tilde{\gamma} _6+12 g_2^2 \tilde{\gamma} _6\right.\\
%%%%
&\left.-12 y_b^2 \tilde{\gamma} _6-12 y_t^2 \tilde{\gamma} _6-4 y_{\tau }^2 \tilde{\gamma} _6-2 \tilde{\gamma} _6^2-\tilde{\gamma} _6 \bar{g}_1^2-\tilde{\gamma} _6 \bar{g}_2^2-3 \tilde{\gamma} _6 \bar{g}_3^2-3 \tilde{\gamma} _6 \bar{g}_4^2-4 y_{\tau } z_l \bar{g}_2 \bar{g}_{14}\right)\\
%%%%
&+m_L^2 \left(\frac{18 g_1^4}{5}-\frac{12}{5} g_1^2 \tilde{\gamma} _{20}-12 g_2^2 \tilde{\gamma} _{20}+4 z_e^2 \tilde{\gamma} _{20}-2 \tilde{\gamma} _{20}^2-12 \tilde{\gamma} _{29}^2-12 \tilde{\gamma} _{30}^2+\tilde{\gamma} _{20} \bar{g}_{12}^2+3 \tilde{\gamma} _{20} \bar{g}_{13}^2\right.\\
%%%%
\end{align*}
\begin{align*}
%%%%
\phantom{\beta_{m^2_E}=}
&\left.-4 z_e z_l \bar{g}_{12} \bar{g}_{14}\right)+M_1^2 \left(z_l^2 \left(2 \bar{g}_2^2+2 \bar{g}_{12}^2\right)-8 y_{\tau } z_l \bar{g}_2 \bar{g}_{14}-8 z_e z_l \bar{g}_{12} \bar{g}_{14}+4 y_{\tau }^2 \bar{g}_{14}^2+4 z_e^2 \bar{g}_{14}^2\right.\\
%%%%
&\left.+3 \bar{g}_1^2 \bar{g}_{14}^2\!+\!3 \bar{g}_2^2 \bar{g}_{14}^2\!+\!\bar{g}_5^2 \bar{g}_{14}^2\!+\!8 \bar{g}_8^2 \bar{g}_{14}^2\!+\!2 \bar{g}_{10}^2 \bar{g}_{14}^2+3 \bar{g}_{12}^2 \bar{g}_{14}^2+16 \bar{g}_{14}^4\right)+m_e^2 \left(\frac{1363 g_1^4}{50}+15 g_2^2 z_l^2\right.\\
%%%%
&-6 z_l^4\!-\frac{\tilde{\gamma} _6^2}{2}\!-\frac{\tilde{\gamma} _{11}^2}{6}\!-\frac{4 \tilde{\gamma} _{15}^2}{3}\!-\frac{\tilde{\gamma} _{18}^2}{3}-\frac{\tilde{\gamma} _{20}^2}{2}-10 \tilde{\gamma} _{21}^2+6 \tilde{\gamma} _{29}^2+6 \tilde{\gamma} _{30}^2+z_l^2 \left(-9 z_d^2-9 z_{q^*}^2-16 \tilde{\gamma} _{21}\right.\\
%%%%
&\left.-\frac{3 \bar{g}_2^2}{4}-\frac{9 \bar{g}_4^2}{4}-\frac{3 \bar{g}_{12}^2}{4}-\frac{9 \bar{g}_{13}^2}{4}\right)+4 y_{\tau } z_l \bar{g}_2 \bar{g}_{14}+4 z_e z_l \bar{g}_{12} \bar{g}_{14}-8 \tilde{\gamma} _{21} \bar{g}_{14}^2-\frac{3}{4} \bar{g}_1^2 \bar{g}_{14}^2-\frac{3}{4} \bar{g}_2^2 \bar{g}_{14}^2\\
%%%%
&-\frac{1}{4} \bar{g}_5^2 \bar{g}_{14}^2-2 \bar{g}_8^2 \bar{g}_{14}^2-\frac{1}{2} \bar{g}_{10}^2 \bar{g}_{14}^2-\frac{3}{4} \bar{g}_{12}^2 \bar{g}_{14}^2-3 \bar{g}_{14}^4+y_{\tau }^2 \left(-3 z_l^2-3 \bar{g}_{14}^2\right)-z_e^2 \left(3 z_l^2+3 \bar{g}_{14}^2\right)\\
%%%%
&\left.+g_1^2 \left(3 z_l^2+\frac{96 \tilde{\gamma} _{21}}{5}+3 \bar{g}_{14}^2\right)\right)+\mu ^2 \left(4 y_{\tau }^2 z_l^2-\frac{216 g_1^4}{25}-24 g_2^2 z_l^2+20 z_l^4+z_l^2 \left(24 z_d^2\!+\!12 z_q^2\right.\right.\\
%%%%
&\left.+12 z_u^2\!+\!24 z_{q^*}^2\!+\!\bar{g}_1^2\!+\!2 \bar{g}_2^2\!+\!3 \bar{g}_3^2+6 \bar{g}_4^2+\bar{g}_{12}^2+3 \bar{g}_{13}^2\right)-8 y_{\tau } z_l \bar{g}_2 \bar{g}_{14}-8 z_e z_l \bar{g}_{12} \bar{g}_{14}\!+2 \bar{g}_1^2 \bar{g}_{14}^2\\
%%%%
&\left.+2 \bar{g}_2^2 \bar{g}_{14}^2\!+\!z_e^2 \left(8 z_l^2\!+\!8 \bar{g}_{14}^2\right)\right)\!+\!\mu  \left(\!-M_2 \left(12 z_l^2 \bar{g}_3 \bar{g}_4+6 \tilde{\gamma} _6 \bar{g}_3 \bar{g}_4\right)\!+4 c_l z_e \bar{g}_1 \bar{g}_{14}\!+\!M_1 \left(-4 z_l^2 \bar{g}_1 \bar{g}_2\right.\right.\\
%%%%
&\left.\left.\left.+8 y_{\tau } z_l \bar{g}_1 \bar{g}_{14}+\bar{g}_1 \bar{g}_2 \left(-2 \tilde{\gamma} _6-8 \bar{g}_{14}^2\right)\right)\right)\right\}.
\end{align*}
%%%%%%%%%%%%%%%%%%%%%%%%%%%%%%%%%%%%%%%%%%%%%%%%%%%%%%%%%%%%%%%%%%%%%%%%%%%%%%%%%
%%%%%%%%%%%%%%%%%%%%%%%%%%%%%%%%%%%%%%%%%%%%%%%%%%%%%%%%%%%%%%%%%%%%%%%%%%%%%%%%%
%%%%%%%%%%%%%%%%%%%%%%%%%%%%%%%%%%%%%%%%%%%%%%%%%%%%%%%%%%%%%%%%%%%%%%%%%%%%%%%%%
\subsubsection{Higgs anomalous dimension}
%%%%%%%%%%%%%%%%%%%%%%%%%%%%%%%%%%%%%%%%%%%%%%%%%%%%%%%%%%%%%%%%%%%%%%%%%%%%%%%%%
%%%%%%%%%%%%%%%%%%%%%%%%%%%%%%%%%%%%%%%%%%%%%%%%%%%%%%%%%%%%%%%%%%%%%%%%%%%%%%%%%
%%%%%%%%%%%%%%%%%%%%%%%%%%%%%%%%%%%%%%%%%%%%%%%%%%%%%%%%%%%%%%%%%%%%%%%%%%%%%%%%%
\begin{align*}
\gamma_H=&\gamma_H^{MES}+\frac{1}{(16\pi^2)^2}\left\{\bar{g}_2 \bar{g}_{10} y_b z_{q^*}-\frac{1}{4} \bar{g}_{10}^2 y_b^2-3 \bar{g}_{11}^2 y_b^2+\bar{g}_2^2 \left(-\frac{\bar{g}_{10}^2}{16}-\frac{3 \bar{g}_{14}^2}{16}-\frac{3 \bar{g}_{12}^2}{32}-\frac{1}{16} 3 z_e^2\right.\right.\\
%%%%
&\left.-\frac{3 z_l^2}{16}-\frac{9 z_{q^*}^2}{16}\right)+\bar{g}_4^2 \left(-\frac{9 \bar{g}_{13}^2}{32}-\frac{1}{16} 9 z_e^2-\frac{9 z_l^2}{16}-\frac{27 z_{q^*}^2}{16}\right)+y_{\tau }^2 \left(-\frac{3 \bar{g}_{14}^2}{4}-\frac{3 \bar{g}_{12}^2}{16}-\frac{9 \bar{g}_{13}^2}{16}\right.\\
%%%%
&\left.-\frac{3 z_e^2}{2} -\frac{3 z_l^2}{4}\right)+y_{\tau } \left(\bar{g}_2 \left(\bar{g}_{14} z_l-\frac{1}{2} \bar{g}_{12} z_e\right)+\frac{3}{2} \bar{g}_4 \bar{g}_{13} z_e\right)+\left(-\frac{\bar{g}_{10}^2}{16}-\frac{3 \bar{g}_{14}^2}{16}-\frac{3 \bar{g}_{12}^2}{32}\right) \bar{g}_1^2\\
%%%%
&\left.-\frac{9}{32} \bar{g}_3^2 \bar{g}_{13}^2+\frac{\tilde{\gamma }_4^2}{24}+\frac{\tilde{\gamma }_5^2}{16}+\frac{\tilde{\gamma }_6^2}{8}+\frac{3 \tilde{\gamma }_{23}^2}{16}-\frac{9}{4} y_b^2 z_{q^*}^2+\frac{121 g_1^4}{400}+\frac{11 g_2^4}{16}-\frac{9}{4} z_{q^*}^2 y_t^2\right\}.
\end{align*}
%%%%%%%%%%%%%%%%%%%%%%%%%%%%%%%%%%%%%%%%%%%%%%%%%%%%%%%%%%%%%%%%%%%%%%%%%%%%%%%%%
%%%%%%%%%%%%%%%%%%%%%%%%%%%%%%%%%%%%%%%%%%%%%%%%%%%%%%%%%%%%%%%%%%%%%%%%%%%%%%%%%
%%%%%%%%%%%%%%%%%%%%%%%%%%%%%%%%%%%%%%%%%%%%%%%%%%%%%%%%%%%%%%%%%%%%%%%%%%%%%%%%%
%%%%%%%%%%%%%%%%%%%%%%%%%%%%%%%%%%%%%%%%%%%%%%%%%%%%%%%%%%%%%%%%%%%%%%%%%%%%%%%%%
%%%%%%%%%%%%%%%%%%%%%%%%%%%%%%%%%%%%%%%%%%%%%%%%%%%%%%%%%%%%%%%%%%%%%%%%%%%%%%%%%

\section{Final comments \label{sec:finalcomments}}

This paper presents results for the beta functions of the low energy theories that result from decoupling heavy sparticles in the following hierarchical SUSY scenarios: Split SUSY and minimal as well as nonminimal Effective SUSY realizations.

 What is the expected range of validity of the low energy effective theories? The answer can be estimated by means of considerations about fine-tuning, at least for the Effective SUSY scenarios, which still allow for naturalness. The low-energy effective theories by themselves do not solve the hierarchy problem associated with the sensitivity of the Higgs mass parameters with respect to potential new Physics at higher scales; rather, this problem is fixed by the corresponding  supersymmetric ultraviolet completions. The knowledge of the extra matter content of the latter can be used to estimate upper values for the cutoffs of the effective theories for a given amount of fine-tuning. Rather than using quadratic divergences, which are renormalization-scheme dependent --and in fact zero in dimensional regularization-- contributions to fine-tuning can be crudely estimated by the size of finite threshold corrections due to additional heavy particles not present in the low energy theory. These finite threshold 
corrections take the form $$\delta_M m^2_H\sim \frac{\tilde\lambda}{16 \pi^2}M^2,$$
where $M$ is the mass of a potential new heavy state and $\tilde\lambda$ parametrizes its coupling to the Higgs field (e.g. $\tilde\lambda\sim y^2$ for the MSSM quartic scalar interactions, $y$ denoting a Yukawa coupling).

As such, if $\delta m^2_H< 50 {\rm GeV}^2$,  particles coupling with the top Yukawa should not be heavier than about 600 GeV, which is the usual expectation for the stop. This of course automatically renders Split SUSY scenarios fine-tuned, so that naturalness arguments do not allow to fix the cutoff; rather, one may obtain bounds by demanding that the gluinos decay sufficiently fast so as to avoid constraints from heavy isotopes, which results in a bound of about $\Lambda \lesssim 10^{13}$ GeV \cite{Giudice:2004tc}. Concerning Effective SUSY scenarios, the supersymmetric ultraviolet completion adds to the low energy fields the sbottom (in minimal scenarios), as well as the first and second generation scalars; these fields couple to the Higgs with  Yukawas that are around a factor 40 or more smaller than the top Yukawa, so that the cutoffs for the low energy effective SUSY scenarios can be as heavy as 20-60 TeV without incurring into significant fine-tuning. 

Of course, above the mass scales of the MSSM there could be further heavy states, e.g. a hidden sector, but as long as they belong to supersymmetric multiplets the threshold corrections will cancel up to susy breaking effects.

As it was said in the introduction, the implementation of decoupling allows to resum in the low energy RG flow some of the large finite corrections due to the heavy particles that were integrated out. Due to this, significant deviations are expected between the MSSM RG and the decoupled RG using the beta functions presented in this paper. As an illustration, fig.~\ref{fig:1} shows the 2 loop running of ${m^2_Q}_{33}$ and  ${m^2_L}_{33}$ in a minimal and nonminimal Effective SUSY scenario, respectively, comparing the flows with and without decoupling; the differences between the two reach $\sim 1\,{\rm TeV}^2$ at low energies, which serves as an estimation of the size of the finite corrections caused by the heavy fields in the $\overline{\rm DR}$ scheme with no decoupling implemented. As it was anticipated, these finite corrections become very large and put into question the reliability of perturbation theory. These effects are particularly relevant in the case of very light third generation sparticles in 
Effective SUSY scenarios. The MSSM  $\overline{\rm DR}$ RG effects of the heavy first and third generation sparticles tends to drive the third generation scalar masses tachyonic, as in fig.~\ref{fig:1}, but large finite corrections may drive them positive --however, computing the finite corrections when the tree-level masses are tachyonic may be problematic, and for this reason many spectrum calculators, which use the non-decoupling $\overline{\rm DR}$ RG flow, discard scenarios in which this happens. Allowed scenarios only have positive tree-level masses, but then the finite corrections render them rather large, and so very light masses are not probed. These problems are much ameliorated by using the decoupled flow, which automatically resums these corrections --for example in fig.~\ref{fig:1} the running masses in the decoupled flow stay positive. Therefore the parameter space can be enlarged by implementing decoupling,  particularly in the region of small third generation sparticle masses, which is of particular interest due to the poor current constraints from 
the LHC \cite{Han:2012fw}.

Some phenomenological aspects of Effective SUSY scenarios using the decoupled RG equations of this paper are analyzed in ref.~\cite{Tamarit:2012ry}.

{\bf Update}: After this paper was written, results for an independent calculation of two-loop beta functions for Split SUSY have been given in ref.~\cite{Benakli:2013msa}. Those results coincide with the ones presented here for dimensionless couplings, though differed in the two-loop contributions for the beta functions of gaugino masses in previous versions of this paper. These have been corrected and now there is full agreement with ref.~\cite{Benakli:2013msa}.

%%%%%%%%%%%%%%%%%%%%%%%%%%%%%%%%%%%%%%%%%%%%%%%%%%%%%%%%%%%%%%%%%%%%%%%%%%%%%%%%%
%%%%%%%%%%%%%%%%%%%%%%%%%%%%%%%%%%%%%%%%%%%%%%%%%%%%%%%%%%%%%%%%%%%%%%%%%%%%%%%%%
%%%%%%%%%%%%%%%%%%%%%%%%%%%%%%%%%%%%%%%%%%%%%%%%%%%%%%%%%%%%%%%%%%%%%%%%%%%%%%%%%
%%%%%%%%%%%%%%%%%%%%%%%%%%%%%%%%%%%%%%%%%%%%%%%%%%%%%%%%%%%%%%%%%%%%%%%%%%%%%%%%%
%%%%%%%%%%%%%%%%%%%%%%%%%%%%%%%%%%%%%%%%%%%%%%%%%%%%%%%%%%%%%%%%%%%%%%%%%%%%%%%%%

\section*{Acknowledgements}
The author wishes to thank Sakura Sch\"afer-Nameki, Natalia Toro and Philip Schuster for useful conversations. Research
at the Perimeter Institute is supported in part by the Government of Canada through
NSERC and by the Province of Ontario through MEDT. This work was financed in part by the Spanish Ministry of Science and Innovation through project FPA2011-24568

\bibliographystyle{h-physrev}
\bibliography{susybetafunctions2ref}

\end{document}